%% file: thesis.tex
\documentstyle[11pt,charu,twoside,psfig]{book}
\renewcommand{\thesection}{\thechapter.\arabic{section}}

\newcommand{\bb}{\begin{equation}}
\newcommand{\ee}{\end{equation}}
\newcommand{\bbb}{\begin{eqnarray}}
\newcommand{\eee}{\end{eqnarray}}
\newcommand{\gcc}{g~cm$^{-3}\ $}

\newcommand{\msun}{$M_{\odot}\ $}
\newcommand{\greq}{$\stackrel{>}{ _{\sim}}$}

\newcommand{\lsim}{\raisebox{-0.3ex}{\mbox{$\stackrel{<}{_\sim} \,$}}}
\newcommand{\gsim}{\raisebox{-0.3ex}{\mbox{$\stackrel{>}{_\sim} \,$}}}
\def\beb{\begin{thebibliography}{99}}
\def\eeb{\end{thebibliography}}

\def \msun {M_{\odot}}

\def\fun#1#2{\lower3.6pt\vbox{\baselineskip0pt\lineskip.9pt
  \ialign{$\mathsurround=0pt#1\hfil##\hfil$\crcr#2\crcr\sim\crcr}}}

\begin{document}
\bibliographystyle{alpha}
\baselineskip=18pt
\def\up{\vspace{-0.5\baselineskip}}
\def\dn{\vspace{0.5\baselineskip}}
\include{frontpage}

\include{decln}

\setcounter{page}{1}
\include{dedication}
\newpage
\setcounter{page}{1}
\pagenumbering{roman}
\include{ackn}

\newpage
\setcounter{page}{1}
\pagenumbering{roman}
\tableofcontents
\newpage
\pagenumbering{arabic}
\setcounter{page}{1}
\include{chap1}

\include{chap2}

\include{chap3}

\include{chap4}

\include{chap5}

\include{chap6}

\include{chap7}

\include{chap8}

\include{biblio}

\include{pub}

\end{document}

%% file: frontpage.tex
\thispagestyle{empty}
\begin{center}
{\huge \bf
Temperature Profiles and Spectra of Accretion Disks around 
Rapidly Rotating Neutron Stars} 

\vspace{3cm}

{\Large A Thesis\\ Submitted for the Degree of \\ {\bf Doctor of Philosophy}\\
in the Faculty of Science}

\vspace{3cm}



{\Large by}\\
\vspace{0.5cm}
{\LARGE \bf Sudip Bhattacharyya}

\vspace{1cm}

\vfill

\centerline{\psfig{figure=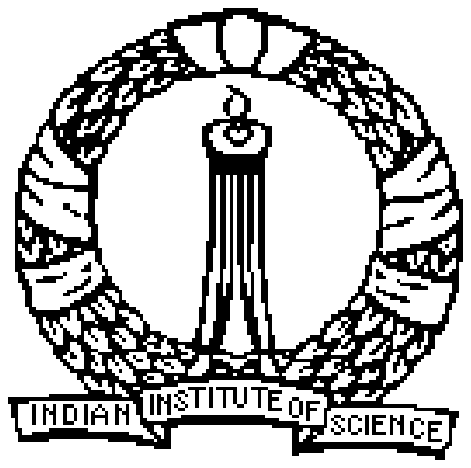,width=2.5cm,angle=0.15}}

\vskip0.5cm

{\Large Department of Physics\\Indian Institute of Science \\ Bangalore 
INDIA \\ \vspace{0.5cm}2001}

\end{center}

%% file: decln.tex
\vspace{1cm}
\begin{center}
{\LARGE Declaration}
\end{center}
\thispagestyle{empty}
\vspace{1cm}

I  hereby declare that the work reported in this thesis 
titled ``Temperature Profiles and Spectra of Accretion Disks around
Rapidly Rotating Neutron Stars''
is entirely original and has been carried out by me independently
in the Department of Physics, Indian Institute of Science,
and Indian Institute of Astrophysics, 
under the Joint Astronomy Programme,
under the supervisions of Dr. Pijushpani Bhattacharjee,
Indian Institute of Astrophysics, Bangalore 560~034 and Dr. Arnab Rai 
Choudhuri, Indian Institute of Science, Bangalore 560~012.
I further declare that this work has not formed
the basis for the award of any degree, diploma, fellowship,
associateship or similar title of any University or Institution.

\begin{flushright}
\vspace*{2cm}
(Sudip Bhattacharyya)
\end{flushright}

\begin{flushright}

Department of Physics\\

Indian Institute of Science\\

Bangalore 560 012, India.\\

\end{flushright}

\pagebreak

%% file: dedication.tex
\thispagestyle{empty}
\vspace*{11cm}
\begin{flushright}
{\huge To the memory of Professor Bhaskar Datta}
\end{flushright}


\pagebreak

%% file: ackn.tex
\vspace{1cm}
\begin{center}
{\LARGE Acknowledgments}
\end{center}
\thispagestyle{empty}
\vspace{1cm}

I thank late Dr. Bhaskar Datta for his invaluable help and guidance in the early years of
my ph.d. work. His sad and untimely demise on 3rd December, 1999 has been the rudest shock
I have ever received in my life.

My heartful thanks are to Dr. Pijush Bhattacharjee for stepping in as my official advisor after
Dr. Datta's death and for constant encouragement and academic help that he has extended towards
me in the later years of my ph.d. work.

I wish to thank Dr. Arnab Rai Choudhuri for the academic discussions that we have had 
and his support throughout my ph.d. years. I also thank Dr. Chanda J. Jog for her help.

I am grateful to Dr. Dipankar Bhattacharya for having helped me in all possible ways after
the demise of Dr. Bhaskar Datta. It is with his help that I have been able to complete my
ph.d. work in reasonable time.

Dr. Arun V. Thampan has stood beside me from the early days of my ph.d. work. I have used his
numerical code extensively. It is a pleasure to thank him for his support.

I thank my collaborators Dr. Ranjeev Misra and Dr. Ignazio Bombaci for their cooperation
academically and otherwise.

I would like to acknowledge the Director, Indian Institute of Astrophysics, for all the 
facilities provided. I am indebted to the faculty and scientific staff of IIA for everything
that they have done for me. I thank the staff of the Director's office, the Librarian and the 
library staff, the computer center staff and the administrative staff for trying their best
to ensure that the road towards attaining my ph.d. degree has been smooth.

I thank the Chairman, the faculty and the students of physics department, IISc. I acknowledge the
staff of the physics office, especially Rakma, for their help. I also thank the Chairman of
SERC, IISc and the staff there for all the state-of-the-art computation facilities provided.

I wish to thank the Director of Raman Research Institute for all the facilities provided to
me. I acknowledge all the faculty members, the students, the computer center staff,
the library staff and in general all my colleagues in RRI.

I thank Dr. Sreekumar and Vivek for helping me learn X--ray data analysis. I also thank Lolita,
Ishwar and the rest of the staff of Technical Physics Division, ISRO for their support.

The discussions with Drs. Bala Iyer, Paul Wiita, Ajit Kembhavi and Sandip Chakrabarti were
extremely helpful for me and I thank all of them.

I thank all the past and present IISc astronomy students, especially 
Srikanth, Sushan, Rajguru, Dharam, Ashish, 
Bhaswati, Dibyendu, Girish and Resmi for their help and friendship.

I acknowledge my batchmates Subhabrata and Niruj for always being 
understanding and helpful to me.

I am indebted to all the students, past and present, in IIA. I specially want to thank
Rajesh, Manoj, Ravindra, Maheswar, Pavan, Sonjoy, Jana, Mangala, Sridharan, Rajalakshmi, 
Ramchandra, Dilip and Preeti. I also thank Dr. Parthasarathy Joarder for his help.

I would like to thank Banibrata for all our discussions and I sincerely hope that he would 
be able to overcome his present predicament soon.

I am thankful to Ayan for his friendship and company throughout my ph.d. years. I also thank
Apratim, Sutirtha, Sudipta and Sriparna for making my life in IISc so much enjoyable. I will
never forget the many wonderful saturday evenings that we had in Hotel Ratna -- which has 
been like a home outside home for me, with its quality food and general ambience.
The weekend screenings of IISc Film Society and film festivals organized by the Bangalore
Film Society have, besides exposing me to the best cinema of the world, provided the much-needed
breaks from academics.

Last but not the least, I am grateful to my parents and brother for their support throughout
my life. Without them, I would not be what I am now.

%% file: chap1.tex
\markright{Chapter 1}
\def\note #1]{{\bf #1]}}
\chapter{Introduction}

\section{General Introduction}

Astronomy stepped into a new era with the discovery of discrete galactic X--ray
sources (for example, Scorpius X-1) by the rocket-borne Geiger counters in 1962
(Giacconi et al. 1962). Subsequent rocket and Balloon flights confirmed this
result. Before 1970, about 20 X--ray sources had been identified, with most of
them believed to be galactic sources. The binary nature of these galactic X--ray
sources was established (Schreier et al. 1972; Tananbaum et al. 1972) 
by the first astronomy satellite {\it Uhuru} (launched
by NASA in 1970). The same satellite also discovered binary X--ray pulsars.
Later X--ray and optical observations confirmed that there are two types of X--ray
emitting binary systems (see for example, Bhattacharya \& van den Heuvel 1991 and the
references therein): 
(1) high mass X--ray binaries (HMXBs), with identified optical counterparts 
associated with very massive and luminous 
(late O or early B supergiants) stars, and (2) low mass X--ray binaries (LMXBs), 
associated with objects for which the optical counterparts, if identified, are 
associated with low mass (M or K spectral type) stars (see section 1.2 for 
discussions). Both types of X--ray emitting systems show different kinds of 
spectral and temporal behaviors. In our work, we are interested in LMXBs and
briefly describe the major observational properties exhibited by them.

Some of the LMXBs show X--ray bursts (Grindlay et al. 1976; Belian, Conner \& Evans
1976). Two types of bursts have been identified: (1)Type I: 
the recurring time of the burst is several hours and a distinct spectral softening
occurs during burst decay (timescales of 10 sec. to a few minutes). The origin of 
such a burst is believed to be the thermonuclear flashes taking place on the neutron
star surface (Joss 1978) and the subsequent spectral softening may be caused by the 
cooling of the stellar surface after the burst. 
The majority of the bursting sources show this type of behavior. (2) Type II: these bursts, seen
for the sources 4U 1730-335 (the Rapid Burster), Cir X-1, GRO J17444-28, are 
repetitive and the timescale is smaller than that for type I bursts (for the 
Rapid Burster, the burst interval is $\sim 7$ sec.). Spasmodic accretion 
may cause such rapid variations in X--ray intensity (Lewin et al. 1976).
 
Although most of the LMXBs are persistent in their X--ray luminosities, some of them 
show transient behavior in the timescale of days to weeks. The X--ray luminosity 
may vary upto $10^4$ times for such a source. Such transience may be caused by an 
instability of the accretion disk (in section 1.2 accreion disk is defined) 
or of the mass transfer process (see Tanaka \& Shibazaki 1996 and
Campana et al. 1998 for reviews). The periodic dipping activity (Walter et al. 1982;
White \& Swank 1982) seen in some sources 
is believed to be due to the obscuration of the central accretor by a portion of 
the accretion disk.
The partial eclipses, observed for a few X--ray sources, may be caused by partial
occultation of matter at some portion of the system.

The X--ray satellite {\it EXOSAT} discovered time variabilities (frequency range 
$\sim 6-60$ Hz) in the X--ray 
intensities of several X--ray sources. As the corresponding power spectra can be
fitted by broad Lorentzian profiles, such temporal behaviors are called Quasi Periodic
Oscillations (QPOs). The origin of QPOs is not yet clear and there exist several
models, though none of them can explain it fully. The most popular model is the 
beat-frequency model (Alpar \& Shaham 1985), in which the QPO frequency is believed
to be the difference between the (fixed) spin frequency of the neutron star and the
(variable) Keplerian frequency of matter moving in the accretion disk at the 
Alf\'{v}en radius. 

In 90's, the X--ray satellite {\it RXTE} discovered QPOs with very high frequencies
(from a few hundred Hz to more than a kHz). These are called kHz QPOs. The origin
of them is unknown, but beat-frequency model (instead of Alf\'{v}en radius, disk 
inner edge radius is taken for the simplest form of the model) 
is the most popular among all the existing models.

It has been seen that both the QPO and the kHz QPO are strongly corelated with
the spectral behaviors of the sources. Such corelation can be studied very well with
the help of color-color diagram (CD) and hardness-intensity diagram (HID) 
(see van der Klis 1995 for a description). There are six very luminous 
X--ray sources, that trace Z-like curves (with three branches: horizontal branch, 
normal branch and flaring branch) in CD and HID. These are called Z sources. On the
other hand, there are several comparatively less luminous sources, which are called
atoll sources. An atoll source has a clustered branch (island state) and an upwardly 
curved branch (banana state) in CD and HID. Most of the Z and atoll sources display
QPO and kHz QPO. The position of an LMXB in CD (or HID) determines its spectral 
nature. This position is also strongly corelated with the nature of QPOs and kHz QPOs.
Therefore the relation between the temporal and the spectral behavior can be 
studied using CD (and HID). For the details of this corelation, see van der Klis (1995; 2000).

A few LMXBs show regular X--ray pulsations. The frequency of such an X--ray pulsar
is believed to be the rotational frequency of the central accretor (like radio 
pulsars). The LMXB SAX 
J1808.4-3658 shows milli-second X--ray pulsation, which supports
the conjecture that LMXBs are progenitors of milli-second radio pulsars.

In our work, We do not try to explain the temporal behavior of the LMXBs; rather
we calculate the accretion disk spectrum considering the full effect of general
relativity and rapid rotation of the neutron star.
In section 1.2, we briefly describe the nature of the X--ray binaries. We explain
the Newtonian accretion disk in section 1.3 and mention the effect of the inclusion of
Schwarzschild metric in section 1.4. In section 1.5, we describe the plan of the 
thesis. 

\section{X--ray Binaries}

An X--ray Binary is a binary stellar system with a compact primary star (black 
hole, neutron star or strange star) and a secondary (or companion) star (main sequence
star, blue super-giant star, red sub-giant star or white dwarf) rotating around each
other. These are galactic sources that emit a substantial part of their energy
in X--rays. The typical luminosities of the strongest sources among 
these systems are in the range $10^{34}-
10^{38}$ erg s$^{-1}$. The source of this energy is the gravitational energy
release, as the transfer of matter occurs from the secondary to the primary. Due to
the deep potential well of the compact primary, a considerable part $(\sim 20$\% for 
a neutron star) of the rest mass of the transfered matter is converted to energy, 
which is eventually emitted in the form of X--rays (as this matter, near the compact
star, is expected to be very hot: temperature $\sim 10^6$~K for neutron stars). 

It is believed that there are two main reasons behind this mass transfer (Frank, King
\& Raine 1992): (1) the companion star may increase in radius, or the binary separation
shrink, to the point where the gravitational pull of the primary star can remove the 
outer layer of its envelope (Roche-lobe overflow); (2) a substantial amount of mass 
of the secondary star may be ejected in the form of stellar wind and a part of it
may be gravitationally captured by the primary. 

In our work, we consider only the Roche-lobe overflow. The mechanism of such a overflow
is as follows (see, for example, Kopal 1959; Tsesevich 1973; Frank et al. 1992).
In a coordinate system co--rotating with the binary, there is a pear-shaped
equipotential (combination of both the gravitational and the centrifugal forces) 
surface around each component. If one goes outwards from each of the mass centers, 
at a certain value of the potential, these two surfaces touch each other at the 
first Lagrangian point $L_1$, located on the connecting line of the centers of two 
components. This critical equipotential surface through $L_1$ is called the {\it 
Roche-lobe}. When the secondary
star fills its Roche lobe, matter from its outer layer flows in a narrow jet towards 
the primary due to the unbalanced pressure at $L_1$ (where the net gravity vanishes).
Due to initial angular momentum, this matter can not flow radially towards the 
compact star, rather it takes a circuital path and eventually forms a disk (because 
of angular momentum redistribution and friction among the particles). Such a disk
is called an accretion disk.





X--ray binaries belong to two categories as described below.
These systems, with neutron stars 
as the central accretors, are the most luminous galactic X--ray
sources (first suggested by Zeldovich \& Guseynov 1965). In our work, we consider
only such systems.

\subsection{High Mass X--ray Binary (HMXB)}

These sources contain early type (massive) stars (generally, O or B type) as 
the companions. So the
optical spectra of HMXBs are dominated by the spectra of their secondary components. We
observe relatively hard X--ray spectra ($kT \gsim 15$~keV in exponential fit; Jones 
1977) and
$L_{\rm opt}/L_{\rm X} > 1$ for these systems $(L_{\rm opt}$ and $L_{\rm X}$ are 
the optical and X--ray luminosities respectively). 
They show regular X--ray pulsations, but no X--ray bursts, which indicates 
that their surface dipole field
strengths are typically of the order of $10^{11}$ to $10^{13}$ Gauss 
(see for example, Taam \& van den Heuvel 1986). HMXBs are found
to be concentrated in the galactic plane. Hence they form a young stellar population
(age $< 10^7$ yrs.), which is consistent with the fact that they contain early type
secondary stars.

As they contain very bright secondary components, which can be easily detected, one
can determine their orbital period, observing the regular X--ray eclipses. 
Such eclipses are very frequent in these systems. With this and the Doppler 
radial velocity curves of the
neutron star (or pulsar) and its companion, and the light curve, the mass of each
component, as well as the average radius of the companion star can be determined 
(Rappaport \& Joss 1983; Rappaport \& Joss 1984).

\subsection{Low Mass X--ray Binary (LMXB)}

These systems contain late type (low mass) companion stars, 
which can not be detected easily
(in fact, normal companions have been detected for very few cases). So the determination
of the masses of the components is not possible for most of the cases. We observe
softer X--ray spectra ($kT \lsim 10$ keV in exponential fit; van Paradijs 1989) and
$L_{\rm opt}/L_{\rm X} < 0.1$ for them. LMXBs show X--ray bursts for many cases, but
regular X--ray pulsations for very few cases. They concentrate in the galactic center
and globular clusters. They are old systems (age $\sim (5-15)\times 
10^9$ yrs.). The neutron stars in these systems generally have weak surface magnetic
fields (Bhattacharya \& van den Heuvel 1991). We study these systems in our work.

\section{Newtonian Accretion Disk}

In this section, we briefly discuss the accretion disks around neutron stars 
in Newtonian 
formalism. In many cases, the disk flow is confined very closely to the
orbital plane and one can regard the disk as a two-dimensional gas flow
to a first approximation. This {\it thin disk approximation} has proved very 
successful (Frank et al. 1992). The present interest in accretion disks 
has been developed from the encouraging results of comparison between 
the theory and observations of close binary systems.

A non-relativistic, incompressible fluid around an unmagnetized star should be
governed by the Navier-stokes equation (see, for example, Landau \& Lifshitz 1987):

\begin{eqnarray}
\frac{\partial \vec{v}}{\partial t} + (\vec{v}.\vec{\nabla})\vec{v} & = &
- \frac{1}{\rho} \vec{\nabla} P - \vec{\nabla} \Phi + \nu {\nabla}^2 \vec{v}.
\end{eqnarray}

\noindent Here $\rho$ is the mass density, $\vec{v}$ is the velocity, $P$ is 
the pressure, $\nu$ is the kinematic viscosity of the fluid and $\Phi$ is the
gravitational potential of the central star. 

However, we directly use the laws of conservation of mass and angular momentum, to
derive the basic governing equation for thin disk.
We assume that the disk is Keplerian, i.e., the angular speed of a particle
in the disk is 

\begin{eqnarray}
\Omega_{\rm K}(r) & = & \mbox{\Huge (}\frac{GM}{r^3} \mbox{\Huge )}^{1/2}
\end{eqnarray}

\noindent where $G$ is the gravitational constant, $M$ is the mass of the central
star and $r$ is the radial coordinate.
The corresponding linear speed is $v_{\phi} = r \Omega_{\rm K}(r)$.
In addition to $v_{\phi}$, the gas is assumed to possess a small radial drift
speed $(v_r)$ 
towards the star. This is because, due to friction among the particles in the
disk, most of the angular momentum is taken away by a small number of particles
and most of the particles move inwards losing their angular momentum (see Frank et al.
1992 for detailed discussion). We characterize the disk by its surface density 
$\Sigma(r, t)$ $(t$ is time), 
which is the mass per unit surface area of the disk, given by
integrating the gas density $\rho$ in the z-direction (i.e., perpendicular 
to the plane of the disk). Now the conservation laws
for mass and angular momentum (in combination with the expression for viscous torque
in the disk; see Frank et al. 1992) give

\begin{eqnarray}
\frac{\partial \Sigma}{\partial t} & = & \frac{3}{r} \frac{\partial}{\partial r}
\mbox{\Huge \{}r^{1/2} \frac{\partial}{\partial r} (\nu \Sigma r^{1/2}) \mbox{\Huge \}}
\end{eqnarray}

\noindent as the basic equation governing the time evolution of surface density in a 
Keplerian disk.

As the radial structure in a thin disk changes in the timescales $\sim t_{\rm visc}
\sim r^2/\nu$, and the external conditions (for example, mass accretion rate) in 
many systems change on timescales longer than $t_{\rm visc}$, a steady-state 
approximation $(\frac{\partial}{\partial t} = 0)$ 
for the disk should be more or less valid.
Therefore from the mass conservation law, we get $r \Sigma v_r$ = constant, i.e., if
$\dot M$ is the (constant) mass accretion rate, then
we get

\begin{eqnarray}
\dot M & = & 2 \pi r \Sigma (-v_r)
\end{eqnarray}

\noindent since $v_r < 0$, as the particles move inwards.

The rotation rate of the star is expected to be slower than the break-up speed at its
equator, i.e.,

\begin{eqnarray}
\Omega_{\rm *} & < & \Omega_{\rm K}(R)
\end{eqnarray}

\noindent where $\Omega_{\rm *}$ is the stellar rotation rate and $R$ is the 
radius of the star. Therefore, very near
the surface of the star the angular speed of the disk particles should decrease
and attain the value $\Omega_{\rm *}$ at $r = R$. This small region, in which
angular speed decreases, is called the boundary layer. The width $(b)$ of the boundary
layer is much smaller than $R$, as shown in Frank et al. (1992).

Now combining Eq. (1.4), the law of conservation of angular momentum (with the 
condition $\frac{\partial}{\partial t} = 0$) and the condition $b << R$, one can
derive the equation 

\begin{eqnarray}
\nu \Sigma & = & \frac{\dot M}{3 \pi} 
\mbox{\Huge \{}1 - \mbox{\Huge (}\frac{R}{r} \mbox{\Huge )}^{1/2} \mbox{\Huge \}}
\end{eqnarray}

\noindent using which we get the viscous dissipation per unit disk face area as
(see Frank et al. 1992 for derivation)

\begin{eqnarray}
F(r) & = & \frac{3 G M \dot{M}}{8 \pi r^3} \mbox{\Huge \{}1 - \mbox{\Huge 
(}\frac{R}{r} \mbox{\Huge )}^{1/2} \mbox{\Huge \}}.
\end{eqnarray}

Therefore, the disk energy flux comes out to be independent of viscosity.
This is a very important result, as we can try to understand the values of $M$, $\dot 
M$ and $R$ for a particular source by fitting the observational data, without 
having much idea about the physical nature of the disk viscosity.

Energy comes out also from the thin boundary layer (around the equator of the
neutron star), as the matter hits the stellar surface. The source of both boundary 
layer luminosity $(L_{\rm BL})$ and disk luminosity $(L_{\rm D})$ is the 
gravitational potential energy release. It is easy to see that, for Newtonian case,

\begin{eqnarray}
L_{\rm BL} = L_{\rm D} = \frac{G M \dot{M}}{2 R}.
\end{eqnarray}

It is very important to calculate the disk temperature profile, as the disk spectrum
can be calculated from it. Fitting this theoretical spectrum to the observed one
for a source, one can hope to constrain the values of its parameters. If the disk
is optically thick in the z-direction, each element of it is expected to radiate
roughly as a blackbody. The temperature profile $(T(r))$ 
of such a blackbody disk is given by

\begin{eqnarray}
\sigma T^4(r) & = & F(r)
\end{eqnarray}

\noindent where $\sigma$ is the Stefan-Boltzmann constant. Now using Eq. (1.7), we get

\begin{eqnarray}
T(r) & = & \mbox{\Huge [}\frac{3 G M \dot{M}}{8 \pi r^3 \sigma} \mbox{\Huge \{}1 - 
\mbox{\Huge(}\frac{R}{r} \mbox{\Huge )}^{1/2} \mbox{\Huge \}}\mbox{\Huge ]}^{1/4}.
\end{eqnarray}

\noindent In the next section, we will see how this expression is modified if we
consider the general relativistic formalism.

\section{Effects of Schwarzschild Space-Time}

Most of the X--rays from an LMXB come from a region which is very close to the
compact star. The gravity is so strong in this region that Newtonian theory does not
provide an adequate description. The correct theory that can describe the motion of the 
particles near the compact star is expected to be general relativity. In this section,
we briefly describe the effects of this theory for non-rotating neutron stars (i.e.,
the Schwarzschild space-time). General relativity introduces some new effects, not found 
in the Newtonian framework. One of these is the existence of an innermost
stable circular orbit (ISCO), that we discuss below.

For a non-rotating neutron star the configuration is spherically symmetric and the 
corresponding space-time geometry (outside the star) 
is described by the Schwarzschild metric (see, for 
example, Misner, Thorne \& Wheeler 1973):

\begin{eqnarray}
ds^{2} & = & g_{\lambda \beta} dx^{\lambda} dx^{\beta}
        ~~~(\lambda, \beta = 0,1,2,3) \nonumber \\
       & = & -\mbox{\Huge(}1-\frac{2GM}{c^2r}\mbox{\Huge)} dt^2 + 
\mbox{\Huge(}1-\frac{2GM}{c^2r}\mbox{\Huge)}^{-1} dr^2 + r^2(d\theta^2+\sin^2\theta
d\phi^2)
\end{eqnarray}
 
\noindent where we have used the $(-$ $+$ $+$ $+)$ convention. 
In the above metric, $r$ is
the radial coordinate, while $\theta$ and $\phi$ are the polar and azimuthal
coordinates respectively. The quantity $M$ is the mass of the star.

It is to be noted that the structure of a neutron star can not be calculated using
Newtonian theory, as the gravity is too strong inside it. Therefore, general relativity
must be explicitly included in constraining the 
equation of state of the neutron star (discussed in latter
chapters). We can formulate the structure equations for a non-rotating neutron star
using the perfect fluid assumption (for the stellar material)
and Einstein's field equations. These are 

\begin{eqnarray}
\frac{dP}{dr} & = & -G \frac{(m+4\pi r^3 P/c^2)(\rho+P/c^2)}{r^2 
\{1-(2Gm/c^2r)\}}
\end{eqnarray}

\begin{eqnarray}
\frac{dm}{dr} & = & 4 \pi r^2 \rho
\end{eqnarray}

\noindent where $P$, $\rho$ and $m$ are the pressure, mass-energy density of the 
system and the mass contained in a radius $r$ respectively and all 
of them are functions of $r$. Here we use the line element (Misner, Thorne \& Wheeler 1973)

\begin{eqnarray}
ds^{2} & = & - e^{2\Phi} dt^2 +
\mbox{\Huge(}1-\frac{2Gm}{c^2r}\mbox{\Huge)}^{-1} dr^2 + r^2(d\theta^2+\sin^2\theta
d\phi^2)
\end{eqnarray}

\noindent with the source function $\Phi$ given by

\begin{eqnarray}
\frac{d\Phi}{dr} & = & \frac{m+4\pi r^3 P/c^2}{r \{r - (2Gm/c^2)\}}.
\end{eqnarray}

\noindent For $r \ge R$ $(R$ is the radius of the star), the pressure vanishes and hence
the line element (1.14) becomes identical with the Schwarzschild line element (1.11).
Eq. (1.12) is the 
Tolman-Oppenheimer-Volkoff (TOV) equation (Oppenheimer \& Volkoff 1939). Eqs. (1.12)
and (1.13) can be solved to get the structure parameters of the neutron star
if its equation of state (i.e., $P$ as a function of $\rho$) is known. As the 
neutron star is degenerate (except a thin outer shell), the temperature does not
enter in the equation of state.

In the Schwarzschild metric, the specific energy $\tilde{E}$ and the specific
angular momentum $l$ are constants of motion (see Shapiro \& Teukolsky 1983).
The equations of motion of a particle (confined to the equatorial plane) in this 
metric are (Thampan 1999):

\begin{eqnarray}
\frac{dt}{d\tau} & = & \tilde{E} \mbox{\Huge(}1-\frac{r_{\rm g}}{r}\mbox{\Huge)}^{-1}
\end{eqnarray}

\begin{eqnarray}
\frac{d\phi}{d\tau} & = & \frac{l}{r^2}
\end{eqnarray}

\begin{eqnarray}
\frac{dr}{d\tau} & = & \tilde{E}^2-\tilde{V}^2
\end{eqnarray}

\noindent where we have used geometric units $c=G=1$. In the above equations, 
$r_{\rm g}$ is the Schwarzschild radius $(2GM/c^2)$, $\tau$ represents the proper 
time and $\tilde{V}$ is the effective potential given by

\begin{eqnarray}
\tilde{V}^2 & = & \mbox{\Huge(}1-\frac{r_{\rm g}}{r}\mbox{\Huge)} 
\mbox{\Huge(}1+\frac{l^2}{r^2}\mbox{\Huge)}.
\end{eqnarray}
 
The conditions for circular orbits and the extremum of energy are $\tilde{E}^2=
\tilde{V}^2$ and $\tilde V_{,r} = 0$ respectively. The radius $(r_{\rm orb})$ 
of the ISCO (as 
mentioned earlier) can be calculated from the equation $\tilde V_{,rr} = 0$. Here
a comma followed by a variable as subscript to a quantity, represents the derivative
of the quantity with respect to the variable. There can be no stable circular
orbit inside the ISCO. Therefore the accretion disk can exist upto the 
ISCO and then the matter quickly (i.e., the radial speed increases enormously) 
falls on the surface of the central star. However, if the stellar radius is greater
than the radius of the ISCO, the disk will be extended upto the surface of the star.
Therefore, the radius $(r_{\rm in})$ of the inner edge of the disk is $R$ 
$(r_{\rm orb})$ for $R > r_{\rm orb}$ $(r_{\rm orb} > R)$.
It is to be noted that throughout our work, we assume that the magnetic field of 
the neutron star is too week to affect the accretion flow.

The specific disk luminosity $(E_{\rm D})$ is given by the energy difference 
between a particle (of unit mass) at infinity and the same particle at $r = 
r_{\rm in}$. The specific boundary layer luminosity $(E_{\rm BL})$ is defined by the
same kind of energy difference, but with the particle positions $r = r_{\rm in}$ and 
$r = R$ (i.e., the particle sitting on the surface of the star).

For Schwarzschild metric, the energy flux of the disk is given by (Yamada \& Fukue 
1993; Novikov \& Thorne 1973)

\begin{eqnarray}
F(r)
={3\mu\dot M\over 8\pi r^{{5\over 2}} (r-3\mu)}\cdot \mbox{\Huge [}\sqrt
{r}-\sqrt{r_{\rm in}}
+{\sqrt{3\mu}\over 2}\cdot \ln{(\sqrt{r}+\sqrt{3\mu})(\sqrt{r_{\rm in}}-\sqrt{3\mu})\over
(\sqrt{r}-\sqrt{3\mu})(\sqrt{r_{\rm in}}+\sqrt{3\mu})} \mbox{\Huge ]}
\end{eqnarray}

\noindent With $\mu={GM\over c^2}$ and $c=1$. The disk temperature profile is then
calculated from Eq. (1.9).

\section{Plan of the Thesis}

In this thesis, we calculate the temperature profile and the spectrum
of an accretion disk around a rapidly rotating neutron star. In Chapter 2, we
give the formalism for the structure calculation of a fast rotating star.
We also describe different sequences possible for such a star. We also give
the details of luminosity calculation and mention the neutron star equations of state
used in this work.

In Chapter 3, we calculate the disk temperature profiles for different values of
stellar rotation rate and for all the chosen equations of state. We compare these
results with those for Newtonian and Schwarzschild cases and point out the 
importance of incorporating the effects of general relativity and rapid rotation
in the accretion disk calculations.

We compare the theoretical results (calculated in Chapter 3) with the observational
({\it EXOSAT}) data for five LMXB sources and constrain several properties for
these systems in Chapter 4. We also discuss possible constraints on the neutron star
equation of state.

In Chapter 5, we compute the general relativistic spectrum of an accretion disk around
a rotating neutron star. We show that the effect of light-bending is very important
at higher energies.

We fit the calculated (in Chapter 5) spectrum with an analytical function in Chapter
6. Here we suggest a method to distinguish between a Newtonian spectrum and a general
relativistic spectrum observationally.

It has been proposed that the central accretors of atleast some of the LMXBs are
strange stars (and not neutron stars). In order to try to answer this question, in 
Chapter 7, we
calculate the values of several properties (including the disk temperature profile)
of a rotating strange star and compare them with those of a rotating neutron star.

In Chapter 8, we give a summary of our work and discuss the future prospects. We also
mention the main conclusions of the thesis.

%% file: chap2.tex
\markright{Chapter 2}
\def\note #1]{{\bf #1]}}
\chapter{Formalism for Rapidly Rotating Neutron Stars}

\section{Introduction}

The necessary condition for disk accretion is that the accreted matter
must have intrinsic angular momentum. Because of this property, matter coming 
out of the companion star can not fall on the surface of the neutron star 
radially, but moves almost in circular orbit and forms a disk.
The specific angular momentum of this matter is much higher than that
of a neutron star. As a result, when it hits the star, the stellar 
angular momentum increases, making the star rotate faster in general. 
Therefore accreting neutron stars are expected to be rapidly rotating due 
to such accretion induced angular momentum transfer. This was the reason 
that the LMXBs were speculated to be the progenitors of millisecond 
radio pulsars for long time (Bhattacharya \& van den Heuvel 1991). 
Recently such speculation has been confirmed with the discovery of a 
millisecond pulsar (SAX J1808.4--3658) in an LMXB (Wijnands \& van der Klis 
1998). SAX J1808.4--3658 shows periodic pulsations ($P = 2.49$ ms) in 
X--rays, which proves that accretion can actually spin up the central accretor
very effectively. We also observe kHz QPO and Burst Oscillation (van der Klis 
2000). The frequency separation $(\sim 300$ Hz) between the two simultaneously
observed peak of kHz QPO is equal to the rotational frequency of the central
star according to the beat-frequency model. The Burst Oscillation frequency
is also believed to be close to (or integer multiple of) the stellar
angular frequency (van der Klis 2000). 
These indicate the rapid rotation of the accreting neutron star. 

It is therefore essential to construct equilibrium sequences for rapidly 
rotating neutron stars, considering the full effect of general relativity.
The Schwarzschild metric is no longer valid even outside a rotating neutron 
star, as the relativistic effect of dragging of inertial frames in the 
vicinity of the star will be important. This will affect the luminosity and 
the spectra of the accretion disk. Therefore to model the observed spectra 
more accurately, we need to compute the metric coefficients around rapidly 
rotating neutron stars.

Relativistic models of slowly rotating neutron stars were constructed by 
Hartle \& Thorne (1968).
Their formalism is valid for strong gravitational fields, but only in the 
limit of slow rotation (neglects terms higher than 
$O(\Omega_*^2/\Omega^{2}_{\rm ms}))$ compared to the 
critical angular speed for centrifugal break--up $(\Omega_{\rm ms})$. 
Similar calculations, using the same formalism, 
were performed by Datta \& Ray (1983), to construct models based on a variety 
of proposed equations of state. An extensive study of the properties of these 
models has been made by Datta, Kapoor \& Ray (see, for example, 
Datta 1988 and references therein).
For a description of the structure calculation of slowly rotating neutron 
stars, see Thampan (1999). 

A formalism, appropriate for a rapidly rotating neutron star, should be exact 
in its treatment of $\Omega_*$. The metric coefficients for such models are to 
be calculated numerically, unlike the case of slowly rotating models, where 
the metric coefficients have analytic expressions. The exact models are the 
solutions of Einstein's equations for the stationary gravitational field 
in axisymmetry, coupled to the equation of hydrostatic equilibrium. Such 
models have previously been constructed by several authors, including 
Bonazzola \& Schneider (1974), Butterworth (1976) (for polytropic EOS) and 
Friedman, Ipser \& Parker (1986) (for realistic EOS). An alternative approach 
using spectral methods was developed by Bonazzola et al. (1993) and used 
for many realistic EOS by Salgado et al. (1994a; 1994b).
However, we follow the procedure used by Cook, Shapiro \& Teukolsky (1994), 
based on a formalism due to Komatsu, Eriguchi \& Hachisu (1989).
For a comparison of different formalisms, see Stergioulas (1998).

For our preliminary study, we calculate different sequences. These are mainly 
of two types: evolutionary sequence and limit sequence. We call sequences 
along which the rest mass $M_{\rm 0}$ is held fixed evolutionary sequences. 
An isolated neutron star is expected to evolve along such a sequence, as 
it slowly loses energy and angular momentum via, for example, electromagnetic 
or gravitational radiation. The set of all evolutionary sequences is 
naturally divided into two groups: the normal sequences and the supramassive 
sequences. Normal evolutionary sequences are those that terminate at one end 
with a nonrotating, spherically symmetric solution. Supramassive sequences 
do not contain such static solution.

The set of equilibrium solutions for a given EOS forms a two parameter family. 
The boundary of the set of stable equilibrium solutions is formed by the 
four limits. The first limit is the static limit, where $\Omega_* 
\rightarrow 0$ and total angular momentum $J \rightarrow 0$. Models
on the static limit sequence are solutions of the TOV equations for 
spherically symmetric models, described in Chapter 1. 

The second limit is the mass--shed limit, which is reached when the 
gravitational attraction at the stellar equator is not sufficient to keep 
matter bound to the surface. For the case of general differential rotation, 
mass--shed limit occurs when 

\begin{eqnarray}
\frac{1}{2} \frac{\partial}{\partial r} (\rho + \gamma) - 
\frac{\tilde v}{1-\tilde v^2} \frac{\partial \tilde v}{\partial r} + F(\Omega_*) 
\frac{\partial \Omega_*}{\partial r} & = & 0
\end{eqnarray}

\noindent at the equator (Cook et al. 1994). Here the relation $F(\Omega_*)=u^tu_{\phi}$ 
specifies the rotation law, $\tilde v$ is the proper velocity of matter at the 
equator with respect to a zero-angular-momentum-observer (ZAMO) and $\rho$ \& 
$\gamma$ are the metric coefficients (see section 2).

The third limit is the stability limit, where an equilibrium solution is 
marginally stable to quasi--radial perturbations. The stability limit 
sequence begins at the maximum--mass point on the static limit sequence and 
usually terminates near the maximum--mass point on the mass--shed limit 
sequence. The intermediate points lie on supramassive evolutionary sequences 
where the stability condition 

\begin{eqnarray}
\mbox{\Huge (}\frac{\partial J}{\partial \epsilon_c}\mbox{\Huge )}_{M_{\rm 0}}
 & < & 0 
\end{eqnarray}

\noindent is marginally satisfied (Cook et al. 
1994). Here $\epsilon_c$ is the central total energy density.

Finally, there is the low--mass limit, below which a neutron star cannot 
form. However, we have not attempted to determine this limit, as it is of 
minimal importance.

In our work, we, in general, choose gravitational mass ($M$) and $\Omega_*$
as the independent parameters for a given EOS. The reason is that these are 
the quantities that can be observationally measured. Therefore we construct 
gravitational mass sequences (i.e., $M$ is kept constant) and study the 
values of different quantities for equilibrium configurations.

In section 2.2 and 2.3, we describe the procedure for structure calculation 
of a rapidly rotating neutron star, considering the full effect of general 
relativity and the corresponding luminosity--calculation--procedure 
respectively. A description of equations of state is given in section 2.4. 
We show our results in section 2.5 and give concluding remarks in section 2.6.

\section{Structure Calculation}

We assume that the space--time in and around a rotating neutron star is 
stationary, axisymmetric, asymptotically flat and reflection--symmetric 
(about the equatorial plane). The metric may be written in the form (Bardeen 
1970) 

\begin{eqnarray}
ds^{2} & = & g_{\lambda \beta} dx^{\lambda} dx^{\beta}
        ~~~(\lambda, \beta = 0,1,2,3) \nonumber \\
       & = & - e^{\gamma + \rho} dt^{2}
             + e^{2 \alpha} (d\bar{r}^{2} + \bar{r}^{2} d \theta^{2})
             + e^{\gamma - \rho} \bar{r}^{2} sin^{2} \theta (d \phi -
            \omega dt)^{2}
\end{eqnarray}

\noindent where the metric potentials $\gamma$, $\rho$, $\alpha$, and the
angular speed ($\omega$) of zero-angular-momentum-observer (ZAMO) with 
respect to infinity, are all functions of the quasi--isotropic radial 
coordinate ($\bar{r}$) and polar angle ($\theta$).
$\bar{r}$ is related to the Schwarzschild--like radial 
coordinate ($r$) through the equation $r= \bar{r}e^{(\gamma-\rho)/2}$ 
(see Misner, Thorne \& Wheeler 1974). Here we use geometric units 
$c = G = 1$.

We assume that the matter source is a perfect fluid with a stress--energy 
tensor given by 

\begin{eqnarray}
T^{\mu \nu} & = & (\epsilon + P) u^{\mu} u^{\nu} + P g^{\mu \nu}
\end{eqnarray}

\noindent where $\epsilon$ is the total energy--density, $P$ is the pressure 
and $u^{\mu}$ is the matter four--velocity, given by (Cook et al. 1994)

\begin{eqnarray}
u^{\mu} & = & \frac{e^{-(\gamma + \rho)/2}}{(1 - \tilde v^2)^{1/2}} 
(1, 0, 0, \Omega_*)
\end{eqnarray}
  
\noindent Here $\Omega_* \equiv u^3/u^0$ is the angular speed  
and the proper velocity $\tilde v$ of the matter, relative to ZAMO, is given 
by 

\begin{eqnarray}
\tilde v & = & (\Omega_* - \omega) \bar r \sin{\theta} e^{-\rho}
\end{eqnarray}

\noindent The tilde over a variable represents the corresponding 
dimensionless quantity. For example, we use $\tilde r \equiv \kappa^{-1/2} 
\bar r$, $\tilde t \equiv \kappa^{-1/2} ct$, $\tilde \omega \equiv 
\frac{1}{c} \kappa^{1/2} \omega$, $\tilde \Omega_* \equiv \frac{1}{c} 
\kappa^{1/2} \Omega_*$, $\tilde \epsilon \equiv \frac{G}{c^2} \kappa \epsilon$, 
$\tilde P \equiv \frac{G}{c^4} \kappa P$, $\tilde J \equiv \frac{G}{c^3} 
\kappa^{-1} J$ and $\tilde M \equiv \frac{G}{c^2} \kappa^{-1/2} M$, 
where the fundamental length scale $\kappa^{1/2}$ is given by 
$\kappa \equiv \frac{c^2}{G \epsilon_o}$, with $\epsilon_o = 
10^{15} \mbox{g cm}^{-3}$. 

For an axisymmetric and equatorial plane symmetric configuration, the 
computational domain in spherical polar coordinates covers 
$0 \le r \le \infty$ and $0 \le \theta \le \pi/2$. For numerical convenience, 
we make a change of variables ($r \rightarrow s$ and $\theta \rightarrow 
\mu$) given by 

\begin{eqnarray}
\tilde{r} & = & \tilde{r_{\rm e}} \frac{s}{1-s} ~ ; \nonumber \\ 
\theta  & = & \cos^{-1} \mu
\end{eqnarray}

\noindent where $\bar{r_{\rm e}}$ is the quasi--isotropic radial coordinate of the 
equator. It is easy to see that $s$ and $\mu$ vary in the range 
$0 \le s \le 1$ \& $0 \le \mu \le 1$ and at the equator $s = 0.5$.

For these variables, the Einstein field equations projected on to the frame 
of reference of a ZAMO yield three elliptic equations for the metric 
potentials $\rho, \gamma$ \& $\omega$ and two linear ordinary differential 
equations for the metric potential $\alpha$ (Bardeen \& Wagoner 1971; 
Butterworth \& Ipser 1976; Komatsu et al. 1989). The elliptic 
equations are of the form (Thampan 1999): 

\begin{eqnarray}
\tilde \Delta[\rho e^{\gamma /2}] & = & \tilde S_{\rho}(s, \mu) 
\end{eqnarray}

\begin{eqnarray}
\mbox{\Huge (}\tilde \Delta + \frac{(1-s)^3}{s} \frac{\partial}{\partial s} - 
\frac{(1-s)^2 \mu}{s^2} \frac{\partial}{\partial \mu}\mbox{\Huge )} ~ 
\gamma e^{\gamma/2} & = & \tilde S_{\gamma}(s, \mu)
\end{eqnarray}

\begin{eqnarray}
\mbox{\Huge (}\tilde \Delta + \frac{2 (1-s)^3}{s} 
\frac{\partial}{\partial s} - 
\frac{2 (1-s)^2 \mu}{s^2} \frac{\partial}{\partial \mu}\mbox{\Huge )} ~ 
\hat{\omega} e^{(\gamma-\rho)/2} & = & \tilde S_{\hat{\omega}}(s, \mu) 
\end{eqnarray}

\noindent where the elliptic differential operator $\tilde \Delta$ is 
given by 

\begin{eqnarray}
\tilde \Delta & = & (1-s)^4 
\mbox{\Huge (}\frac{\partial^2}{\partial s^2}\mbox{\Huge )} 
- 2 (1-s)^3 \mbox{\Huge (}\frac{\partial}{\partial s}\mbox{\Huge )}^2 
+ \frac{2 (1-s)^3}{s} 
\mbox{\Huge (}\frac{\partial}{\partial s}\mbox{\Huge )} \nonumber \\ 
 &   & + \frac{(1-s)^2 (1-\mu^2)}{s^2} 
\mbox{\Huge (}\frac{\partial^2}{\partial \mu^2}\mbox{\Huge )} 
- \frac{(1-s)^2 \mu}{s^2} 
\mbox{\Huge (}\frac{\partial}{\partial \mu}\mbox{\Huge )} 
+ \frac{(1-s)^2}{s^2 (1-\mu^2)} 
\mbox{\Huge (}\frac{\partial^2}{\partial \phi^2}\mbox{\Huge )}
\end{eqnarray}

\noindent The effective sources $\tilde S$'s are defined as (Cook et al. 1994)

\begin{eqnarray}
\tilde S_{\rho}(s, \mu) & = & e^{\gamma/2} \mbox{\Huge [}8 \pi e^{2 \alpha} 
\tilde r_{\rm e}^2 (\tilde \epsilon + \tilde P) 
\mbox{\Huge (}\frac{s}{1-s}\mbox{\Huge )}^2 
\frac{1+\tilde v^2}{1-\tilde v^2} \nonumber \\
 &   & + \mbox{\Huge (}\frac{s}{1-s}\mbox{\Huge )}^2 (1-\mu^2) e^{-2 \rho}
\{[s (1-s) \hat \omega_{,s}]^2 + (1-\mu^2) \hat \omega_{,\mu}^2\} \nonumber \\
 &   & + s (1-s) \gamma_{,s} - \mu \gamma_{,\mu} + \frac{\rho}{2} 
\mbox{\Huge \{}16 \pi e^{2 \alpha} \tilde r_{\rm e}^2 \tilde P 
\mbox{\Huge (}\frac{s}{1-s}\mbox{\Huge )}^2 \nonumber \\
 &   & - s (1-s) \gamma_{,s} \mbox{\Huge (}\frac{s (1-s)}{2} \gamma_{,s} 
+ 1\mbox{\Huge )} - \gamma_{,\mu} \mbox{\Huge (}\frac{1-\mu^2}{2} 
\gamma_{,\mu} - \mu\mbox{\Huge )\}]}
\end{eqnarray}
 
\begin{eqnarray}
\tilde S_{\gamma}(s, \mu) & = & e^{\gamma/2} \mbox{\Huge [}16 \pi e^{2 \alpha}
 \tilde r_{\rm e}^2 \tilde P\mbox{\Huge (}\frac{s}{1-s}\mbox{\Huge )}^2 \nonumber \\
 &   &  + \frac{\gamma}{2} \mbox{\Huge \{}16 \pi e^{2 \alpha} \tilde r_{\rm e}^2 
\tilde P\mbox{\Huge (}\frac{s}{1-s}\mbox{\Huge )}^2 - 
\frac{s^2 (1-s)^2}{2} \gamma_{,s}^2 - \frac{1-\mu^2}{2} 
\gamma_{,\mu}^2\mbox{\Huge \}]}
\end{eqnarray}
 
\begin{eqnarray}
\tilde S_{\hat{\omega}}(s, \mu) & = & e^{(\gamma-2 \rho)/2} 
\mbox{\Huge [}-16 \pi e^{2 \alpha} \frac{(\hat \Omega_* - \hat \omega)}{1 - 
\tilde v^2} \tilde r_{\rm e}^2 (\tilde \epsilon + \tilde P) \mbox{\Huge 
(}\frac{s}{1-s}\mbox{\Huge )}^2 \nonumber \\
 &   & + \hat \omega \mbox{\Huge \{}-8 \pi e^{2 \alpha} \tilde r_{\rm e}^2 
\frac{(1+\tilde v^2) \tilde \epsilon + 2 \tilde v^2 \tilde P}{1-\tilde v^2} 
\mbox{\Huge (}\frac{s}{1-s}\mbox{\Huge )}^2 \nonumber \\
 &   & - s (1-s) \mbox{\Huge (}2 \rho_{,s} + \frac{1}{2} 
\gamma_{,s}\mbox{\Huge )} + \mu \mbox{\Huge (}2 \rho_{,\mu} + \frac{1}{2}
\gamma_{,\mu}\mbox{\Huge )} \nonumber \\
 &   & + \frac{s^2 (1-s)^2}{4} (4 \rho_{,s}^2 - \gamma_{,s}^2) + 
\frac{1-\mu^2}{4} (4 \rho_{,\mu}^2 - \gamma_{,\mu}^2) \nonumber \\
 &   & - (1-\mu^2) e^{-2 \rho} \mbox{\Huge (}s^4 \hat{\omega}_{,s}^2 + 
\frac{s^2 (1-\mu^2)}{(1-s)^2} \hat{\omega}_{,\mu}^2\mbox{\Huge )\}]}
\end{eqnarray}

\noindent where 

\begin{eqnarray}
\hat \omega & \equiv &  \tilde r_{\rm e} \tilde \omega ~ ; \nonumber \\
\hat \Omega_* & \equiv &  \tilde r_{\rm e} \tilde \Omega_* 
\end{eqnarray}

The differential equation for $\alpha$ with respect to $s$ does not provide 
any new information (see Butterworth \& Ipser 1976). Here we use the 
differential equation for $\alpha$ with respect to $\mu$ 

\begin{eqnarray}
\alpha_{,\mu} & = & -\frac{1}{2} (\rho_{,\mu} + \gamma_{,\mu}) - 
\{(1-\mu^2) [1+s (1-s) \gamma_{,s}]^2 + 
[-\mu+(1-\mu^2)\gamma_{,\mu}]^2\}^{-1} \nonumber \\
 &   & \times \mbox{\Huge [}\frac{1}{2} \{s (1-s) [s (1-s) \gamma_{,s}]_{,s}  
+ s^2 (1-s)^2 \gamma_{,s}^2 - [(1-\mu^2) \gamma_{,\mu}]_{,\mu} \nonumber \\
 &   & - \gamma_{,\mu} [-\mu+(1-\mu^2) \gamma_{,\mu}]\}  
[-\mu+(1-\mu^2) \gamma_{,\mu}] + \frac{1}{4} [s^2 (1-s)^2 (\rho_{,s}+
\gamma_{,s})^2 \nonumber \\
 &   & - (1-\mu^2) (\rho_{,\mu}+\gamma_{,\mu})^2] 
[-\mu+(1-\mu^2) \gamma_{,\mu}] - s (1-s) (1-\mu^2) \nonumber \\
 &   & \times \mbox{\Huge (}\frac{1}{2} (\rho_{,s}+\gamma_{,s}) 
(\rho_{,\mu}+\gamma_{,\mu}) + \gamma_{,s\mu} + \gamma_{,s} 
\gamma_{,\mu}\mbox{\Huge )} [1+s (1-s) \gamma_{,s}] \nonumber \\
 &   & + s (1-s) \mu \gamma_{,s} [1+s (1-s) \gamma_{,s}] + \frac{1}{4} 
(1-\mu^2) e^{-2 \rho} \nonumber \\
 &   & \times \mbox{\Huge \{}2 \frac{s^3}{1-s} (1-\mu^2) \hat{\omega}_{,s} 
\hat{\omega}_{,\mu} [1+s (1-s) \gamma_{,s}] \nonumber \\
 &   & - \mbox{\Huge (}s^4 \hat{\omega}_{,s}^2 - \frac{s^2}{(1-s)^2}
(1-\mu^2) \hat{\omega}_{,s}^2\mbox{\Huge )} [-\mu+(1-\mu^2) 
\gamma_{,\mu}]\mbox{\Huge \}]}
\end{eqnarray}

\noindent with the initial condition that $\alpha = (\gamma-\rho)/2$ 
at $\mu = 1$.

In the formalism given by Komatsu et al. (1989), the elliptical differential 
equations are converted to integral equations (so that the boundary 
conditions can be handled easily) using Green's function approach. 
Therefore, the three metric potentials $\rho, \gamma$ \& $\omega$ can be 
written as 

\begin{eqnarray}
\rho(s, \mu) & = & -e^{-\gamma/2} \sum_{n=0}^{\infty} P_{2 n}(\mu) 
\mbox{\Huge [}\mbox{\Huge (}\frac{1-s}{s}\mbox{\Huge )}^{2 n+1} \int_0^s
\frac{ds' s'^{2 n}}{(1-s')^{2 n+2}} \int_0^1 d\mu' P_{2 n}(\mu') 
\tilde S_{\rho}(s', \mu') \nonumber \\
 &   & + \mbox{\Huge (}\frac{s}{1-s}\mbox{\Huge )}^{2 n} \int_s^1 
\frac{ds' (1-s')^{2 n-1}}{s'^{2 n+1}} \int_0^1 d\mu' P_{2 n}(\mu') 
\tilde S_{\rho}(s', \mu')\mbox{\Huge ]}
\end{eqnarray}

\begin{eqnarray}
\gamma(s, \mu) & = & -\frac{2 e^{-\gamma/2}}{\pi} \sum_{n=1}^{\infty}
\frac{\sin[(2 n-1) \theta]}{(2 n-1) \sin \theta} 
\mbox{\Huge [}\mbox{\Huge (}\frac{1-s}{s}\mbox{\Huge )}^{2 n} \nonumber \\
 &   & \times \int_0^s \frac{ds' s'^{2 n-1}}{(1-s')^{2 n+1}} 
\int_0^1 d\mu' \sin[(2 n-1) \theta'] \tilde S_{\gamma}(s', \mu')
+ \mbox{\Huge (}\frac{s}{1-s}\mbox{\Huge )}^{2 n-2} \nonumber \\
 &   & \times \int_s^1 \frac{ds' (1-s')^{2 n-3}}{s'^{2 n-1}} 
\int_0^1 d\mu' \sin[(2 n-1) \theta'] 
\tilde S_{\gamma}(s', \mu')\mbox{\Huge ]} 
\end{eqnarray}

\begin{eqnarray}
\hat \omega(s, \mu) & = & -e^{(2 \rho-\gamma)/2} \sum_{n=1}^{\infty} 
\frac{P^1_{2 n-1}(\mu)}{2 n (2 n-1) \sin \theta} \mbox{\Huge [}\mbox{\Huge 
(}\frac{1-s}{s}\mbox{\Huge )}^{2 n+1} \nonumber \\
 &   & \times \int_0^s \frac{ds' s'^{2 n}}{(1-s')^{2 n+2}} 
\int_0^1 d\mu' \sin \theta' P^1_{2 n-1}(\mu') 
\tilde S_{\hat \omega}(s', \mu') + \mbox{\Huge 
(}\frac{s}{1-s}\mbox{\Huge )}^{2 n-2} \nonumber \\
 &   & \times \int_s^1 \frac{ds' (1-s')^{2 n-3}}{s'^{2 n-1}} 
\int_0^1 d\mu' \sin \theta' P^1_{2 n-1}(\mu') 
\tilde S_{\hat \omega}(s', \mu')\mbox{\Huge ]}
\end{eqnarray}

\noindent where $P_n(\mu)$ are the Legendre polynomials, $P_n^m(\mu)$ 
are the associated Legendre polynomials and $\sin(n\theta)$ is a function of 
$\mu$ through $\theta = \cos^{-1} \mu$.

The equation of hydrostatic equilibrium for a barytropic fluid is 

\begin{eqnarray}
h(\tilde P) - h_{\rm p} ~ \equiv ~ \int_{\tilde P_{\rm p}}^{\tilde P} 
\frac{d\tilde P}{\tilde \epsilon + \tilde P} & = & \ln u^t - \ln u_{\rm p}^t 
- \int_{\tilde \Omega_{\rm *, c}}^{\tilde \Omega_*} F(\tilde \Omega_*) d\tilde \Omega_*
\end{eqnarray}

\noindent where $h(\tilde P)$ is the dimensionless specific enthalpy as a 
function of pressure. $\tilde P_{\rm p}, u_{\rm p}^t$ and $h_{\rm p}$ are 
the dimensionless values of pressure, t-component of the four--velocity and the 
specific enthalpy at the pole. $\tilde \Omega_{\rm *, c}$ is the (dimensionless)
central value of the angular speed, which on the rotation axis is constant 
and equal to its value at the pole. $F(\tilde \Omega_*) = u^t u_{\phi}$ 
is obtained from an integrability condition on the equation of 
hydrostatic equilibrium. Choosing the form of this function fixes the 
rotation law for the matter. Following Komatsu et al. (1989), we set it to 

\begin{eqnarray}
F(\tilde \Omega_*) & = & A^2 (\tilde \Omega_{\rm *, c} - \tilde \Omega_*)
\end{eqnarray}

\noindent where $A$ is a rotation constant such that rigid rotation is 
achieved in the limit $A \rightarrow \infty$. An appropriately chosen value 
of $h_{\rm p}$ defines the surface of the star.

Integrating Eq. (2.20), we obtain

\begin{eqnarray}
h(\tilde P) - h_{\rm p} & = & \frac{1}{2} [\gamma_{\rm p}+\rho_{\rm p}
-\gamma-\rho-\ln (1-\tilde v^2)+A^2 (\tilde \Omega_* - \tilde 
\Omega_{\rm *, c})^2]
\end{eqnarray}

\noindent where $\gamma_{\rm p}$ and $\rho_{\rm p}$ are the values of the 
metric potentials at the pole. Therefore for the center and the equator of 
a rigidly rotating neutron star, we get

\begin{eqnarray}
h(\tilde P_{\rm c}) - h_{\rm p} - \frac{1}{2} [\gamma_{\rm p}+\rho_{\rm p} 
-\gamma_{\rm c}-\rho_{\rm c}] & = & 0
\end{eqnarray}

\noindent and

\begin{eqnarray} 
(\gamma_{\rm p}+\rho_{\rm p}-\gamma_{\rm e}-\rho_{\rm e})
- \ln [1 - (\tilde \Omega_{\rm *, e} - \tilde \omega_{\rm e})^2 
\tilde r_{\rm e}^2 e^{-2 \rho_{\rm e}}] & = & 0 
\end{eqnarray}
 
\noindent where the subscripts p, e and c denote the values at the pole, 
equator and center respectively.

We follow the formalism of Komatsu et al. (1989) to compute the equilibrium 
configurations of a rapidly rotating neutron star. 
For a given equation of state (EOS), we take 
the maximum energy density $\epsilon_{\rm c}$ and the ratio 
$(\tilde r_{\rm p}/\tilde r_{\rm e})$ of the coordinate radii 
at the pole and equator as the inputs. An equilibrium solution for given 
values of the configuration parameters is obtained iteratively in the 
following way. Let $\tilde r'_{\rm e}$ and the metric potentials 
$\rho', \gamma'$ \& $\alpha'$ be values of the current approximate 
solution. Then $\rho', \gamma'$ \& $\alpha'$ are first scaled (divided) 
by $(\tilde r'_{\rm e})^2$ to obtain $\hat \rho, \hat \gamma$ \& $\hat \alpha$.
A new value for $\tilde r_{\rm e}$ 

\begin{eqnarray}
\tilde r^2_{\rm e} & = & \frac{2 [h(\tilde P(\tilde \epsilon_{\rm c})) 
- h_{\rm p}]}{\hat \gamma_{\rm p}+\hat \rho_{\rm p}-\hat \gamma_{\rm c}
-\hat \rho_{\rm c}}
\end{eqnarray}

\noindent is obtained using Eq. (2.23). Using Eq. (2.24), we compute the 
value of $\hat \Omega_{\rm *, c}$ as 

\begin{eqnarray}
\hat \Omega_{\rm *, c} & = & \hat \omega_{\rm e} + 
e^{\rho_{\rm e}} \mbox{\Large [}1 - e^{(\gamma_{\rm p}+\rho_{\rm p}-
\gamma_{\rm e} -\rho_{\rm e})}\mbox{\Large ]}^{1/2}
\end{eqnarray}

\noindent Now $\hat \rho, \hat \gamma$ \& $\hat \alpha$ 
are rescaled (multiplied) 
by the new value of $\tilde r^2_{\rm e}$. Using these values, we solve 
Eq. (2.23) to obtain the new matter--energy distribution, namely 
$\tilde \epsilon, \tilde P, \tilde v$ etc. Finally, Eqs. (2.16)--(2.19) 
are solved for the new values of the metric potentials. These steps are 
repeated until the value of $\tilde r_{\rm e}$ converges to within a 
tolerance of $10^{-5}$. For a detailed description of the numerical 
procedure, see Cook et al. (1994); Datta et al. (1998) and 
Thampan (1999).

Once $\tilde r_{\rm e}$ converges, the metric potentials $\rho, \gamma, 
\hat \omega$ and $\alpha$ together with the density $(\tilde \epsilon)$ 
and pressure $(\tilde P)$ profiles can be used to compute the structure
parameters with the following formulae (Cook et al. 1994). The total mass 
$M$ is 

\begin{eqnarray}
M & = & \frac{4 \pi \kappa^{1/2} c^2 \tilde r^3_{\rm e}}{G} 
\int_0^1 \frac{s^2 ds}{(1-s)^4} \int_0^1 d\mu \,e^{2 \alpha+\gamma} \nonumber \\
 &   & \times \mbox{\Huge \{}\frac{\tilde \epsilon+\tilde P}{1-\tilde v^2} 
\mbox{\Huge [}1+\tilde v^2+\frac{2 s \tilde v}{1-s} (1-\mu^2)^{1/2} 
\hat \omega e^{-\rho}\mbox{\Huge ]}+2 \tilde P\mbox{\Huge \}}
\end{eqnarray}

\noindent The total rest (baryonic) mass $M_0$ of the system is given by 

\begin{eqnarray}
M_0 & = & \frac{4 \pi \kappa^{1/2} m_{\rm B} c^2 \tilde r^3_{\rm e}}{G} 
\int_0^1 \frac{s^2 ds}{(1-s)^4} \int_0^1 d\mu \,e^{2 \alpha+(\gamma
-\rho)/2} \frac{\tilde n}{(1-\tilde v^2)^{1/2}}
\end{eqnarray}

\noindent where $\tilde n$ is the dimensionless baryonic number density 
and $m_{\rm B}$ is the mass per baryon. The total proper mass $M_p$ 
of the system represents the energy stored in the configuration excluding 
gravitational potential energy and rotational kinetic energy. It is 
defined as

\begin{eqnarray}
M_p & = & \frac{4 \pi \kappa^{1/2} c^2 \tilde r^3_{\rm e}}{G} 
\int_0^1 \frac{s^2 ds}{(1-s)^4} \int_0^1 d\mu \,e^{2 \alpha+(\gamma
-\rho)/2} \frac{\tilde \epsilon+\tilde P}{(1-\tilde v^2)^{1/2}}
\end{eqnarray}

\noindent The total angular momentum $J$ of the system is given by 

\begin{eqnarray}
J & = & \frac{4 \pi \kappa c^3 \tilde r^4_{\rm e}}{G}
\int_0^1 \frac{s^3 ds}{(1-s)^5} \int_0^1 d\mu \,(1-\mu^2)^{1/2} 
e^{2 \alpha+\gamma-\rho} (\tilde \epsilon+\tilde P) 
\frac{\tilde v}{1-\tilde v^2}
\end{eqnarray}

\noindent The moment of inertia $I$ is obtained by the prescription 

\begin{eqnarray}
I & = & \frac{J}{\Omega_*}
\end{eqnarray}

\noindent The total rotational kinetic energy $T$ of the system is defined by 

\begin{eqnarray}
T & = & \frac{2 \pi \kappa^{1/2} c^2 \tilde r^3_{\rm e}}{G}
\int_0^1 \frac{s^3 ds}{(1-s)^5} \int_0^1 d\mu \,(1-\mu^2)^{1/2} 
e^{2 \alpha+\gamma-\rho} (\tilde \epsilon+\tilde P) 
\frac{\tilde v \hat \Omega_*}{1-\tilde v^2}
\end{eqnarray}

\noindent Then the gravitational binding energy $W$ of the star is given by 

\begin{eqnarray}
W & = & M_p + T - M
\end{eqnarray}

\noindent The circumferential radius $R$ at the equator is defined by 

\begin{eqnarray}
R & = & \kappa^{1/2} \tilde r_{\rm e} e^{(\gamma_{\rm e}
-\rho_{\rm e})/2}
\end{eqnarray}

\noindent where the subscript `e' denotes evaluation at the equator.

\section{Luminosity Calculation}

We calculate the luminosities of the accretion disk and the boundary layer 
using the test particle approach, i.e., we determine the amount of 
gravitational energy release by a test particle, as it spirals in.
Since the chosen metric (given by Eq. 2.3) is stationary and axisymmetric, 
the energy and angular momentum of this particle are constants of motion. 
As we consider a geometrically thin disk, the particle is always confined 
to the equatorial plane. Then using the standard Lagrangian technique, 
the equations of motion of the particle can be written as (Thampan \& Datta 
1998)

\begin{eqnarray}
\dot t ~ \equiv ~ \frac{dt}{d\tau} & = & e^{-(\gamma+\rho)} (\tilde E - 
\omega l)
\end{eqnarray}

\begin{eqnarray}
\dot \phi ~ \equiv ~ \frac{d\phi}{d\tau} 
 ~ \equiv ~ \Omega_{\rm tp} \dot t & = & e^{-(\gamma+\rho)} \omega 
(\tilde E - \omega l) + \frac{l}{\bar r^2 e^{(\gamma-\rho)}}
\end{eqnarray}

\begin{eqnarray}
\dot{\bar r}^2 ~ \equiv ~ e^{2 \alpha+\gamma+\rho} 
\mbox{\Huge (}\frac{d\bar r}{d\tau}\mbox{\Huge )}^2 & = & 
\tilde E^2 - \tilde V^2
\end{eqnarray}

\noindent Here $\tau$ is the proper time, $\Omega_{\rm tp}$,  
$\tilde E$ and $l$ are the angular speed, specific
energy and specific angular momentum of the test particle respectively and
$\tilde V$ is the effective potential given by 

\begin{eqnarray}
\tilde V^2 & = & e^{\gamma+\rho} \mbox{\Huge [}1+
\frac{l^2/\bar r^2}{e^{\gamma-\rho}}\mbox{\Huge ]} + 
2 \omega \tilde E l - \omega^2 l^2
\end{eqnarray}

\noindent The conditions for circular orbits, extremum of energy and 
minimum of energy are respectively: 

\begin{eqnarray}
\tilde E^2 & = & \tilde V^2 \\
\tilde V_{,\bar r} & = & 0 \\
\tilde V_{,\bar r\bar r} & > & 0
\end{eqnarray}

\noindent For marginally stable orbits, 

\begin{eqnarray}
\tilde V_{,\bar r\bar r} & = & 0
\end{eqnarray}

\noindent In our notation, a comma followed by one `$\bar r$' represents a 
first order partial derivative with respect to $\bar r$ and so on.

Using Eqs. (2.35), (2.36) and (2.39), the condition for circular orbits can
be written as 

\begin{eqnarray}
\tilde E - \omega l & = & \frac{e^{(\gamma+\rho)/2}}{\sqrt{1-\tilde 
v_{\rm tp}^2}} \\
l  & = & \frac{\tilde v_{\rm tp} \tilde r e^{(\gamma-\rho)/2}}{\sqrt{1-\tilde 
v_{\rm tp}^2}} 
\end{eqnarray}

\noindent where $v_{\rm tp}$, the proper velocity (in the equatorial plane) 
of the test particle relative to ZAMO, is given by 

\begin{eqnarray}
\tilde v_{\rm tp} & = & (\Omega_{\rm tp} - \omega) \bar r e^{-\rho}
\end{eqnarray}

\noindent Conditions (2.40) and (2.42) yield respectively,

\begin{eqnarray}
\tilde v_{\rm tp} & = & \frac{e^{-\rho} \bar r^2 \omega_{,\bar r} \pm 
[e^{-2 \rho} \bar r^4 \omega_{,\bar r}^2 + 2 \bar r (\gamma_{,\bar r}+
\rho_{,\bar r}) + \bar r^2 (\gamma_{,\bar r}^2-\rho_{,\bar r}^2)]^{1/2}}{2+
\bar r(\gamma_{,\bar r}-\rho_{,\bar r})}
\end{eqnarray}

\begin{eqnarray}
\tilde V_{,\bar r\bar r} & \equiv & 2 \mbox{\Huge [}\frac{\bar r}{4} 
(\rho_{,\bar r}^2-\gamma_{,\bar r}^2) - \frac{1}{2} e^{-2 \rho} 
\omega_{,\bar r}^2 \bar r^3 - \rho_{,\bar r} + \frac{1}{\bar r}\mbox{\Huge ]}
\tilde v_{\rm tp}^2 \nonumber \\
 &   & + [2+\bar r (\gamma_{,\bar r}-\rho_{,\bar r})] \tilde v_{\rm tp} 
\tilde v_{{\rm tp},\bar r} - e^{-\rho} \omega_{,\bar r} \bar r \tilde 
v_{\rm tp}\nonumber \\
 &   & + \frac{\bar r}{2} (\gamma_{,\bar r}^2-\rho_{,\bar r}^2) - 
e^{-\rho} \bar r^2 \omega_{,\bar r} \tilde v_{{\rm tp},\bar r} ~ = ~ 0
\end{eqnarray}

\noindent where we have made use of Eq. (2.46) and its derivative with 
respect to $\bar r$ in order to eliminate the second order derivatives in 
Eq. (2.47). The zero of $\tilde V_{,\bar r\bar r}$ gives the radius 
$(r_{\rm orb})$ of the innermost stable circular orbit (ISCO) and the 
corresponding $\tilde v_{\rm tp}$ yields $\tilde E$ and $l$.
In Eq. (2.46), the positive sign refers to the co--rotating particles and 
the negative sign to the counter--rotating particles. In our work we 
consider only the co--rotation case.

In a circular orbit, the Keplerian angular speed of the test particle is 
denoted by $\Omega_{\rm K}$. Using Eq. (2.45), we get the $\Omega_{\rm K}$ 
profile as

\begin{eqnarray}
\Omega_{\rm K}(\bar r) & = & e^{\rho(\bar r)} \frac{\tilde 
v_{\rm tp}(\bar r)}{\bar r} + \omega(\bar r) 
\end{eqnarray}

\noindent where $\tilde v_{\rm tp}$ is given by Eq. (2.46).
The value of $\Omega_{\rm K}$ in an orbit at the surface of the 
neutron star puts a firm upper limit on the angular speed the star can attain 
(Friedman et al. 1986) and hence the boundary layer luminosity, when the star 
attains this maximum $\Omega_*$, should be zero (Sunyaev \& Shakura 1986).

Depending on the chosen EOS and the values of $M$ and $\Omega_*$, the 
equatorial radius ($R$) of the neutron star can be greater than or less than 
$r_{\rm orb}$. The accretion luminosities are different for these two 
cases (Kluzniak \& Wagoner 1985; Sunyaev \& Shakura 1986; Datta et al. 1995). 
These quantities can be calculated in the following way.

For $R > r_{\rm orb}$, the disk extends upto the surface of the weak 
magnetic field neutron star. The energy of a test particle at infinite 
distance from the star is equal to its rest mass $m_0$. Now the specific disk 
luminosity $(E_{\rm D})$ is equal to the gravitational energy release by 
the particle in the unit of $m_0$. Therefore

\begin{eqnarray}
E_{\rm D} & = & 1-\tilde E_{\rm K}(r=R)
\end{eqnarray}

\noindent where $E_{\rm K}(r=R)$ is the specific energy of the particle 
in Keplerian orbit at the surface, obtained by solving Eq. (2.43), (2.44) 
and (2.46). 

The specific boundary layer luminosity $(E_{\rm BL})$ is equal to the 
energy loss (in the unit of $m_0$) by the particle in the boundary layer 
(a very narrow gap near the neutron star surface). Therefore 

\begin{eqnarray}
E_{\rm BL} & = & \tilde E_{\rm K}(r=R) - \tilde E_0
\end{eqnarray}

\noindent where $\tilde E_0$ is the energy of the particle `at rest' on the 
stellar surface (the particle will be moving with the velocity 
$\tilde v_{\rm tp} = \tilde v_*$ of the stellar fluid at the surface, 
where $\tilde v_*$ is obtained by substituting into Eq. (2.6) all the 
relevant parameters for $r = R$) and is calculated by solving Eqs. (2.43) 
and (2.44) for $\tilde E$ at $r = R$ and $\tilde v_{\rm tp} = \tilde v_*$.

For $R < r_{\rm orb}$, the accretion disk does not touch the surface 
of the star. The specific disk luminosity is given by 

\begin{eqnarray}
E_{\rm D} & = & 1 - \tilde E_{\rm orb}
\end{eqnarray}

\noindent Consequently the specific boundary layer luminosity is 

\begin{eqnarray}
E_{\rm BL} & = & \tilde E_{\rm orb} - \tilde E_0
\end{eqnarray}

\noindent Here $\tilde E_{\rm orb}$ is the specific energy of the particle 
in ISCO, calculated using the Eqs. (2.43), (2.44) and (2.46) for 
$r = r_{\rm orb}$.

\section{Equation of State}

For calculating the structure of a neutron star, we need to know
its equation of state (EOS), i.e., the pressure $P$ as a function
of the matter--energy density $\epsilon$. The outer crust of the star 
is expected to be made of $^{56}$Fe, as its binding energy per nucleon 
is the lowest among all the atoms. As we proceed towards the center,
the density increases enormously and the matter becomes degenerate.
At the nuclear density $(\epsilon_0 = 2.4 \times 10^{14} \mbox{g cm}^{-3})$, 
the nucleii dissolve and all the nucleons form a single
huge nucleus. The composition of matter upto this density is fairly well 
understood. For densities $\epsilon > \epsilon_0$, we have to rely on 
extrapolation from known nuclear properties under terrestrial conditions. 
The goodness of such extrapolations is checked by how well it reproduces the 
values of parameters like compression modulus of equilibrium nuclear matter, 
the nuclear saturation density, symmetry energy etc. (for which 
experimental estimates are available).   

\nopagebreak
\begin{figure}[h]
\psfig{file=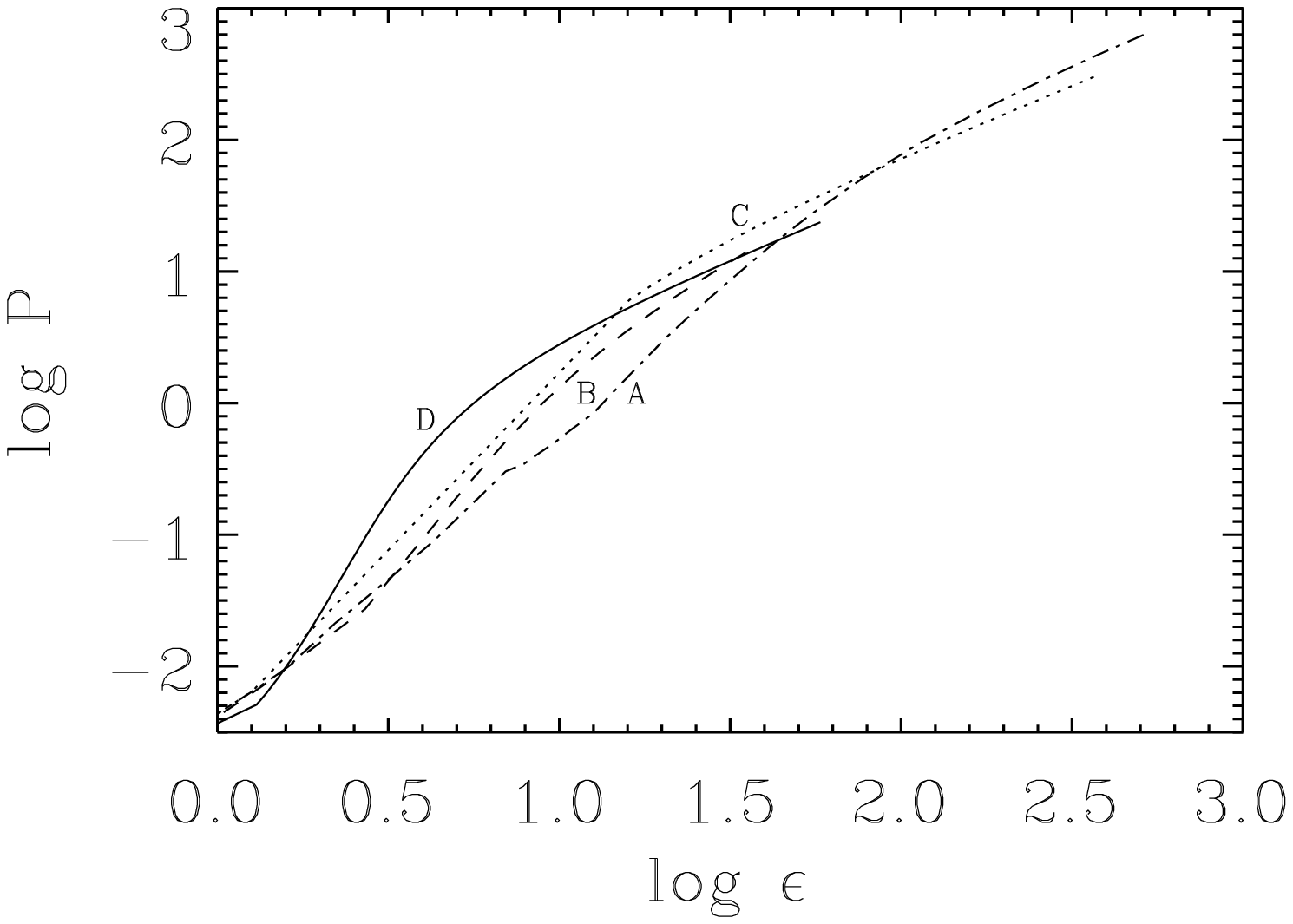,width=15 cm}
\caption{Logarithmic plot of pressure vs. matter density for the
EOS models used here. The density and pressure are in units of
$1.0\times10^{14}$~\gcc and $(1.0\times10^{14})~c^2$~cgs respectively.}
\end{figure}

The structure of neutron stars depends sensitively on the EOS at high 
densities. Although the main composition of degenerate matter at densities
higher than $\epsilon_0$ is expected to be dominated by neutrons, 
significant admixtures of other elementary particles (such as pions, kaons 
and hyperons) are not ruled out. In the literature, many EOS models are 
available. The various formalisms used in deriving these models give rise to 
a substantial spread in their qualitative features. Which of these is the 
correct EOS model is therefore a fundamental question of physics. It is 
hoped that a theoretical computation of quantities of astrophysical interests 
using representative EOS models and subsequent comparison with observations 
will provide an answer to this question. This is one of the main motivations 
for the work presented in this thesis.

For excellent reviews on neutron star EOS models, we refer Canuto (1974), 
Canuto (1975) and Baym \& Pethick (1975) (also see Shapiro \& Teukolsky 1983). 
In this thesis, we have studied luminosities, disk temperature profiles and 
spectra for certain representative EOS models. An important quantity that 
characterises EOS models is the stiffness parameter, defined as $S = 
d\log P/d\log \epsilon$. For higher values of $S$, the EOS model is 
stiffer. For every EOS, there exists a maximum possible stable mass 
$(M_{\rm max})$. The stiffer the EOS, the higher is the value of $M_{\rm max}$.

For our calculations, we choose four EOS models of widely varying 
stiffness parameters. This ensures sufficient generality of our results.
We describe below the salient features of these models.

(A) {\it Pandharipande (hyperonic matter)}: One of the early attempts to 
derive nuclear EOS with admixture of hyperons is due to Pandharipande (1971), 
who assumed the hyperonic potentials to be similar to the nucleon--nucleon 
potentials, but altered suitably to represent the different isospin states. 
The many--body method adopted is based on the variational approach of 
Jastrow (1955). The two--body wave function was taken as satisfying a 
simplified form of the Bethe--Goldstone equation, in which terms representing 
the Pauli exclusion principle were omitted but simulated by imposing a 
`healing' constraint on the wave function. This model is soft, i.e., the 
value of $S$ is comparatively low. 
The nonrotating $M_{\rm max}$ for this EOS is $1.41 ~\msun$.

(B) {\it Baldo, Bombaci \& Burgio (AV14 + 3bf)}:  Baldo, Bombaci \& Burgio 
(1997) have given a microscopic EOS for asymmetric nuclear matter, 
derived from the Brueckner--Bethe--Goldstone many--body theory with explicit 
three--body terms. The three--body force parameters are adjusted to give a 
reasonable saturation point for nuclear matter. 
This model is intermediate in stiffness with nonrotating 
$M_{\rm max} = 1.79 ~\msun$.

(C) {\it Walecka (neutrons)}: The EOS model of Walecka (1974) corresponds to 
pure neutron matter and is based on a mean--field theory with exchange of 
scaler and (isoscalar) vector mesons representing the nuclear interaction.
It is a stiff EOS model with nonrotating $M_{\rm max} = 2.28 ~\msun$.

(D) {\it Sahu, Basu \& Datta}: Sahu, Basu \& Datta (1993) gave a field 
theoretical EOS for neutron--rich matter in beta equilibrium based on the 
chiral sigma model. The model includes an isoscalar vector field generated 
dynamically and reproduces the empirical values of the nuclear matter 
saturation density and binding energy and also the isospin symmetry 
coefficient for asymmetric nuclear matter. The energy per nucleon of 
nuclear matter according to these authors is in very good agreement, upto 
about four times the equilibrium nuclear matter density, with estimates 
inferred from heavy--ion collision experimental data.
This model is the stiffest among all the EOS models we have considered.
The nonrotating $M_{\rm max}$ for this EOS is $2.59 ~\msun$.

The pressure--density relationship of the above EOS models is illustrated 
in Fig 2.1. The composite EOS for the entire span of neutron star densities 
is constructed by joining one of the selected high 
density EOS models to that of Negele \& 
Vautherin (1973) for the density range $10^{14} - 5 \times 10^{10} 
\mbox{g cm}^{-3}$, Baym, Pethick \& Sutherland (1971) for densities down to 
$\sim 10^3 \mbox{g cm}^{-3}$ and Feynman, Metropolis \& Teller (1949) for 
densities less than $10^3 \mbox{g cm}^{-3}$.

\section{The Results}

\nopagebreak
\begin{figure}[h]
\psfig{file=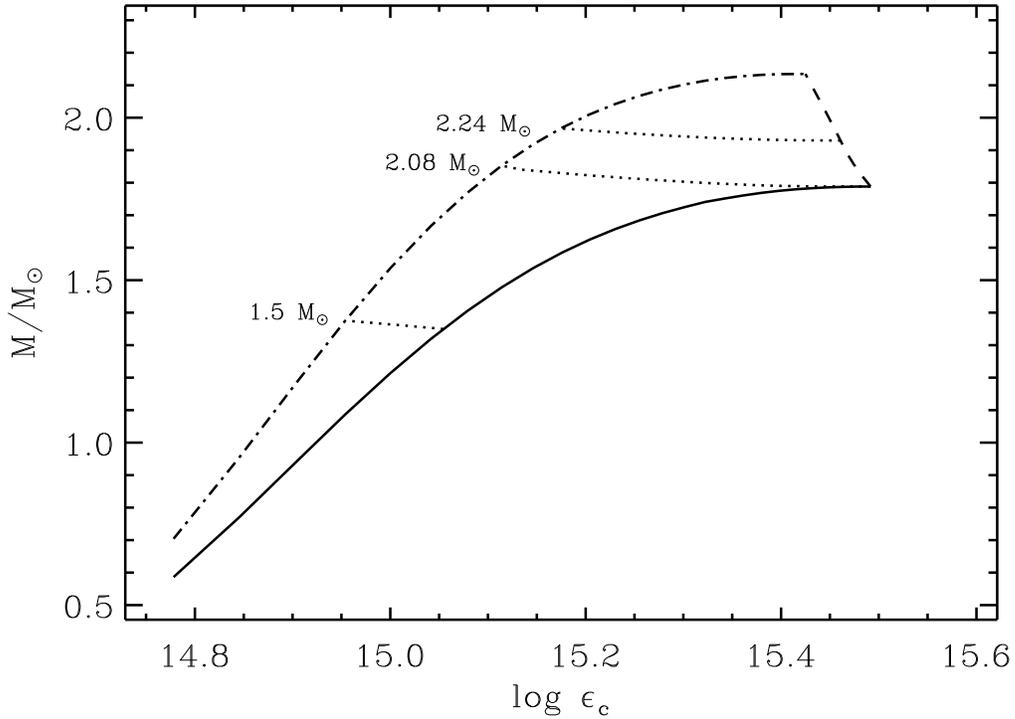,width=15 cm}
\caption{Semi--logarithmic plot of gravitational mass vs. central matter
density (in unit of $10^{14}$ g cm$^{-3})$ for the EOS model B. The solid line
is for static limit, dash--dot line is for mass-shed limit and the dashed line
is for radial-instability limit. The stable equilibrium configurations
occur only in the region, bound by these three limits (see the text). 
The dotted lines are evolutionary sequences with the values of corresponding 
$M_{\rm 0}$ written.}
\end{figure}

\nopagebreak
\begin{figure}[h]
\psfig{file=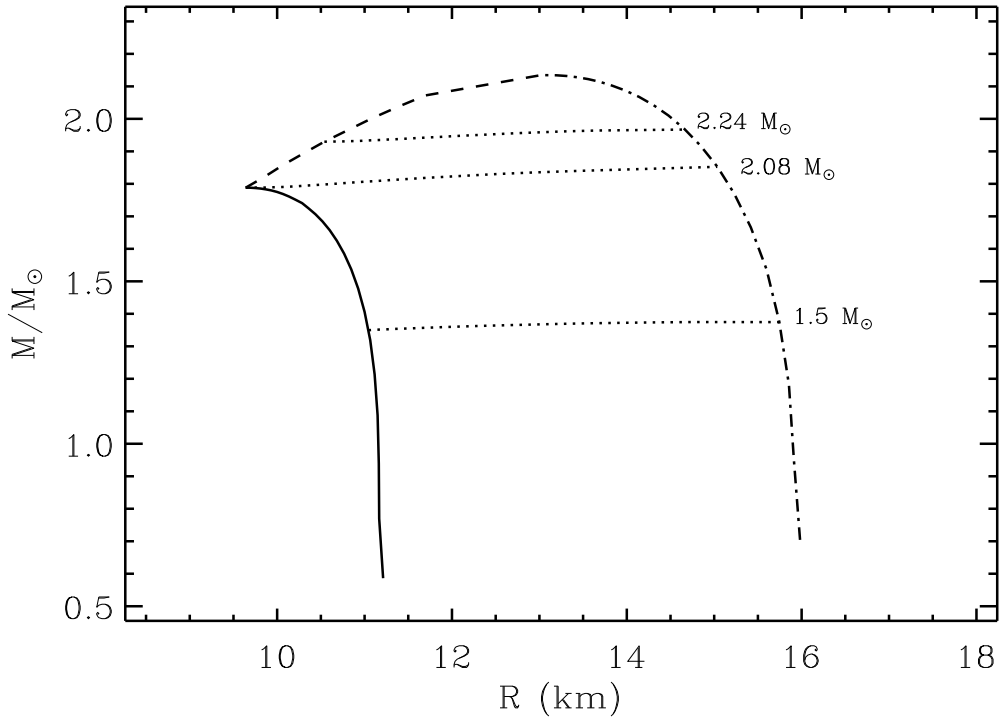,width=15 cm}
\caption{Plot of gravitational mass vs. equatorial radius for the EOS model B.
All the lines are as described in Fig. 2.2.}
\end{figure}

\nopagebreak
\begin{figure}[h]
\psfig{file=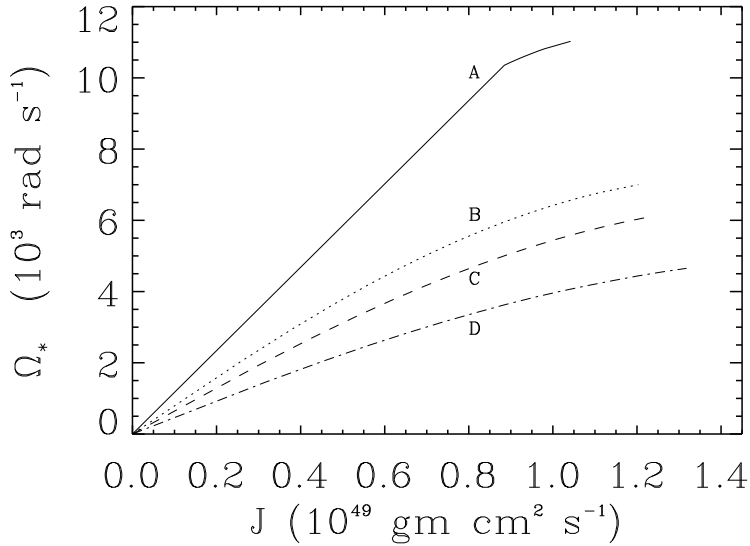,width=12 cm}
\caption{Angular speed $(\Omega_{\rm *})$ as a function of total angular
momentum $(J)$. The curves are for different EOS models (mentioned in the 
figure) and are for a fixed gravitational mass $(M = 1.4~\msun)$.}
\end{figure}

\nopagebreak
\begin{figure}[h]
\psfig{file=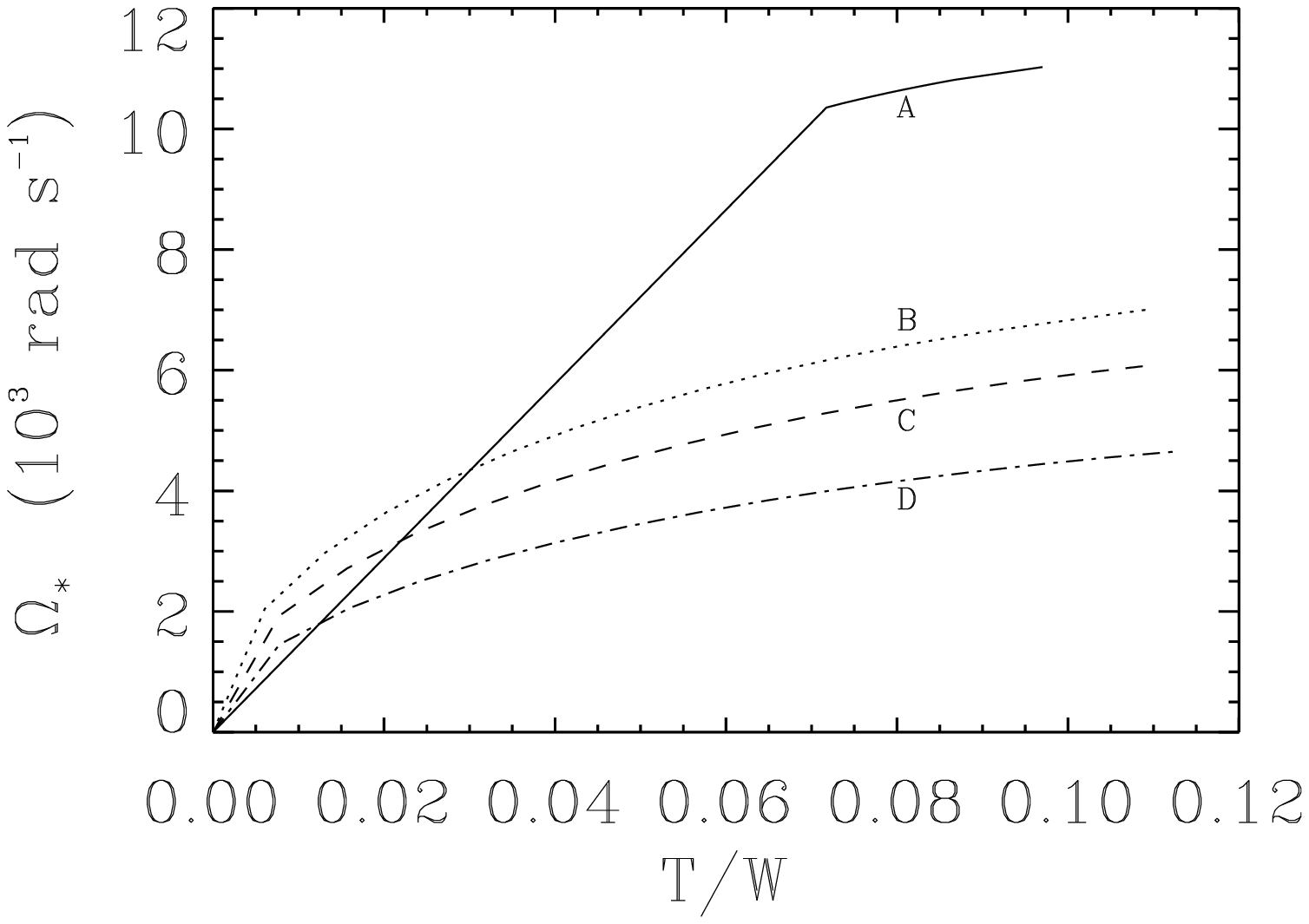,width=12 cm}
\caption{Angular speed $(\Omega_{\rm *})$ as a function of the ratio of 
rotational kinetic energy and gravitational binding energy $(T/W)$. 
Curve labels have the same meaning as in Fig. 2.4.}
\end{figure}

We calculate the equilibrium configurations of neutron stars with $\Omega_*$ 
ranging from 0 to the mass--shed limit value. In Fig 2.2, we illustrate 
the nonrotating limit, mass--shed limit and radial instability limit for the 
EOS model (B). Several evolutionary sequences are also shown.

In Fig 2.3, the above three limits are illustrated in $M-R$ space, for the 
same EOS model. It is to be noted that $M-R$ diagram has in general a 
negative slope for neutron stars (as we will see in Chapter 7, the slope 
is positive for strange stars).
 
We construct gravitational mass sequences for all the chosen EOS models. 
We take $M = 1.4 ~\msun$ (the canonical mass) for the purpose of illustration.
In Fig 2.4, we plot $\Omega_*$ vs. $J$ from the nonrotating limit to the 
mass--shed limit. We see that both $J$ and $\Omega_*$ increase monotonically. 
We also notice that for softer EOS, higher value of $\Omega_*$ can be achieved 
at the mass--shed limit, but the corresponding value of $J$ is smaller. 

In Fig 2.5, we plot $\Omega_*$ vs. $T/W$ with other specifications same as in 
Fig 2.4. Here we always present the absolute value of $W$.
As we will elaborate in Chapter 7, the higher the value of $T/W$, the greater
is the possibility for the star to be a subject of triaxial instability.
As we see from the figure, for a stiffer EOS, the value of $T/W$ is higher, 
but the maximum value does not exceed $0.12$. 
(for strange stars, it is $0.25-0.30$, see Chapter 7).

The value of $r_{\rm orb}$ compared to that of $R$ has profound effect on the 
disk luminosity, temperature profile and spectrum. We will illustrate the 
variations of $r_{\rm in}$ (radius of disk inner edge) and luminosities with
$\Omega_*$ (for $M = 1.4 ~\msun$) in the next chapter.
  
\section{Concluding Remarks}

It is expected that the accreting neutron stars are rapidly rotating 
because of the huge amount of angular momentum, transfered to them by 
the accreted matter. The very short pulsation period $(P = 2.49 \mbox{ms})$ 
of SAX J1808.4--3658 strengthens this speculation. Therefore we 
compute the equilibrium configurations for rapidly rotating neutron stars, 
considering the full effect of general relativity. Then using the structure 
parameters and metric coefficients for these configurations, we calculate 
general relativistically correct values for luminosities, disk temperature
profiles and disk spectra as functions of $\Omega_*$. Comparing these model
spectra with the observed ones will help to constrain neutron star 
structure parameters, as well as the EOS. In the subsequent chapters, 
we will elaborate the importance of rapid--rotation--calculation, by
showing that the results for such calculation is considerably different 
from those for Schwarzschild or Newtonian case.

%% file: chap3.tex
\markright{Chapter 3}
\def\note #1]{{\bf #1]}}
\chapter{Calculation of Disk Temperature Profile}


\section{Introduction}

The soft X--ray spectra of luminous low--mass X--ray binaries (LMXBs) are
believed to originate in geometrically thin accretion disks around neutron
stars with weak surface magnetic fields (see for e.g. White 1995).
An important parameter in modeling these
spectra is the maximum value of the effective temperature in the accretion
disk. The effective temperature profile in the disk can be estimated
(assuming the disk to radiate from its surface like a blackbody) if one
knows the accretion energy released in the disk.
In a Newtonian treatment, the innermost region of an accretion disk
surrounding a neutron star with weak magnetic field will extend rather
close to the neutron star surface. The amount of energy released in the
disk will be one--half of the total accretion energy, the other half being
released in the thin boundary layer between the disk's inner edge and the
neutron star's surface. This then gives the disk effective temperature
$(T_{\rm eff})$ varying with the radial distance $(r)$ as
$T_{\rm eff} \propto r^{-3/4}$
and the maximum effective temperature $(T^{\rm max}_{\rm eff})$ will depend on
the (nonrotating) neutron star mass $(M)$ and radius $(R)$
as $T_{\rm eff}^{\rm max} \propto (M \dot{M}/R^3)^{1/4}$,
where $\dot{M}$ is the steady state mass accretion rate. The value of
$(T^{\rm max}_{\rm eff})$ in the disk, in this approach, occurs at a radial
distance $1.36~R$.

Mitsuda et al. (1984) parameterized the disk
spectrum by the maximum temperature of the disk, using the above
formalism and assuming the mass of the neutron star is equal to
$1.4~\msun$. These authors assumed that the inner parts of the disk
do not contribute to the X--ray spectrum, and suggested a multi--color
spectrum for the X--ray emission from the disk. It was shown by these
authors, that the observed spectra of Sco X--1, 1608--52, GX 349+2 and
GX 5--1,
obtained with the {\it Tenma} satellite, can be well fitted with the
sum of a multi--color spectrum and a single blackbody spectrum (presumably
coming from the boundary layer). White, Stella \& Parmar (1988) (WSP)
suggested
that the simple blackbody accretion disk model should be modified
to take into account the effects of electron scattering.
Using {\it EXOSAT} observations, these authors compared the spectral
properties of the
persistent emission from a number of X--ray burst sources
with various X--ray emission models.
This work suggests that either the neutron star (in each system
considered) rotates close to equilibrium with the Keplerian
disk, or that most of the boundary layer emission is not represented
by a blackbody spectrum.

For accretion disks around compact objects, one possibility is that of
the accretion disk not being Keplerian in nature.  For e.g. Titarchuk,
Lapidus \& Muslimov (1998) have formulated a boundary problem in which
the Keplerian accretion flow in the inner disk is smoothly adjusted
to the neutron star rotation rate.  The generality of such a formulation
permits application even to black holes, but only for certain assumed inner
boundary conditions.  These authors demonstrate that there exists a
transition layer (having an extent of the order of the neutron star radius)
in which the accretion flow is sub-Keplerian.  An attractive feature of this
formalism is that it allows super-Keplerian motion at the outer boundary of
the transition layer, permitting the formation of a hot blob that ultimately
bounces out to the magnetosphere. This formalism (Titarchuk \& Osherovich
1999; Osherovich \& Titarchuk 1999a; Osherovich \& Titarchuk 1999b;
Titarchuk, Osherovich \& Kuznetsov 1999) therefore provides a mechanism
for the production of high frequency quasi--periodic oscillations (QPOs)
observed in the X--ray flux from several LMXBs.  Such effects, when taken
into account, can modify the Newtonian disk temperature profile
(Chakrabarti \& Titarchuk 1995).

There are several other effects which will modify the Newtonian disk
temperature profile, such as the effects of general relativity and of
irradiation of the disk by the central neutron star. The wind mass loss
from the disk and the residual magnetic field near the disk's inner edge
may also play a part in modifying the effective temperature (Knigge 1999).
Czerny, Czerny \& Grindlay (1986) calculated LMXB disk spectra assuming
that a disk radiates locally as a blackbody with the energy flux detemined
by viscous forces, as well as irradiation by the boundary layer, and took
into account relativistic effects, some of them in an approximate way.
The possible effects of general relativity were also discussed by
Hanawa (1989), using the Schwarzschild (nonrotating) metric, assuming that
the neutron star radius is
less than the radius of the innermost stable circular orbit
($r_{\rm in} = 6 G M/c^2$), which they identified
as the disk inner boundary. The color temperature
was assumed to be higher than the effective temperature by a factor
of 1.5.  It was found by Hanawa (1989) that the observations are
consistent with a geometrically thin, optically thick accretion
disk, whose inner edge is at $r=r_{\rm in}$, $r$ being the Schwarzschild
radial coordinate.

An important dynamical aspect of disk accretion on to a weakly magnetized
neutron star is that the neutron star will get spun up to its equilibrium
period, which is of the order of milliseconds (see Bhattacharya \& van den
Heuvel 1991, and refereces therein).
The effect of rotation is to increase the equatorial radius of the
neutron star, and also to relocate the innermost stable circular orbit (for a
corotating disk) closer
to the stellar surface (as compared to the Schwarzschild case).
These effects will be substantial for rapid rotation rates in a fully
general relativistic treatment that includes rotation. Therefore, for
accreting neutron stars with low magnetic fields,  the
stellar radius can be greater or less than the radius of the
innermost stable orbit, depending on the neutron star equation of state
and the spacetime geometry.  The effect of magnetic field will be to
constrain the location of the inner--edge of the accretion disk to the
magnetospheric (Alf\'{v}en) radius. In such a case, $r_{\rm in}$ would lose
the astrophysical relevance as discussed here.  However, this will be so
only if the magnetic field strength ($B$) is large.  The problem addressed
in this paper refer to LMXBs which contain old neutron stars which are
believed to have undergone sufficient magnetic field decay (Bhattacharya \&
Datta 1996).  Clearly, for low magnetic field case, a number of different
disk geometries will be possible if general relativistic effects of rotation
are taken into account. These structural differences influence the effective
temperature profile and the conclusions derived by Czerny, Czerny \& Grindlay
(1986) and Hanawa (1989) are likely to be modified.

In this chapter, we attempt to highlight the effects of general relativity 
and rotation of the neutron star on the accretion disk temperature profile. 
For simplicity (unlike Titarchuk, Lapidus \& Muslimov 1998), we assume the
accretion disk to be fully Keplerian, geometrically thin and optically thick.
We construct gravitational mass sequences for the chosen EOS models and 
calculate the luminosities and temperature profiles for equilibrium 
configurations corresponding to different $\Omega_*$ values.

In section 3.2, we will describe the procedure for disk temperature profile 
calculation. We will show the results in section 3.3 and summarise the 
content of the chapter in section 3.4.

\section{The Effective Temperature of the Disk}

\subsection{Effects of General Relativity and Rotation}

The effective temperature in the disk (assumed to be optically thick) is
given by

\begin{eqnarray}
T_{\rm eff} & = & (F/\sigma)^{1/4}
\end{eqnarray}

\noindent where $\sigma$ is the Stephan--Boltzmann constant and $F$ is
the X--ray energy flux per unit surface area. We use the formalism
given by Page \& Thorne (1974), who gave the following general relativistic
expression for $F$ emitted from the surface of an (geometrically thin and 
non--self--gravitating) accretion disk around a rotating black hole:

\begin{eqnarray}
F(r) & = & \frac{\dot{M}}{4 \pi r} f(r)
\end{eqnarray}

\noindent where

\begin{eqnarray}
f(r) & = & -\Omega_{{\rm K},r} (\tilde{E} - \Omega_{\rm K} l)^{-2}
\int_{r_{\rm in}}^{r} (\tilde{E} - \Omega_{\rm K} l) l_{,r} dr.
\end{eqnarray}

\noindent Here $r_{\rm in}$ is the disk inner edge radius, $\tilde{E}$,
$l$
are the specific energy and specific angular momentum of a test particle
in a Keplerian orbit and $\Omega_{\rm K}$ is the Keplerian angular velocity at
radial distance $r$. In our notation, a comma followed by a variable as
subscript to a quantity, represents
a derivative of the quantity with respect to the variable. We 
use the geometric units $c=G=1$.
Eq. (3.3) is valid for a spacetime described by a stationary, axisymmetric, 
asymptotically flat and reflection--symmetric (about the equatorial
plane) metric. Our metric (2.3) satisfies all these conditions.

For accreting neutron stars located within the disk inner edge,
the situation is analogous to the black hole binary case, and the
above formula, using a metric describing a rotating neutron
star, can be applied directly for our purpose. However, unlike the black
hole binary case, there can be situations for neutron star binaries
where the  neutron star
radius exceeds the innermost stable circular orbit radius.
In such situations, the boundary condition, assumed by Page \& Thorne
(1974), that the torque vanishes
at the disk inner edge will not be strictly valid.
Use of Eq. (3.1) will then be an approximation. This
will affect the temperatures close to the disk inner edge, but not
the $T_{\rm eff}^{\rm max}$ to any significant degree (see section 3.4
for discussion).

In order to evaluate $T_{\rm eff}$ using Eq. (3.1), we need to
know the radial profiles of $\tilde{E}$, $l$ and $\Omega_{\rm K}$.
For this purpose, first we construct gravitational mass sequences 
starting from the static limit all the way upto the mass--shed limit.
Then the radial profiles are calculated using Eqs. (2.43), (2.44) and (2.48).

Eq. (3.1) gives the effective disk temperature $T_{\rm eff}$
with respect to an observer comoving with the disk.  From the observational
viewpoint this temperature must be modified, taking into account the
gravitational redshift and the rotational Doppler effect.
In order to keep our analysis tractable, we use
the expression given in Hanawa (1989) for this modification :

\begin{eqnarray}
1+z = (1-\frac{3M}{r})^{-1/2}.
\end{eqnarray}

\noindent This equation is a special case of Eq. (5.3) with the inclination 
angle $i = 0$ and Schwarzschild metric used. Such assumptions make the 
calculation easier, but does not affect the general conclusion of Chapter 4.
With this correction for $(1+z)$, we define a temperature relevant for
observations ($T_{\rm obs}$) as:

\begin{eqnarray}
T_{\rm obs} = \frac{1}{1+z} T_{\rm eff}
\end{eqnarray}

\subsection{Disk Irradiation by the Neutron Star}

For luminous LMXBs, there can be substantial irradiation of the disk surface
by the radiation coming from the  neutron star boundary layer.  The radiation
temperature at the surface of a disk irradiated by a central source is given
by  (King, Kolb \& Burderi 1996)
\begin{eqnarray}
T_{\rm irr}(r) & = &
\left(\frac{\eta \dot{M}c^2(1-\beta)}{4\pi \sigma r^2} \frac{h}{r} (n-1) \right)
^{1/4}
\label{eq: tirr}
\end{eqnarray}
\noindent where $\eta$ is the efficiency of conversion of accreted rest mass
to energy, $\beta$ is the X--ray albedo, $h$ is the
half--thickness of the disk at $r$ and $n$ is given by the relation
$h \propto r^n$.  For actual values of $\beta$, $h/r$ and $n$, needed for
our computation here, we choose the same values (i.e., 0.9, 0.2 and 9/7 respectively)
as given in King, Kolb \& Burderi (1996). It is to be noted that the constant value
taken for $h/r$ is an approximation, as $n \ne 1$ . However, it does not
change the relative feature 
(which may be important for disk instability) of $T_{\rm irr}(r)$ and $T_{\rm eff}(r)$ 
much.
Although Eq. (3.6) is derived based on Newtonian
considerations, corrections due to general relativity (including that of
rapid rotation) will be manifested through the factor $\eta$.  We have made a
general relativistic evaluation
of $\eta$ for various neutron star rotating configurations, corresponding
to realistic neutron star EOS models, as described in Thampan \& Datta (1998).
Since $T_{\rm irr}(r) \propto r^{-1/2}$ and
$T_{\rm eff}(r) \propto r^{-3/4}$,   $T_{\rm irr}$ will dominate over
$T_{\rm eff}$ only at large distances.
The net effective
temperature of the disk will be given by  (see Vrtilek et al. 1990)
\begin{eqnarray}
T_{\rm disk}(r) & = & (T_{\rm eff}^4(r) + T_{\rm irr}^4(r))^{1/4}
\label{eq: tdisk}
\end{eqnarray}
For the modeling of X--ray sources presented in Chapter 4, we find that 
$T_{\rm irr}$ does not play any significant role.  
However, since this quantity has
consequences for the disk instability, we calculate it using
Eq. (3.6) and illustrate it for the rotating neutron star
models considered here.

\section{The Results}

\begin{table}
\begin{center}
\caption{Centrifugal mass-shed limit $(\Omega_{\rm ms})$, the neutron star
radius $(R)$, the disk inner edge radius $(r_{\rm in})$, specific gravitational
energy release in the boundary layer $(E_{\rm BL})$ and in the disk
$(E_{\rm D})$, their ratio $E_{\rm BL}/E_{\rm D}$, the maximum effective
temperature $(T_{\rm eff}^{\rm max})$,
the radial location $(r_{\rm eff}^{\rm max})$ in the disk corresponding to
$T_{\rm eff}^{\rm max}$, $T_{\rm obs}^{\rm max}$ (see text) and the
radial location $(r_{\rm obs}^{\rm max})$ corresponding to this.  These
values are listed for two values of $M$ for all EOS models considered
here (except for EOS model (A), where the maximum neutron star mass is less than
$1.78~\msun$, so only $M=1.4~\msun$ is considered). The number following
the letter $E$ represents powers of $10$. The values of $E_{\rm BL}$ \& 
$E_{\rm BL}/E_{\rm D}$ corresponding to $\Omega_{\rm *} = \Omega_{\rm ms}$ are 
expected to be zero and the small values given here are the measure of 
numerical error.}
\hspace{-2.0cm}
\psfig{file=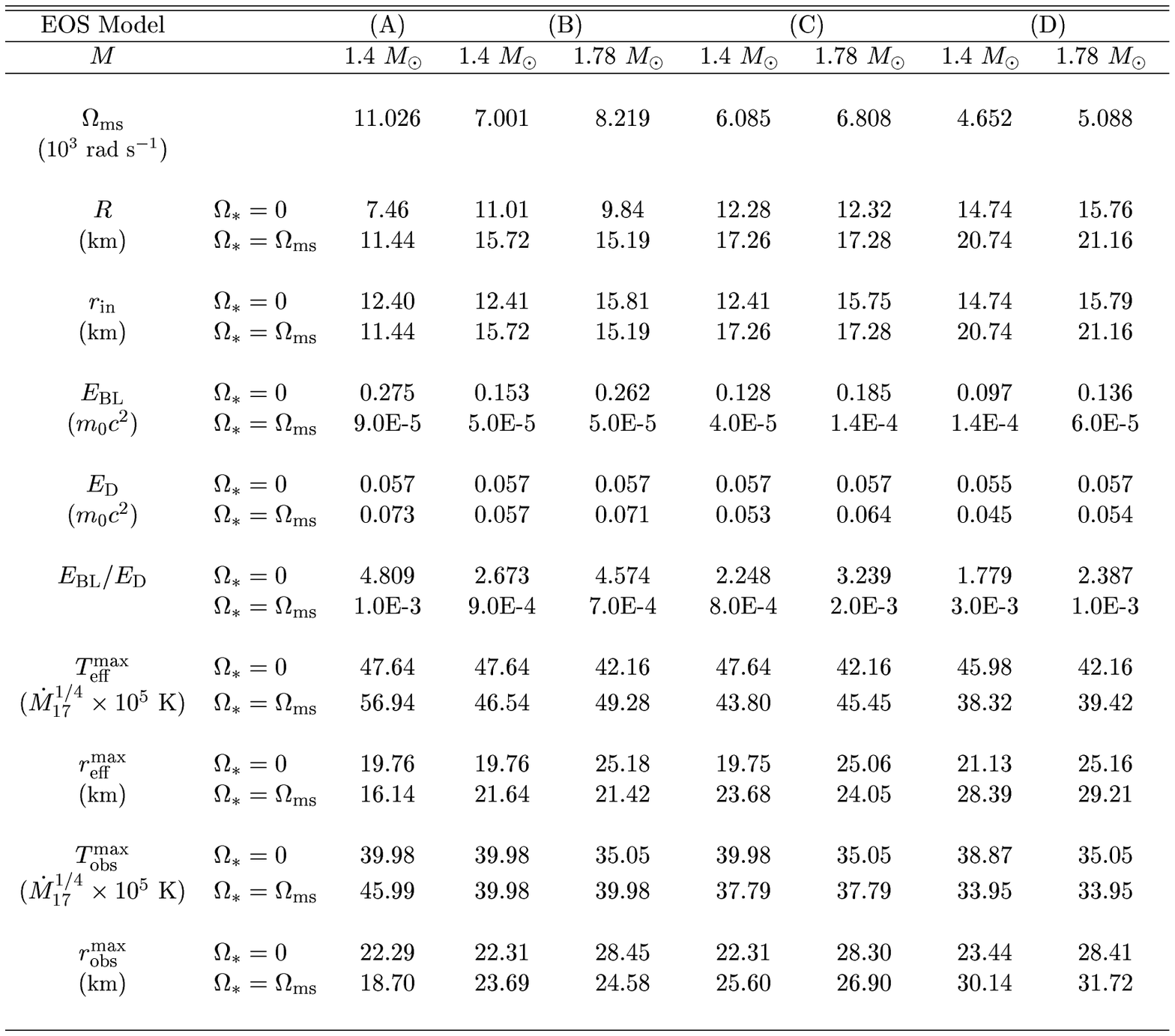,width=18 cm}
\end{center}
\end{table}

We have calculated the disk temperature profiles for rapidly rotating,
constant gravitational mass sequences of neutron stars in general relativity.
For our purpose here, we choose two values for the gravitational mass, namely,
$1.4~\msun$ and $1.78~\msun$, the former being the canonical mass for neutron
stars (as inferred from binary X--ray pulsar data), while the latter is the
estimated mass for the neutron star in Cygnus X--2 (Orosz \& Kuulkers 1999), 
that we use in Chapter 4.

\nopagebreak
\begin{figure}[h]
\psfig{file=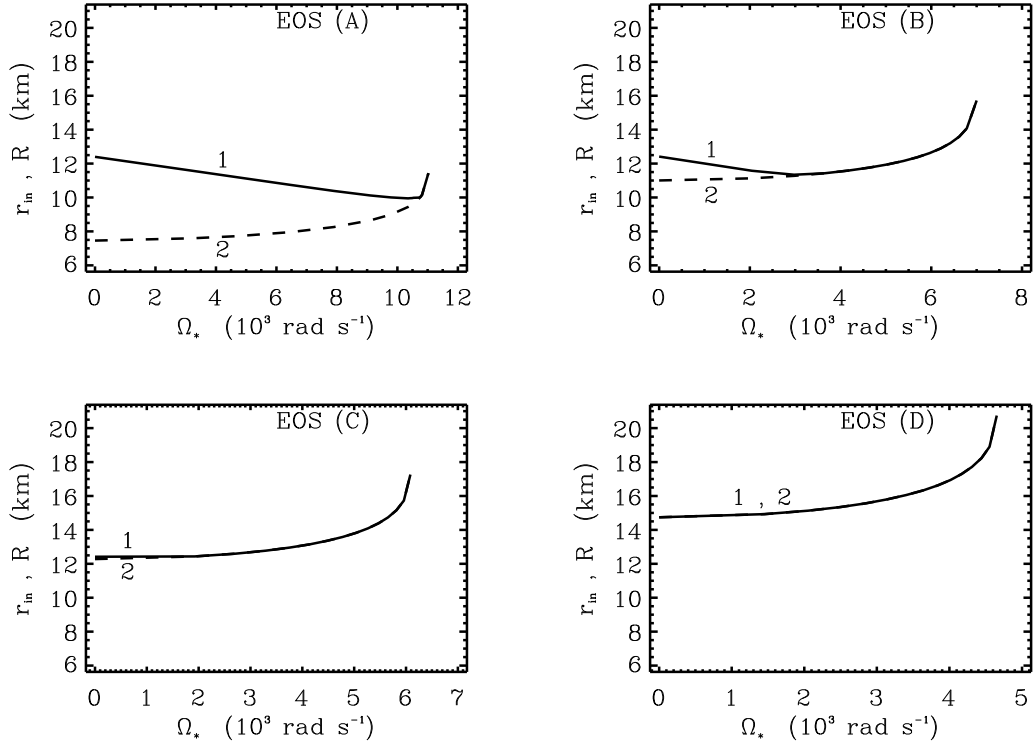,width=15 cm}
\caption{Disk inner edge radius ($r_{\rm in}$; curve 1) and neutron star
radius ($R$; curve 2),
as functions of neutron star angular velocity ($\Omega_*$) for
various EOS models.  The curves are for a fixed gravitational mass
($M=1.4~\msun$) of the neutron star.}
\end{figure}

\nopagebreak
\begin{figure}[h]
\psfig{file=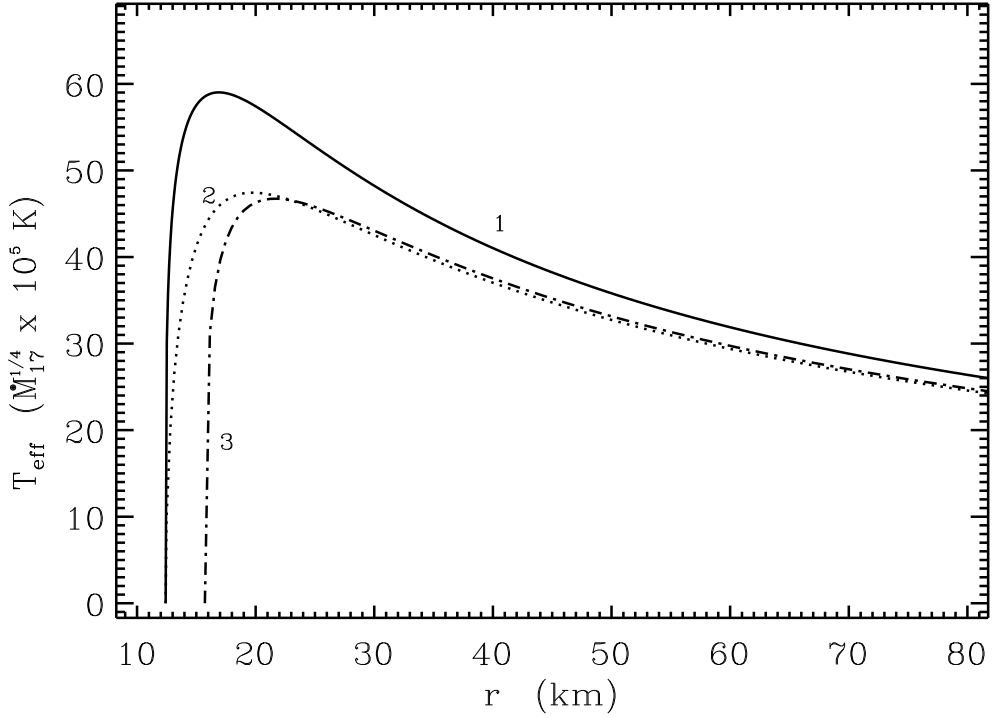,width=15 cm}
\caption{General relativistic corrections to Newtonian temperature profiles
for EOS model (B) and the neutron star gravitational mass $M=1.4~\msun$.
Curve 1 corresponds to the Newtonian case, curve 2 to the
Schwarzschild case and curve 3 to a neutron star rotating at
the centrifugal mass-shed limit, calculated using the
metric (2.3). For curve 1, it is assumed that,
$r_{\rm in}=6 G M/c^2$.
In this and all subsequent figures (except Fig. 3.6) the temperature is 
expressed in units of $\dot M_{17}^{1/4} \times 10^5$~K, where $\dot M_{17}$ 
is the steady state mass accretion rate in units of $10^{17} 
\mbox{g s}^{-1}$.} 
\end{figure}

\nopagebreak
\begin{figure}[h]
\psfig{file=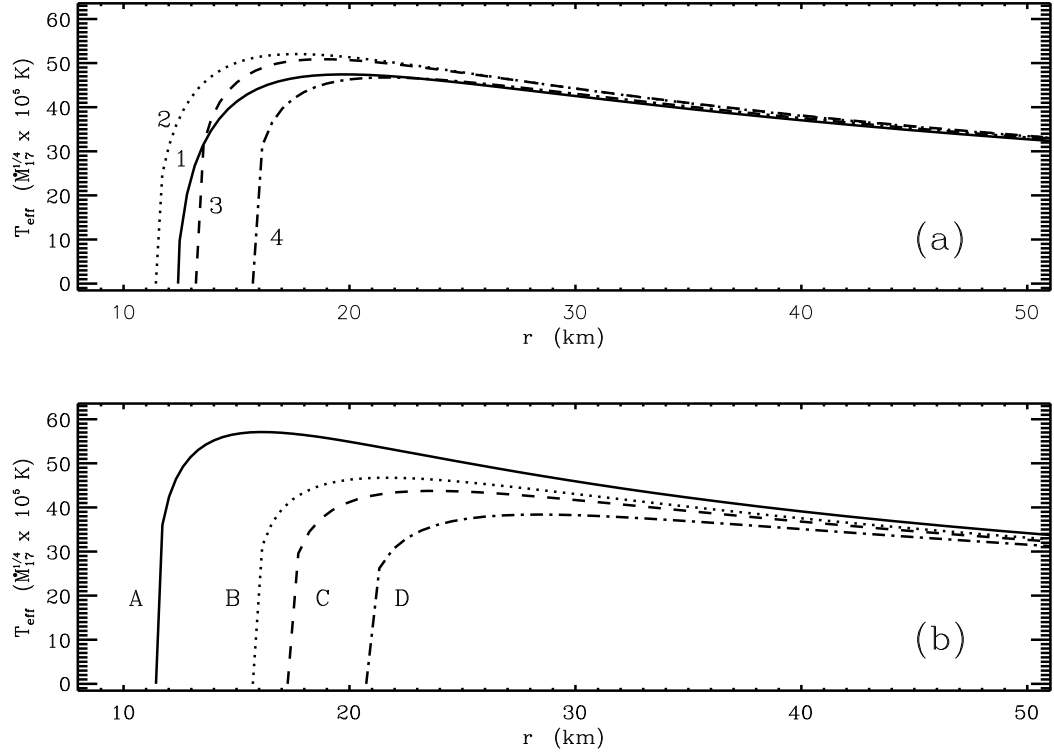,width=15 cm}
\caption{Temperature profiles incorporating the effects of rotation of the
neutron star. The plots correspond to (a) For EOS model (B) 
and an assumed neutron star mass
of $M=1.4~\msun$ for rotation rates: $\Omega_*=0$ (curve 1),
$\Omega_*=3.647\times10^3$~rad~s$^{-1}$ (curve 2),
$\Omega_*=6.420\times10^3$~rad~s$^{-1}$ (curve 3) \&
$\Omega_*=7.001\times10^3$~rad~s$^{-1}= \Omega_{\rm ms}$ (curve 4) and
(b) For the same assumed mass and $\Omega_*=\Omega_{\rm ms}$ for
the EOS models A, B, C and D.}
\end{figure}

\nopagebreak
\begin{figure}[h]
\psfig{file=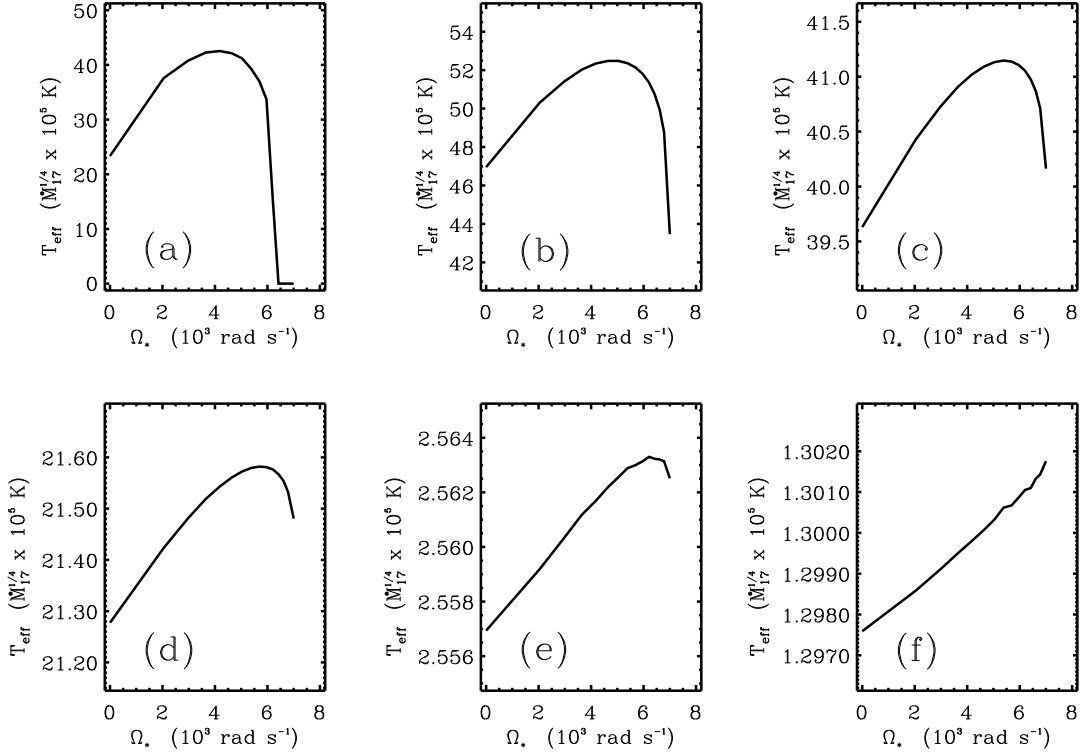,width=15 cm}
\caption{Plots of $T_{\rm eff}$ vs. $\Omega_*$ for
chosen constant radial distances for fixed neutron star mass $M=1.4~\msun$
and EOS model B. The plots correspond to 
(a) $r=13$~km, (b) $r=18$~km, 
(c) $r=35$~km, (d) $r=100$~km, (e) $r=2000$~km
and (f) $r=5000$~km.}
\end{figure}

\nopagebreak
\begin{figure}[h]
\psfig{file=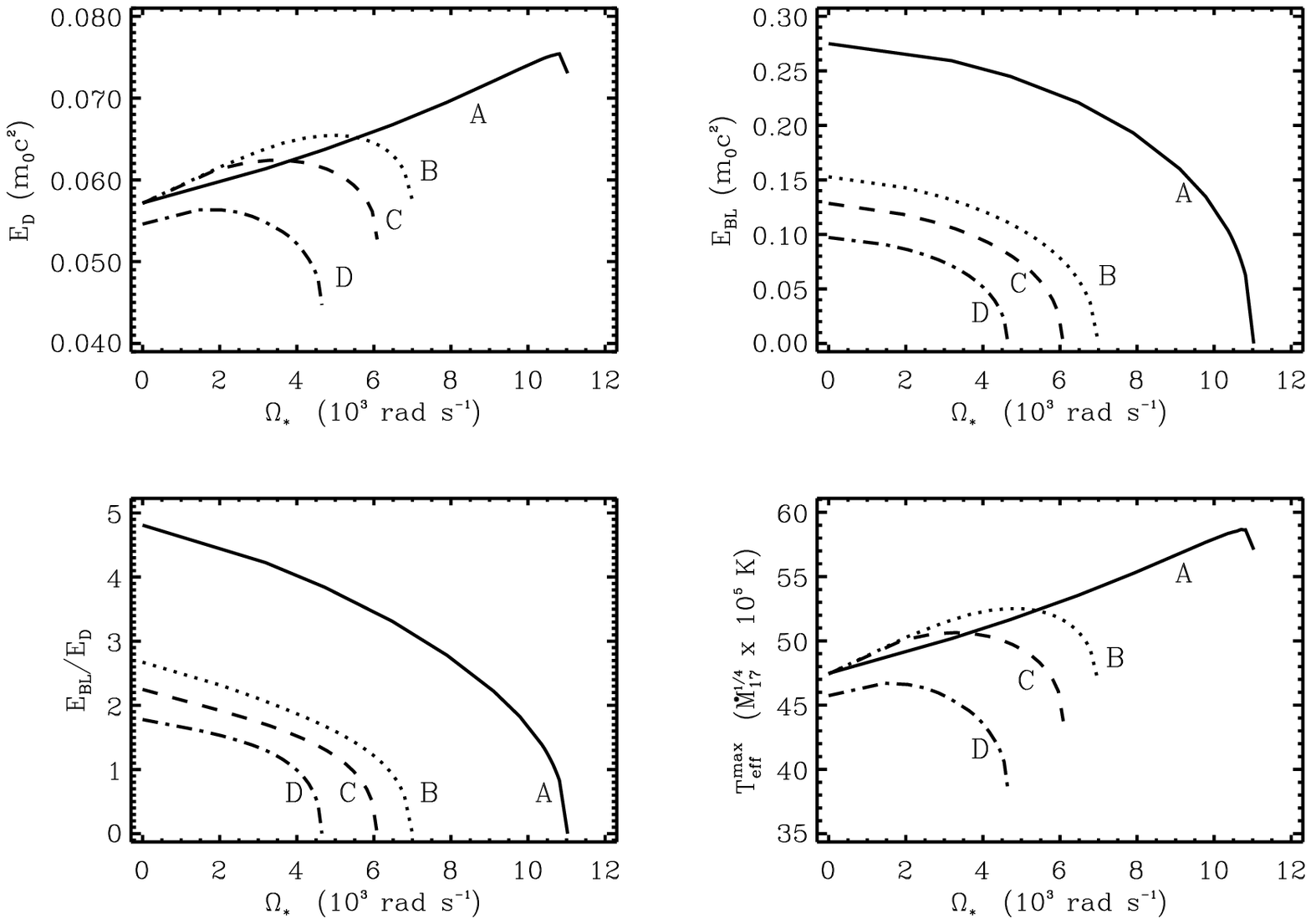,width=15 cm}
\caption{The variations of the $E_{\rm D}$, $E_{\rm BL}$,
$E_{\rm BL}/E_{\rm D}$ and $T_{\rm eff}^{\rm max}$, with
$\Omega_*$ for a chosen neutron star mass value of $1.4~\msun$ for
the four EOS models. The curves have the same significance as Fig. 3.3b.}
\end{figure}

\nopagebreak
\begin{figure}[h]
\psfig{file=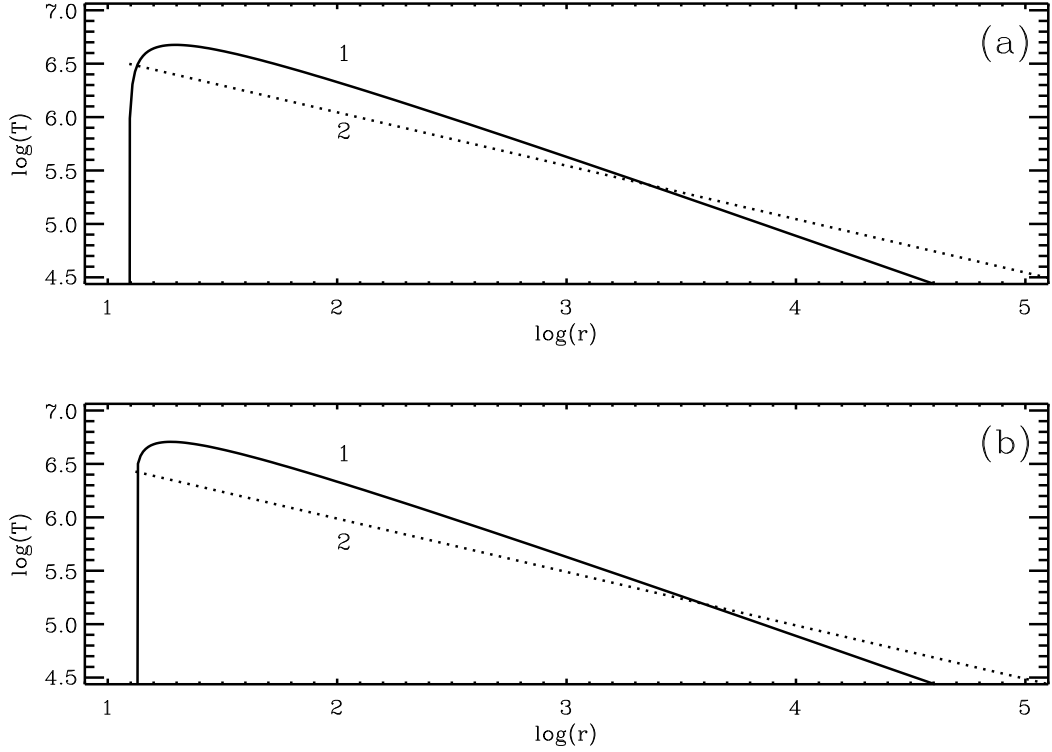,width=15 cm}
\caption{Comparison between the radial profiles of $T_{\rm eff}$ (curve 1) 
and $T_{\rm irr}$
(curve 2), calculated for $\eta=E_{\rm BL} + E_{\rm D}$, 
$\beta=0.9$, $h/r=0.2$ and $n=9/7$
in Eq. (3.6) for two values of neutron star spin rates:
(a) $\Omega_*=0$ and (b) $\Omega_*=6.420\times10^{3}$~rad~s$^{-1}$.
The curves are for a neutron star configuration having
$M=1.4~\msun$, described by EOS model B.
The temperatures are in units of $\dot M_{17}^{1/4}$, and the radial extent 
is in km. For illustrative purposes, we have displayed this comparison in 
a log-log plot.}
\end{figure}

\nopagebreak
\begin{figure}[h]
\psfig{file=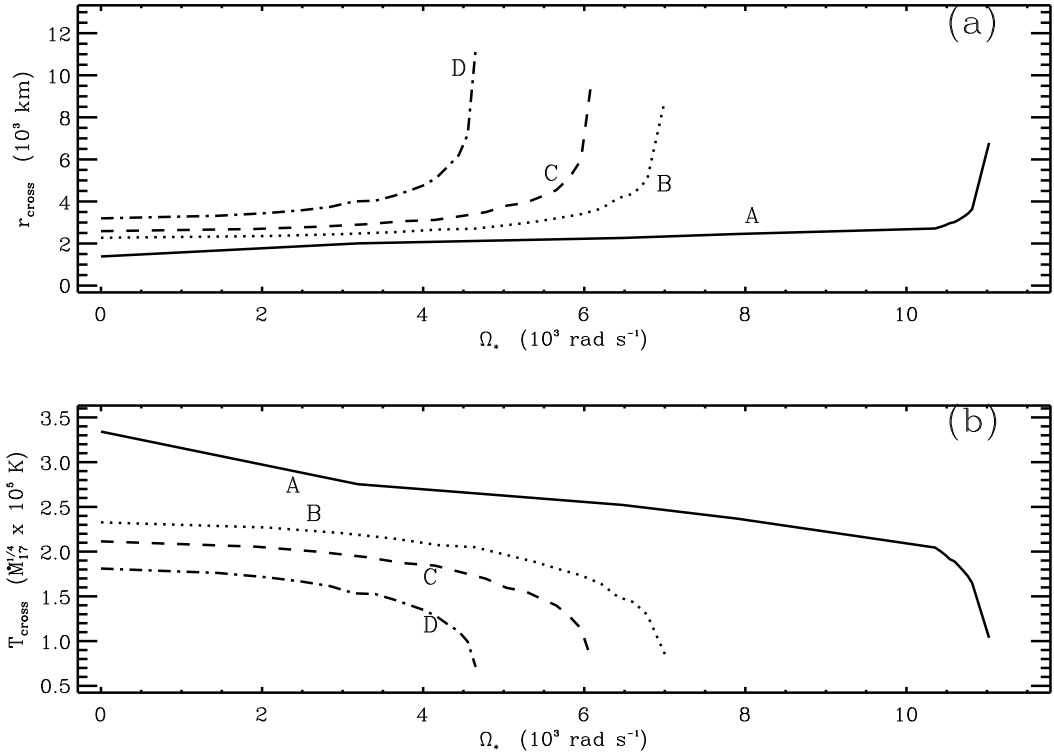,width=15 cm}
\caption{Plots: (a) $r_{\rm cross}$ vs. $\Omega_*$
and (b) $T_{\rm cross}$ vs. $\Omega_*$.  These
are for a fixed neutron star gravitational mass of $M=1.4~\msun$ and for
the different EOS models as in Fig. 3.3b. Here $T_{\rm irr}$ is calculated
for $\eta=E_{\rm BL}+E_{\rm D}$, $\beta=0.9$, $h/r=0.2$ and $n=9/7$.}
\end{figure}

\nopagebreak
\begin{figure}[h]
\psfig{file=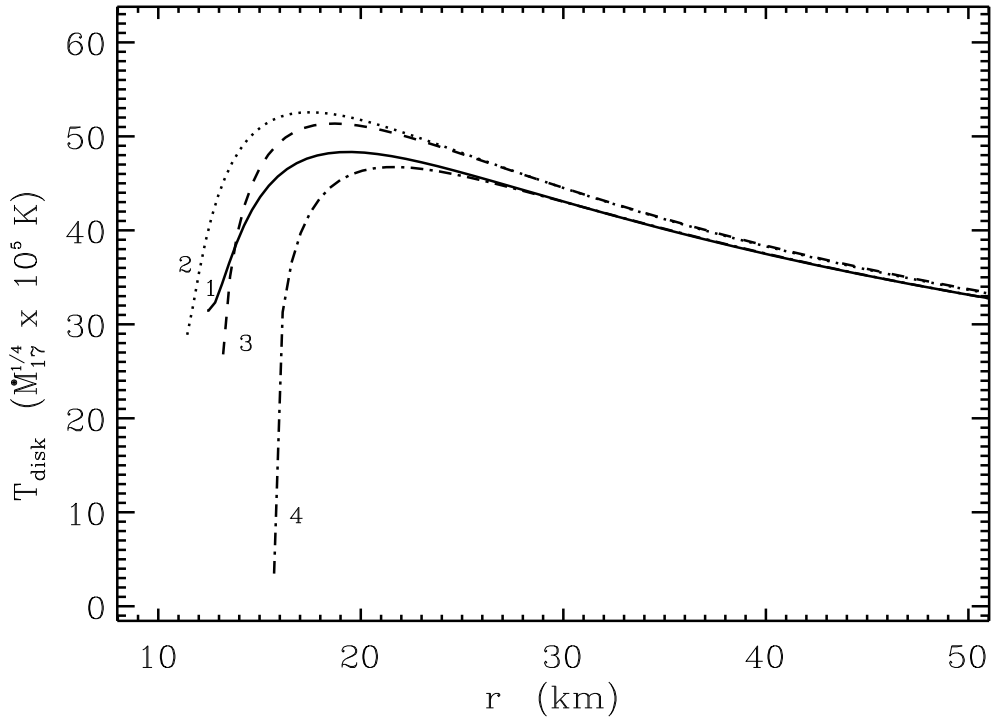,width=15 cm}
\caption{The disk temperature $(T_{\rm disk})$ profiles for
a $M=1.4~\msun$ neutron star corresponding to EOS model (B), having various
rotation rates as in Fig. 3.3a. These curves are obtained for 
$\eta = E_{\rm BL}$, and the same values of 
$\beta$, $h/r$ and $n$ as in Fig. 3.6 are used.}
\end{figure}

\nopagebreak
\begin{figure}[h]
\psfig{file=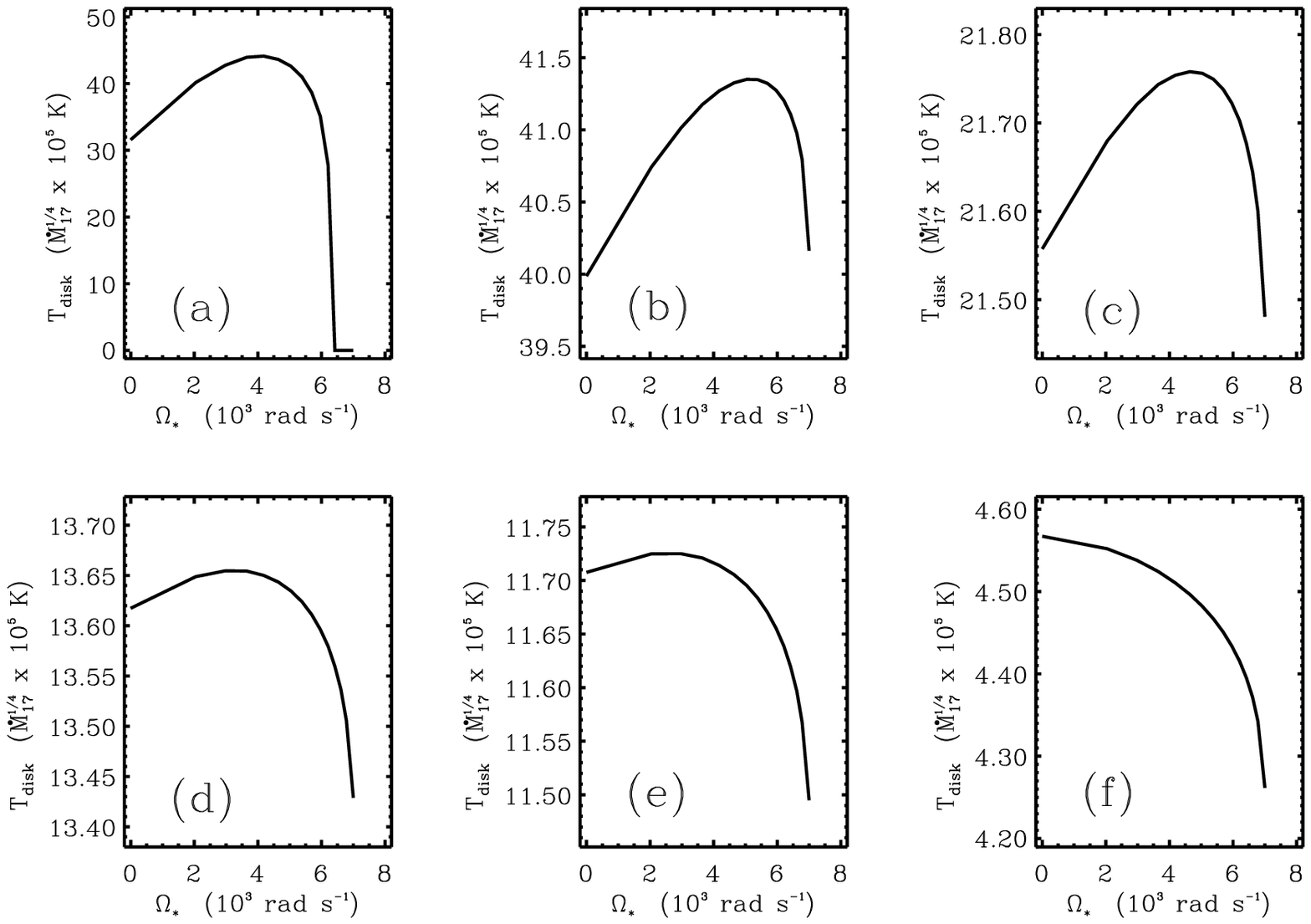,width=15 cm}
\caption{Plots of $T_{\rm disk}$ vs. $\Omega_*$ at
various chosen radial distances: (a) $r= 13$~km, (b) $r= 35$~km,
(c) $r= 100$~km, (d) $r= 200$~km, (e) $r= 250$~km and (f) $r= 1000$~km.
These are for EOS model (B), an assumed gravitational mass value of
$1.4~\msun$. and the same
values of $\eta$, $\beta$, $h/r$ and $n$ as in Fig. 3.8 are used.}
\end{figure}

In Table 3.1, we list the values of the stellar rotation rate at centrifugal
mass-shed limit ($\Omega_{\rm ms}$); the neutron star radius ($R$);
the radius of the inner edge of the disk ($r_{\rm in}$); $E_{\rm BL}$,
$E_{\rm D}$ and the ratio $E_{\rm BL}/E_{\rm D}$; $T_{\rm eff}^{\rm max}$
 \& $T_{\rm obs}^{\rm max}$ and $r_{\rm eff}^{\rm max}$
 \& $r_{\rm obs}^{\rm max}$ for the two mentioned values of
$M$ and for the different EOS models.
The last nine computed quantities
are given for two values of neutron star rotation rate, namely,
the static limit ($\Omega_*=0$) and the centrifugal mass-shed
limit ($\Omega_*=\Omega_{\rm ms}$). $E_{\rm D}$ and $E_{\rm BL}$
are in specific units (i.e. units of rest energy $m_0 c^2$, of the
accreted particle).
The temperatures are expressed in units of $\dot{M}_{17}^{1/4}\times 10^5$~K
(where $\dot{M}_{17} = \dot{M}/10^{17}$~g~s$^{-1}$).
From this Table it may be seen
that for a given neutron star gravitational mass ($M$):
(1) $\Omega_{\rm ms}$ decreases for increasing stiffness of the EOS model.
(2) $R$ is greater for stiffer EOS.
(3) The behavior of $r_{\rm in}$ depends on whether $r_{\rm ms}>R$ or
    $r_{\rm ms}<R$ and hence appears non--monotonic.
(4) $E_{\rm BL}$ for the non--rotating configuration decreases with stiffness
    of the EOS.  For a configuration rotating at the mass-shed limit,
    $E_{\rm BL}$ is insignificant.
(5) In the non--rotating limit, $E_{\rm D}$ remains roughly constant for
    varying stiffness of the EOS model.  However, for the rapidly rotating
    case, the value of $E_{\rm D}$ decreases with increasing stiffness.
(6) The ratio $E_{\rm BL}/E_{\rm D}$ in static limit is highest for the
    softest EOS model.  For the rapidly rotating case, this ratio is
    uniformly insignificant.
(7) $T_{\rm eff}^{\rm max}$ and $T_{\rm obs}^{\rm max}$ decrease with
    increasing stiffness
    of the EOS  models.  However, these values exhibit non--monotonic variation
    with $\Omega_*$ (see Fig. 3.5 for the first parameter).
(8) The rest of the parameters, namely, $r_{\rm eff}^{\rm max}$ and
     $r_{\rm obs}^{\rm max}$ are non--monotonic with respect to
     the EOS stiffness parameter.

In Fig. 3.1, we display the variation of $R$ (the dashed curve) and
$r_{\rm in}$ (the continuous curve) with
$\Omega_*$ for $M=1.4~\msun$ for the four EOS models that we
have chosen.  From this figure it is seen that for a constant gravitational
mass sequence, for both soft and intermediate EOS models, $r_{\rm in} > R$ for
slow rotation rates whereas, for rapid rotation rates $r_{\rm in} = R$.
In other words, for neutron stars spinning very rapidly, the inner edge of
the disk will almost coincide with the stellar surface. It may be noted that
for the stiff EOS models, this condition obtains even at slow rotation
rates of the neutron star.

It is instructive to make a comparison of the temperature profiles
calculated using a Newtonian prescription with that
obtained in a relativistic description using Schwarzschild metric.
This is shown in Fig. 3.2, for the EOS model (B) and $M=1.4~\msun$
(the trend is similar for all the EOS).  The vertical axis in
this figure is $T_{\rm eff}$ (in this and all other figures, the
temperatures are shown in units of $\dot{M}_{17}^{1/4}$) and  the
horizontal axis, the radial distance in km.
This figure underlines the importance of general relativity in determining
the accretion disk temperature profiles;
the Schwarzschild result for $T_{\rm eff}^{\rm max}$ is always less than
the Newtonian result, and
for the neutron star configuration considered here, the overestimate
is almost $25$\%.
For the sake of illustration, we also show the corresponding
curve for a neutron star rotating at the mass-shed limit (curve 4, Fig. 3.3a).
The disk inner edge is at the radius of the
innermost stable circular orbit  for all the cases.
Note that the disk inner edge should be at $R$ for Newtonian case; but we
 have taken $r_{\rm in}=6GM/c^2$ as assumed in Shapiro \& Teukolsky (1983).

The effect of neutron star rotation on the accretion disk temperature, treated
general relativistically, is illustrated in Fig. 3.3a and 3.3b.
Fig. 3.3a corresponds to the EOS model (B). The qualitative features
of this graph are similar for the other EOS models, and are not
shown here.
However, the temperature profiles exhibit a marked dependence on
the EOS.  This dependence is illustrated in Fig. 3.3b, which is
done for a particular value of $\Omega_*= \Omega_{\rm ms}$.
All these temperature profiles have been calculated for a neutron star
mass equal to $1.4~\msun$.
The temperature profiles shown in Fig. 3.3a do not have a monotonic behavior
with respect to $\Omega_*$.  This behavior is a composite of two
underlying
effects: (i) the energy flux emitted from the disk increases with
$\Omega_*$ and (ii) the nature of the dependence of $r_{\rm in}$
(where $T_{\rm eff}$ vanishes : the boundary condition)
on $\Omega_*$ (see Fig. 3.1). This is more clearly brought out in
Fig. 3.4, where we have plotted of
$T_{\rm eff}$ vs. $\Omega_*$ for selected
constant radial distances (indicated in six different panels) and EOS (B).
At large radial distances, the value $T_{\rm eff}$
is almost independent of the boundary condition; hence the
temperature always increases with $\Omega_*$ in Fig. 3.4f.

The variations of $E_{\rm D}$, $E_{\rm BL}$, the ratio
$E_{\rm BL}/E_{\rm D}$ and
$T_{\rm eff}^{\rm max}$ with $\Omega_*$
are displayed in Fig. 3.5 for all EOS models considered here.  All the plots
correspond to $M=1.4~\msun$.  Unlike constant central density neutron star
sequences (Thampan \& Datta 1998), for the constant gravitational mass
sequences, $E_{\rm D}$ does not have a general monotonic behavior with
$\Omega_*$.  $T_{\rm eff}^{\rm max}$ has a behavior akin
to that of $E_{\rm D}$ (because of the reasons mentioned earlier).
$E_{\rm BL}$ decreases with $\Omega_*$, slowly at first but
rapidly as $\Omega_*$ tends to $\Omega_{\rm ms}$. The variation of
$E_{\rm BL}/E_{\rm D}$ with respect to $\Omega_{\ast}$ is similar to
that of $E_{\rm BL}$.

We provide a comparison between the effective temperature
(Eq. 3.1) and the irradiation temperature (Eq. 3.6),
in Fig. 3.6. We have taken $\eta=E_{\rm BL}+E_{\rm D}$.  Fig. 3.6a is for
$\Omega_*=0$ while
Fig. 3.6b is for a higher $\Omega_* = 6420$ rad s$^{-1}$.  The curves are
for the gravitational mass corresponding to $1.4~\msun$ for the EOS model (B).
The irradiation temperature
becomes larger than the effective temperature at some large value of
the radial distance, the ratio of the former to the latter becoming
increasingly large beyond this distance. For $E_{\rm BL}$ small
compared to $E_{\rm D}$ (as will be the case for a rapid neutron
star spin rate), irradiation effects in the inner disk region will
not be significant.
Defining the radial point where
the irradiation temperature profile crosses the effective temperature
profile as $r=r_{\rm cross}$ and the corresponding temperature as
$T_{\rm cross}$, we display plots of $r_{\rm cross}$ and $T_{\rm cross}$
with $\Omega_{\ast}$ respectively in Figs. 3.7a and 3.7b. 
It can be seen that $r_{\rm cross}$ increases with $\Omega_*$,
just as $E_{\rm S}$ does, and hence the irradiation effect decreases
with increasing $\Omega_*$. Therefore $T_{\rm cross}$
decreases with increasing $\Omega_*$.

In Fig. 3.8, we illustrate the disk temperature
($T_{\rm disk}$) profile for EOS model (B) corresponding to
$M=1.4~\msun$ for various values of $\Omega_{\ast}$.
We illustrate the variation of $T_{\rm disk}$ with $\Omega_*$ at
fixed radial points in the disk in Fig. 3.9. The effect of $T_{\rm irr}$
on $T_{\rm disk}$ can be noted in Fig. 3.9f.

\section{Summary and Discussion}

In this chapter, we have calculated the temperature profiles of accretion
disks around rapidly rotating and non--magnetized neutron stars, using
a fully general relativistic formalism.  
The maximum temperature
and its location in the disk are found to differ substantially from their
values corresponding to the Schwarzschild space-time, depending on the
rotation rate of the accreting neutron star.
This shows the importance of the rapid--rotation--calculation.  

A few comments regarding the validity of the Page \& Thorne (1974)
formalism for accreting neutron star binaries are in order here.
Unlike for the case of black holes, neutron stars possess hard
surface that could be located outside the marginally stable orbit.
For neutron star binaries, this gives rise to a possiblity of
the disk inner edge coinciding with
the neutron star surface.  We
have assumed that the torque (and hence the flux of energy) vanishes
at the disk inner edge even in cases where the latter touches
the neutron star surface.  In the case of rapid spin of the neutron
star, the angular velocity of a particle in
Keplerian orbit at disk
inner edge will be close to the rotation rate of the neutron star.
Therefore, the torque between the neutron star surface and the inner
edge of the disk is expected to be negligible.  Independently of
whether or not the neutron star spin is large, Page \& Thorne (1974)
argued that the error in the calculation of $T_{\rm eff}$ will not be
substantial outside a radial
 distance $r_{\rm o}$, where $r_{\rm o}$ is given by $r_{\rm o} - r_{\rm in} =
0.1 r_{\rm in}$. In our calculation, we find that $r_{\rm eff}^{\rm max}$
(which is the most important region for the generation of X--rays) is
greater than $r_{\rm o}$ by several kilometers for all the cases
considered.

Temperature profile is the main ingredient for the calculation of disk 
spectrum. As we have seen that both general relativity and rapid rotation 
have profound effect on the inner disk temperature profile, we expect 
the modeling of hard X--ray spectrum to be very much sensitive to them. 
This we will study in Chapters 5 \& 6.

%% file: chap4.tex
\markright{Chapter 4}
\def\note #1]{{\bf #1]}}
\chapter{Disk Temperature Profile: Implications for Five LMXB Sources}

\section{Introduction}

We have calculated the disk temperature profile for a rapidly rotating
neutron star in the previous chapter. We have also computed the disk
luminosity and the boundary layer luminosity. In this chapter, we compare
our theoretical results with the {\it EXOSAT} data (analysed by White, Stella 
\& Parmar 1988) to constrain different properties of five LMXB sources: 
Cygnus X-2, XB 1820-30, GX 17+2, GX 9+1 and GX 349+2.

XB 1820-30 is an atoll source which shows type I X--ray bursts. Cygnus X-2, GX 17+2 and GX 
349+2 are Z sources, of which the first two
show X--ray bursts. GX 9+1 is an
atoll source. As all of them are LMXBs (van Paradijs 1995), the magnetic
field of the neutron stars are believed to have decayed to low values
$(\sim 10^8$ G; see Bhattacharya \& Datta 1996 and Bhattacharya \&
van den Heuvel 1991). Therefore, we ignore the effect of the
magnetic field on the accretion disk structure in our calculations.

In this chapter, we calculate the allowed ranges of several properties of 
these LMXBs and make general comments on the rotation rates of the neutron 
stars in these systems. We also discuss possible constraints on the neutron 
star equation of state.

In section 4.2, we describe the procedure of comparison of theoretical
values of the parameters with the observed ones. We give the results in 
section 4.3 and the conclusions in section 4.4.

\section{Procedure of Comparison with Observations}

As mentioned in Chapter 3, the X-ray spectrum from an LMXB may have two 
contributions: one from the optically thick disk and the other from the 
boundary layer near the neutron star surface. The spectral shape of the disk
emission depends on the accretion rate. For $\dot {M} << 10^{17}$~g~s$^{-1}$,
the opacity in the disk is dominated by free-free absorption and the spectrum
will be a sum of blackbody spectra from different radii. The local spectrum 
(with respect to a co--moving observer) will be characterized by a temperature 
$T_{\rm eff}(r)$ at that radius. 
The observer at a large distance will see a temperature $T_{\rm obs}(r)$,
which includes the effect of gravitational redshift and Doppler broadening,
as mentioned in section 3.2.1.
At higher accretion rates ($\dot {M} \approx 10^{17}$~g~s$^{-1}$)
the opacity will be dominated by Thomson scattering and the spectrum from
the disk will be that of a modified blackbody (Shakura \& Sunyaev 1973).
However, for still higher accretion
rates Comptonization in the upper layer of the disk becomes important
leading to a saturation of the local spectrum forming a Wien peak.
The emergent spectrum can then be described as a sum of blackbody
emissions but at a temperature different from $T_{\rm obs}$. The
temperature infered by a distant observer from the spectrum is the color temperature
$T_{\rm col}$. In general
$T_{\rm col} = f(r) T_{\rm obs}$ where the function $f$ is called
the color factor
(or the spectral hardening factor), and it depends on the vertical
structure of the disk. Shimura \& Takahara (1995) calculated the color
factor for various accretion rates and masses of the accreting compact
object (black hole) and found that $f\approx$~($1.8$--$2.0$) is nearly
independent of accretion rate and radial distance, for
$\dot{M}\sim\dot{M}_{\rm e}$, where $\dot M_{\rm e} = 1.4\times 
10^{17} M/\msun$~g~s$^{-1}$. These authors find that for accretion rate
$\sim 10$\% of $\dot M_{\rm e}$, $f\approx1.7$.  More recently, however,
from the analysis of high--energy radiation from GRO J1655-40, a black--hole
transient source observed by RXTE, Borozdin et al. (1999) obtain a value of
$f=2.6$, which is higher than previous estimates used in the literature.
With this approximation for $T_{\rm col}$, the spectrum from optically thick
disks with high accretion rates can be represented as a sum of diluted
blackbodies.  The local flux at each radius is
\begin{equation}
F_\nu = \frac{1}{f^4} \pi B_\nu (fT_{\rm eff})
\end{equation}
where $B_\nu$ is the Planck function.
For high accretion rates the boundary layer at the neutron star surface
is expected to be optically thick and an additional single component blackbody
spectrum should be observed.

White et al. (1988) have fitted the observed data for the said LMXB sources 
to several spectral models. One of
the models is a blackbody emission upto the innermost stable circular
orbit of the
accretion disk and an additional blackbody
spectrum to account for the boundary layer emission. The
spectrum from such a disk is the sum of blackbody emission with a
temperature profile
\begin{equation}
T \propto r^{-3/4} (1 - (r_{\rm in}/r)^{1/2})^{1/4}.
\end{equation}
White et al. (1988) have identified this temperature
as the effective temperature which, as mentioned by them, is inconsistent
since the accretion rates for these sources are high. 
However, as mentioned above, identifying this
temperature profile as the
color temperature makes the model consistent if the color factor
is nearly independent of radius. Moreover, the inferred
temperature profile (i.e., $T_{\rm obs} = T_{\rm col}/f$)
is similar to the one developed in previous chapter. Therefore, in this chapter
we assume that the maximum of the best-fit color temperature profile
$T_{\rm col}^{\rm max}$
is related to the maximum temperature $T_{\rm obs}^{\rm max}$ computed in 
previous chapter
by ($T_{\rm col}^{\rm max} \approx f T_{\rm obs}^{\rm max}$).
Shimura \& Takahara (1988)
suggested a value of $1.85$ for the factor $f$, for an assumed neutron
star mass equal to $1.4~\msun$ and $\dot{M}=10 \dot{M}_{\rm e}$.

We compare the best-fit values of the parameters maximum color temperature 
$(T_{\rm col}^{\rm max})$, disk luminosity $(L_{\rm D})$ and boundary layer 
luminosity $(L_{\rm BL})$ with their theoretical values for a given 
neutron star mass, accretion rate ($\dot {M}$), color
factor $f$ and equation of state. However, in order to make allowance for
the uncertainties in the fitting procedure and in the value of $z$, and
also those arising due to the simplicity of the model, we consider a range of
acceptable values for $T_{\rm col}^{\rm max}$, $L_{\rm D}$ and $L_{\rm BL}$.
In particular, we allow for deviations in the best-fit values of 
$T_{\rm col}^{\rm max}$ and luminosities:  
we take two combinations of these, namely,
($10$\%, $25$\%) and ($20$\%, $50$\%), where the first number in parentheses
corresponds to the error in $T_{\rm col}^{\rm max}$ and the second to
the error in the best-fit luminosities.
Note that we neglect the irradiation temperature here,
as $T_{\rm disk} \approx T_{\rm eff}$ at the inner region of the disk
(the region where the disk temperature reaches a maximum).
We obtain a range of consistent values for $\dot {M}$,
$\Omega_{\rm *}$ and $f$ (and hence, allowed ranges of different quantities).
The procedure is as follows.

We can calculate
the different quantities ($E_{\rm D}$, $E_{\rm BL}$,
$T_{\rm obs}^{\rm max}$, $R$, $r_{\rm in}$, etc.) as functions of
$\Omega_{\rm *}$.
Taking the observed (or fitted) values for $T_{\rm col}^{\rm max}$,
$L_{\rm BL}$ and ($L_{\rm BL}+L_{\rm D}$) with the error bars,
we have two limiting
values for each of these quantities. We assume a particular
value for each of $f$ and $\dot {M}$, from which we obtain the corresponding
fitted values of $T_{\rm obs}^{\rm max}$,  $E_{\rm BL}$ and
($E_{\rm BL}+E_{\rm D}$) by the
relations $E_{\rm BL} = L_{\rm BL}/\dot {M}$,
$E_{\rm BL}+E_{\rm D} = (L_{\rm BL}+L_{\rm D})/\dot {M}$ and
$T_{\rm obs}^{\rm max}=T_{\rm col}^{\rm max}/(f {\dot {M}}^{1/4})$
(because here
$T_{\rm obs}^{\rm max}$ is in the unit of ${\dot {M}}^{1/4}$).
By interpolation, we calculate two corresponding
limiting $\Omega_{\rm *}$'s (i.e., the allowed range in $\Omega_{\rm *}$)
for each fitted quantity. We take the common
region of these three ranges, which is the net allowed
range in $\Omega_{\rm *}$. We do this for $\dot {M}$'s in the
range $10^{-13}~M_{\odot} y^{-1}$ to $10^{-6}~M_{\odot} y^{-1}$
(which is reasonable for LMXB's) with logarithmic interval
0.0001, for a particular value of $f$.  If for some $\dot {M}$,
there is no allowed $\Omega_{\rm *}$, then that value of $\dot {M}$ is not
allowed. Thus we get the allowed range of $\dot {M}$ for a particular $f$.
Next we repeat the whole procedure described above for various values
of $f$, in the range 1 to 10. If for some $f$, there is no
allowed $\dot {M}$, then that $f$ is not allowed. Thus we get
an allowed range of $f$. Taking the union of all the allowed
ranges of $\dot {M}$, we get the net allowed range of $\dot {M}$
(and similarly the net allowed range of $\Omega_{\rm *}$) for
a particular EOS, gravitational mass and a set of error bars.
The allowed ranges of $\nu_{\rm in}$, $R$, $r_{\rm eff}^{\rm max}$ etc. then
easily follow, since their general variations with respect to
$\Omega_{\ast}$ are already known.

\section{The Results}

In this chapter, we calculate gravitational mass sequences for different
EOS models (mentioned in Chapter 2) and constrain several properties of 
five LMXB sources. For the neutron star in each of the sources, we assume
$M = 1.4~\msun$ (i.e., the canonical mass value for neutron stars).
For Cygnus X-2, we assume an additional mass value $(1.78)$, which 
is the estimated mass for the neutron star in Cygnus X-2 (Orosz \& Kuulkers 
1999). It may be noted with caution (Haberl \& Titarchuk 1995), that this value
is not confirmed from X--ray burst spectral analysis. We use the value of
$M=1.78~\msun$ for the illustration of our results, and leave the issue for
future confirmation. We take $\cos i = 0.5$ $(i$ is the inclination 
angle of the source) for 
Cygnus X-2 (Orosz \& Kuulkers 1999), while for each of other four sources, 
we use two values for $\cos i$, namely, $0.2$ and $0.8$. These two widely
different values ensure the sufficient generality of our results.

For the source Cygnus X-2, the best spectral fit to the data is when
$T_{\rm col}^{\rm max} = 1.8 \times 10^7 $ K, $L_{\rm D} = 2.1 \times
10^{38}$~ergs~s$^{-1}$
and $ L_{\rm BL} = 2.8 \times 10^{37} $~ergs~s$^{-1}$ (White et al. 1988).
For the other four sources, the best-fit values of the parameters
$T_{\rm col}^{\rm max}$, $L_{\rm D}$ and $L_{\rm BL}$ are
respectively as follow (White et al. 1988):
(1) XB 1820-30: $1.59 \times 10^7$~K,
$(1.49 \times 10^{38}$ ergs s$^{-1}$, $0.37 \times 10^{38}$ ergs
s$^{-1})$ and $2.56 \times 10^{37}$ ergs s$^{-1}$; (2) GX 17+2:
$1.76 \times 10^7$~K, $(6.49 \times 10^{38}$ ergs s$^{-1}$,
$1.62 \times 10^{38}$ ergs s$^{-1})$ and $7.10 \times 10^{37}$ ergs
s$^{-1}$; (3) GX 9+1: $2.25 \times 10^7$~K, $(6.01 \times 10^{38}$
ergs s$^{-1}$, $1.50 \times 10^{38}$ ergs s$^{-1})$ and
$2.50 \times 10^{37}$ ergs s$^{-1}$; (4) GX 349+2:
$2.07 \times 10^7$~K, $(8.54 \times 10^{38}$ ergs s$^{-1}$, $2.14
\times 10^{38}$ ergs s$^{-1})$ and $4.80 \times 10^{37}$ ergs s$^{-1}$.
Here the first term in the bracket is $L_{\rm D}$ for $\cos i = 0.2$
and the second term is that for $\cos i = 0.8$.

We take the distance $(D)$ of the source as 8 kpc (Orosz 
\& Kuulkers 1999) for Cygnus X-2 and 6.4 kpc (Bloser et al. 2000) for 
XB 1820-30. We assume $D = 8$~kpc for both GX 17+2 and
GX 9+1, as their locations are believed to be near the galactic center
(Deutsch et al. 1999; Hertz et al. 1990) and distance of the galactic
center is $7.9 \pm 0.3$ kpc, as concluded by McNamara et al. (2000).
For GX 349+2, we take $D = 9$~kpc (Deutsch et al. 1999).

We display the constrained values with the help of five tables.
In Table 4.1 (for Cygnus X-2), we have taken two values of $M$
(but a fixed value for $i)$, while in each of the other four tables, we use
a fixed value for $M$ (but two values of $i)$.
From these results, we notice that the accretion rates of all
the sources are very high. It is to be noted that here $\dot M$ is 
presented in unit of $\dot M_{\rm e}$ (defined in the previous section). The
Eddington accretion rate $(\dot M_{\rm Edd})$ is $\dot M_{\rm e}/\eta$, with  
$\eta = E_{\rm BL}+E_{\rm D}$.
Therefore, as the actual value of $\eta$ is much less than
1.0 (generally not greater than 0.3 and for rapidly rotating neutron star,
typically less than 0.2), the value of $(\dot M_{\rm Edd})$ is
much higher than $\dot M_{\rm e}$.
 
The EOS model A is the softest in the sample. The maximum
mass of neutron stars (at $\Omega_{\ast}=\Omega_{\rm ms}$) corresponding
to this EOS is 1.63 $M_\odot$. So the constraint results for Cygnus X-2,
using this EOS are done only for $M=1.4~\msun$.

We notice that the allowed ranges (combined for all the cases
considered in each of the tables) of
$\Omega_{\rm *}/\Omega_{\rm ms}$ are $0.97-1.00$, $0.93-1.00$ 
and $0.75-1.00$ for the three sources Cygnus X-2, GX 9+1 and GX 349+2 
respectively (see Table 3.1 for the mass-shed limit values).
Therefore the neutron stars in these three sources
can be concluded to be rapidly rotating in general. In the
next section, we will discuss the significance of the obtained results.

\section{Summary and Discussion}

In this chapter, we have constrained the values of several properties of
five LMXB sources. For all of them, the accretion rates come out to
be very high (always $\ge 0.5 ~\dot M_{\rm e})$. This is
in accord with the fact that these are very luminous sources.

From our results, it can be concluded that the neutron star in Cygnus X-2 is
rotating very rapidly. The rotation rate of the neutron star in each of the 
other four sources is also very
close to the mass-shed limit for $\cos i = 0.2$. This is
because the values of $L_{\rm BL}/L_{\rm D}$ are very low for these
cases (see Chapter 3).
But, for $\cos i = 0.8$, rotation rate can not be constrained
effectively for the sources XB 1820-30 and GX 17+2. Therefore, for
these two sources, no general conclusion (about the values of
$\Omega_{\rm *})$ can be drawn. However,
as mentioned in the previous section, rotation rate can be concluded
to be very rapid for the sources GX 9+1 and GX 349+2.

According to Shimura \& Takahara (1988), the spectrum from the disk can
be represented as a multi-color blackbody only if $\dot{M} >0.1 \dot{M}_{\rm 
e}$, which always comes out to be the case for all the chosen LMXBs. 
Our calculated allowed ranges for $f$ are in accord with the results
$(f \sim 1.7-2.0)$ obtained by Shimura \& Takahara (1995). 
However, if we take the value of $f=2.6$, as reported by Borozdin et al.
(1999), then for Cygnus X-2, one would require an EOS model that is stiffer
than the stiffest used here, or a mass greater than $M=1.78~\msun$ (if one
uses the narrower limits on the luminosity and color temperature). On the
other hand, if one were to use the broader limits, the hardening factor
$f=2.6$ is disallowed only by the softest EOS model. For the other four 
sources, the assumption $f=2.6$ would require a very stiff EOS 
model or a mass greater than
$M = 1.4~\msun$ for most of the cases with $\cos i = 0.2$.

High frequency quasi periodic oscillations (kHz QPO) have been
observed for four (Cygnus X-2, XB 1820-30, GX 17+2 and GX 349+2) of the chosen
sources. The observed maximum kHz QPO frequencies are 1.005 kHz (Cygnus X-2), 
1.100 kHz (XB 1820-30), 1.080 kHz (GX 17+2) and 1.020 kHz (GX 349+2)
(van der Klis 2000). Now, the maximum possible frequency (i.e., the shortest 
time scale) of such a system should be given by the rotational frequency in 
innermost stable circular orbit (ISCO) $(\nu_{\rm in}$; col. 5 of the tables),
unless the model invoked to explain the temporal behavior predicts a substantial
power in the second harmonic, i.e., $\nu_{\rm QPO} \approx 2 \nu_{\rm in}$.
Therefore, the stiffest EOS model D is unfavored for Cygnus X-2, as the 
maximum value of $\nu_{\rm in}$ $(= 0.938$~kHz, Table 4.1)
is less than the observed maximum kHz QPO frequency. For the same
reason, EOS model D is unfavored for $\cos i = 0.2$ for the sources XB 1820-30 and GX 17+2.
It can also be seen from Table 4.3 that 
if we use only the narrower limits on the luminosities and color
temperature, EOS model D (for $\cos i = 0.8)$ and EOS model C
(for $\cos i = 0.2)$ are not likely to be the correct EOS for GX 17+2. 
The same is true for model D for the source GX 349+2. As we also see from Table 
4.4, EOS model C is unfavored for $\cos i = 0.2$ for this source.
Further, the neutron star mass estimate in Cygnus X-2 ($\approx 1.78
M_\odot$, Orosz \& Kuulkers 1998) is not consistent with the soft
EOS model A. Our analysis, therefore, favors neutron star EOS models
which are intermediate in the stiffness parameter values.

We have ignored the magnetic fields of the neutron stars in our
calculations. Therefore, the necessary condition for the validity of
our results is that the Alfv\'{e}n radius $(r_{\rm A})$ be less than the
radius of the inner edge of the disk. This condition will always
be valid if $R > r_{\rm A}$ holds. Here $r_{\rm A}$ is given by
(Shapiro \& Teukolsky 1983),

\begin{eqnarray}
r_{\rm A} & = & 2.9 \times 10^8  {\mbox{\Huge (}{\dot {M}\over \dot
{M}_{\rm e}}\mbox{\Huge )}}^{-2/7} \mu_{30}^{4/7} \mbox{\Huge (}{M\over M_\odot}\mbox{\Huge 
)}^{-3/7}
\end{eqnarray}

\noindent where
$M$ is the mass of the neutron star, $\mu_{30}$ is
the magnetic moment in unit of $10^{30}$ G cm$^3$ and $r_{\rm A}$ is in cm.
With typical values of the parameters for the chosen sources $(R = 10$~km,
$M = 1.4 M_\odot$ and $\dot{M} = 10 \dot {M}_{\rm e})$, the upper
limit of the neutron star surface magnetic field comes out to be
about $2 \times 10^{8}$~G. Therefore, our results are in general valid for
the neutron star magnetic field upto of the order of $10^{8}$~G. This
is a reasonable value for the magnetic field of neutron stars in LMXBs, as
mentioned in section 4.1. However, this estimate of low magnetic field is based on
the assumption of its dipolar form. If the magnetic field geometry contains higher
order components, then the field-strength may be higher than the estimated value.

In our analysis, we have assumed that the boundary layer between the
disk and the neutron star surface does not affect the inner regions of
the disk. This will be a valid approximation when the boundary layer
luminosity is smaller than the disk luminosity,
and the boundary layer extent is small compared to the radius
of the star. The first part of this condition is true for all the chosen
LMXBs, as we see from the previous section. We now show that the second part is
also true for the source Cygnus X-2. The flux received at the earth from this 
region is

\begin{eqnarray}
F_{\rm BL} = \mbox{\Huge (}{2 \pi R \frac{\Delta R}{D^2}}\mbox{\Huge )} \cos i
\mbox{\Huge (}{\sigma T_{\rm BL}^4 \over \pi}\mbox{\Huge )}
\end{eqnarray}

\noindent where $\Delta R$ is the width of the boundary layer, $D = 8$ kpc
is the distance to the source, $i = 60^o$ is the inclination angle and
$T_{\rm BL}$ is the effective temperature. Spectral fitting gives
a best-fit value
for $F_{\rm BL} \approx 4 \times 10^{-9}$ ergs sec$^{-1}$ cm$^{-2}$
and $T_{\rm BL} = T_{\rm col(BL)}/f_{\rm BL}$ = 2.88/$f_{\rm BL}$~keV,
where $f_{\rm BL}$ is the color factor for
the boundary layer and $T_{\rm col(BL)}$ is the color temperature of the
boundary layer. Using these values, we get $\Delta R \approx $ 0.2
$f_{\rm BL}^4$ km,
which is indeed smaller than $R$ provided the boundary layer color
factor $f_{\rm BL}$ is close to unity. This is supported
by the work of London, Taam \& Howard (1986)
and Ebisuzaki (1987), who obtain $f_{\rm BL}\approx 1.5$.
The same conclusion can be drawn for other four chosen sources, as their
parameter values are similar to those for Cygnus X-2.

We have not attempted to model the observed temporal behavior of
the sources, and in particular, the QPO observations. Beat frequency model
identifies the peak separation of the two observed kHz QPOs with the neutron
star spin rate. For example, 
for Cygnus X-2 the observed peak separation is $\Delta \nu = 346
\pm 29$ Hz (Wijnands et al. 1998) which is smaller than the typical rotation
frequencies
calculated here. However, a pure beat-frequency model has been called into
question due to several observations. For instance, $\Delta \nu$ has been
observed to vary by about
40\% for Sco X-1 (van der Klis et al. 1997) and the kHz QPO frequencies
have been found to be correlated with the  break frequency
($\approx$ 20 Hz) of the power spectrum density. 
Inclusion of an alternate model,
where the QPOs are suggested to originate due to non--Keplerian motion
of matter in the disk (Titarchuk \& Osherovich 1999; Osherovich \&
Titarchuk 1999a; Osherovich \& Titarchuk 1999b; Titarchuk, Osherovich \&
Kuznetsov 1999), 
into the framework of the calculations mentioned in this work require
a new formulation within the space--time geometry chosen herein.

In this chapter, we have calculated the allowed ranges of several
quantities for five LMXBs, which give valuable information about
these systems. Such information will be helpful to understand their
temporal behaviors. Besides, LMXBs are believed to be the progenitors of
the millisecond pulsars. This is in accord with our result that
the neutron stars in Cygnus X-2, GX 9+1 and GX 349+2 are rapidly rotating. 
However, the data from the present and the future generation X--ray satellites
({\it Chandra}, {\it XMM}, {\it Constellation-X} etc.) with better spectral
resolutions (compared to those of earlier satellites)
will make better use (i.e., to get best-fit values of $\Omega_{\rm *}$
and to constrain EOS models) of the general relativistic model
presented here.

\begin{table*}
\begin{tabular}{lllllllll}
\hline
\hline
EOS  & $M$ &  & $f$  & $\nu_{\ast}$& $\nu_{\rm in}$ & $R$ &
$r_{\rm eff}^{\rm max}$ & $\dot{M}$   \\
   & $M_\odot$ &  & &  kHz & kHz & km & km & $\dot{M}_{\rm e}$  \\
\hline
(A)  &   1.4 & L  & 1.37[1.16] & 1.753[1.743] & 1.755[1.755] & 11.3[10.7] & 
16.0[15.6] & 11.2[5.8]  \\
  &  &       U  & 1.99[2.56] & 1.755[1.755] & 1.787[1.944] & 11.4[11.4] & 
16.1[16.1] & 22.9[27.5]  \\
\hline
(B)  &   1.4 & L  & 1.53[1.29] & 1.106[1.087] & 1.132[1.123] & 15.2[14.3] & 
21.0[20.0] & 13.8[7.2] \\
  &  &       U  & 2.18[2.74] & 1.112[1.113] & 1.177[1.285] & 15.6[15.6] & 
21.5[21.6] & 27.0[33.5]  \\
\hline
(C)  &   1.4 & L  & 1.57[1.33] & 0.964[0.945] & 0.975[0.971] & 16.8[15.6] & 
23.1[21.7] & 14.9[7.7]  \\
  &  &       U  & 2.24[2.81] & 0.968[0.968] & 1.015[1.134] & 17.2[17.2] & 
23.6[23.7] & 29.3[36.5]  \\
\hline
(D)  &   1.4 & L  & 1.67[1.42] & 0.736[0.719] & 0.745[0.742] & 20.1[18.6] & 
27.6[25.7] & 17.5[9.1]  \\
  &  &       U  & 2.38[2.97] & 0.740[0.740] & 0.779[0.876] & 20.7[20.7] & 
28.3[28.4] & 34.6[42.4]  \\
\hline
(B)  &   1.78 & L  & 1.58[1.33] & 1.303[1.292] & 1.322[1.315] & 14.8[14.2] & 
21.2[20.7] & 8.9[4.7]  \\
  &  &        U  & 2.28[2.91] & 1.307[1.307] & 1.361[1.462] & 15.1[15.1] & 
21.4[21.4] & 17.2[21.4]  \\
\hline
(C)  &   1.78 & L  & 1.65[1.39] & 1.081[1.067] & 1.086[1.085] & 17.1[16.2] & 
23.8[23.0] & 9.8[5.1]  \\
  &  &        U  & 2.39[3.01] & 1.083[1.083] & 1.109[1.209] & 17.3[17.3] & 
24.0[24.1] & 19.3[24.0]  \\
\hline
(D)  &   1.78 & L  & 1.74[1.47]  & 0.806[0.791] & 0.817[0.813] & 20.6[19.2] & 
28.6[27.1] & 11.4[6.0]  \\
  &  &        U  & 2.50[3.15]  & 0.809[0.809] & 0.848[0.938] & 21.1[21.1] & 
29.1[29.2] & 22.2[27.7]  \\
\hline
\end{tabular}
\caption{Observational constraints for various EOS models: (A),
(B), (C), (D) for the source Cygnus X-2.
L and U stand for lower and upper limits. The parameters
are $f$ (color factor), $\nu_{\ast}$ (rotational frequency of the neutron star),
$\nu_{\rm in}$ (rotational frequency in the ISCO),
$R$ (equatorial radius of the neutron star), $r_{\rm eff}^{\rm max}$
(radius where the effective temperature of the disk is maximum) and
$\dot {M}$ (the accretion rate). The limits are for 25\% uncertainty in
luminosities and 10\% uncertainty in the color temperature.
Values in brackets are for 50\% uncertainty in luminosities and 20\% uncertainty
in the color temperature. For EOS model A, the mass of the neutron star cannot
exceed 1.63 $M_\odot$ hence the $1.78 M_\odot$ solution
is not presented. The accretion rate is given in unit of $\dot {M}_{\rm e} =
1.4\times 10^{17} M/\msun$~g~s$^{-1}$, where $M$ is the neutron star mass.}
\end{table*}

\newpage

\begin{table*}

\begin{tabular}{lllllllll}
\hline
\hline
EOS  & $\cos i$ &  & $f$  & $\nu_{\ast}$& $\nu_{\rm in}$ & $R$ &
$r_{\rm eff}^{\rm max}$ & $\dot{M}$   \\
   & &  & &  kHz & kHz & km & km & $\dot{M}_{\rm e}$  \\
\hline
(A)  &   0.2 & L  & 1.31[1.10] & 1.751[1.726] & 1.756[1.756] & 11.2[10.2] &
18.6[18.1] & 7.7[3.8]  \\
  &  &       U  & 1.91[2.41] & 1.755[1.755] & 1.819[2.078] & 11.4[11.4] &
18.7[18.7] & 16.1[20.3]  \\
\hline
(B)  &   0.2 & L  & 1.45[1.30] & 1.103[1.059] & 1.137[1.126] & 15.0[13.7] &
23.0[21.6] & 9.7[4.5] \\
  &  &       U  & 2.07[2.61] & 1.112[1.113] & 1.197[1.372] & 15.5[15.6] &
23.5[23.6] & 19.4[24.4]  \\
\hline
(C)  &   0.2 & L  & 1.49[1.30] & 0.961[0.913] & 0.979[0.973] & 16.5[15.0] &
24.8[23.1] & 10.4[4.9]  \\
  &  &       U  & 2.12[2.71] & 0.968[0.968] & 1.042[1.206] & 17.2[17.2] &
25.5[25.6] & 21.2[26.7]  \\
\hline
(D)  &   0.2 & L  & 1.59[1.40] & 0.735[0.687] & 0.748[0.743] & 19.9[17.7] &
29.1[26.5] & 12.2[5.6]  \\
  &  &       U  & 2.25[2.81] & 0.740[0.740] & 0.795[0.941] & 20.6[20.7] &
30.0[30.1] & 25.0[31.4]  \\
\hline
(A)  &   0.8 & L  & 1.79[1.50] & 1.463[0.000] & 1.822[1.571] & 9.9[7.5] &
18.1[18.1] & 1.2[0.5]  \\
  &  &       U  & 3.06[4.70] & 1.751[1.754] & 2.165[2.165] & 11.2[11.4] &
20.4[22.3] & 4.5[6.4]  \\
\hline
(B)  &   0.8 & L  & 1.94[1.71] & 0.498[0.000] & 1.207[1.152] & 11.3[11.0] &
20.2[20.2] & 1.4[0.9]  \\
  &  &        U  & 3.22[4.20] & 1.102[1.110] & 1.782[1.782] & 14.9[15.4] &
22.9[23.4] & 5.6[7.9]  \\
\hline
(C)  &   0.8 & L  & 1.99[1.70] & 0.175[0.000] & 1.046[0.991] & 12.3[12.3] &
21.5[21.0] & 1.5[1.0]  \\
  &  &        U  & 3.36[4.02] & 0.960[0.966] & 1.573[1.568] & 16.5[17.0] &
24.7[25.4] & 6.1[8.5]  \\
\hline
(D)  &   0.8 & L  & 2.11[1.80]  & 0.000[0.000] & 0.806[0.758] & 14.7[14.7] &
23.1[23.1] & 1.8[1.2]  \\
  &  &        U  & 3.30[3.90]  & 0.733[0.739] & 1.212[1.212] & 19.7[20.5] &
28.9[29.8] & 7.2[9.9]  \\
\hline
\end{tabular}
\caption{Observational constraints for various EOS models : (A),
(B), (C), (D) for the source XB 1820-30. The mass of the neutron star is
assumed to be 1.4 $M_\odot$. Other specifications are same as in Table 4.1.}
\end{table*}

\newpage

\begin{table*}

\begin{tabular}{lllllllll}
\hline
\hline
EOS  & $\cos i$ &  & $f$  & $\nu_{\ast}$& $\nu_{\rm in}$ & $R$ &
$r_{\rm eff}^{\rm max}$ & $\dot{M}$   \\
   & &  & &  kHz & kHz & km & km & $\dot{M}_{\rm e}$  \\
\hline
(A)  &   0.2 & L  & 1.01[1.00] & 1.754[1.748] & 1.756[1.756] & 11.4[11.0] &
18.7[18.5] & 36.1[19.8]  \\
  &  &       U  & 1.43[1.82] & 1.755[1.755] & 1.773[1.869] & 11.4[11.4] &
18.7[18.7] & 68.7[82.6]  \\
\hline
(B)  &   0.2 & L  & 1.12[1.00] & 1.108[1.097] & 1.128[1.122] & 15.3[14.7] &
23.3[22.6] & 44.4[24.4] \\
  &  &       U  & 1.57[1.96] & 1.113[1.113] & 1.163[1.236] & 15.6[15.7] &
23.6[23.6] & 82.6[101.7]  \\
\hline
(C)  &   0.2 & L  & 1.15[1.00] & 0.966[0.954] & 0.974[0.971] & 16.9[16.0] &
25.2[24.2] & 47.5[26.1]  \\
  &  &       U  & 1.62[2.01] & 0.968[0.968] & 1.000[1.091] & 17.2[17.2] &
25.5[25.6] & 88.5[111.5]  \\
\hline
(D)  &   0.2 & L  & 1.22[1.03] & 0.738[0.728] & 0.744[0.742] & 20.3[19.2] &
29.7[28.3] & 55.9[30.7]  \\
  &  &       U  & 1.72[2.13] & 0.740[0.740] & 0.766[0.849] & 20.7[20.7] &
30.1[30.1] & 106.4[131.0]  \\
\hline
(A)  &   0.8 & L  & 1.39[1.20] & 1.702[0.000] & 1.782[1.571] & 9.9[7.5] &
18.1[18.1] & 6.6[2.0]  \\
  &  &       U  & 2.20[3.72] & 1.754[1.755] & 2.166[2.165] & 11.3[11.4] &
18.6[22.3] & 18.5[25.5]  \\
\hline
(B)  &   0.8 & L  & 1.53[1.31] & 1.009[0.000] & 1.172[1.141] & 13.1[11.0] &
21.1[20.2] & 7.7[3.2]  \\
  &  &        U  & 2.31[3.35] & 1.107[1.111] & 1.463[1.782] & 15.2[15.5] &
23.2[23.5] & 22.8[30.7]  \\
\hline
(C)  &   0.8 & L  & 1.61[1.30] & 0.858[0.000] & 1.010[0.983] & 14.3[12.3] &
22.4[21.0] & 8.3[3.6]  \\
  &  &        U  & 2.33[3.20] & 0.965[0.967] & 1.289[1.568] & 16.8[17.1] &
25.1[25.4] & 25.0[32.9]  \\
\hline
(D)  &   0.8 & L  & 1.66[1.41]  & 0.631[0.000] & 0.775[0.750] & 16.8[14.7] &
25.4[23.1] & 9.5[4.4]  \\
  &  &        U  & 2.47[3.10]  & 0.737[0.740] & 1.011[1.212] & 20.2[20.6] &
29.5[30.0] & 29.3[38.6]  \\
\hline
\end{tabular}
\caption{Observational constraints for various EOS models : (A),
(B), (C), (D) for the source GX 17+2.
Other specifications are same as in Table 4.2.}
\end{table*}

\newpage

\begin{table*}

\begin{tabular}{lllllllll}
\hline
\hline
EOS  & $\cos i$ &  & $f$  & $\nu_{\ast}$& $\nu_{\rm in}$ & $R$ &
$r_{\rm eff}^{\rm max}$ & $\dot{M}$   \\
   & &  & &  kHz & kHz & km & km & $\dot{M}_{\rm e}$  \\
\hline
(A)  &   0.2 & L  & 1.33[1.13] & 1.755[1.755] & 1.756[1.756] & 11.4[11.4] &
18.7[18.7] & 36.9[22.8]  \\
  &  &       U  & 1.85[2.25] & 1.755[1.755] & 1.756[1.761] & 11.4[11.4] &
18.7[18.7] & 59.9[72.0]  \\
\hline
(B)  &   0.2 & L  & 1.47[1.24] & 1.112[1.110] & 1.120[1.117] & 15.6[15.4] &
23.6[23.4] & 43.4[27.4] \\
  &  &       U  & 2.04[2.49] & 1.114[1.114] & 1.130[1.147] & 15.7[15.7] &
23.6[23.7] & 73.6[90.6]  \\
\hline
(C)  &   0.2 & L  & 1.51[1.28] & 0.968[0.967] & 0.970[0.970] & 17.2[17.1] &
25.5[25.4] & 47.5[29.3]  \\
  &  &       U  & 2.09[2.56] & 0.968[0.968] & 0.975[0.986] & 17.3[17.3] &
25.6[25.6] & 80.7[99.3]  \\
\hline
(D)  &   0.2 & L  & 1.61[1.36] & 0.740[0.739] & 0.742[0.741] & 20.7[20.5] &
30.1[29.9] & 55.9[34.4]  \\
  &  &       U  & 2.24[2.74] & 0.740[0.740] & 0.745[0.754] & 20.7[20.7] &
30.1[30.1] & 94.9[116.7]  \\
\hline
(A)  &   0.8 & L  & 1.84[1.61] & 1.752[1.728] & 1.756[1.756] & 11.2[10.3] &
18.6[18.1] & 7.9[3.8]  \\
  &  &       U  & 2.69[3.52] & 1.755[1.755] & 1.815[2.064] & 11.4[11.4] &
18.7[18.7] & 16.5[20.3]  \\
\hline
(B)  &   0.8 & L  & 2.05[1.80] & 1.103[1.064] & 1.136[1.126] & 15.0[13.8] &
22.9[21.7] & 9.7[4.6]  \\
  &  &        U  & 2.92[3.72] & 1.112[1.113] & 1.200[1.361] & 15.5[15.6] &
23.5[23.6] & 19.4[24.4]  \\
\hline
(C)  &   0.8 & L  & 2.10[1.80] & 0.961[0.919] & 0.978[0.972] & 16.5[15.0] &
24.8[23.2] & 10.4[5.0]  \\
  &  &        U  & 3.00[3.81] & 0.968[0.968] & 1.041[1.195] & 17.2[17.2] &
25.5[25.6] & 21.2[26.7]  \\
\hline
(D)  &   0.8 & L  & 2.24[1.92]  & 0.734[0.692] & 0.748[0.743] & 19.8[17.8] &
29.0[26.7] & 12.2[5.7]  \\
  &  &        U  & 3.19[4.00]  & 0.740[0.740] & 0.802[0.932] & 20.6[20.7] &
30.0[30.1] & 25.0[31.4]  \\
\hline
\end{tabular}
\caption{Observational constraints for various EOS models : (A),
(B), (C), (D) for the source GX 9+1.
Other specifications are same as in Table 4.2.}
\end{table*}

\newpage

\begin{table*}

\begin{tabular}{lllllllll}
\hline
\hline
EOS  & $\cos i$ &  & $f$  & $\nu_{\ast}$& $\nu_{\rm in}$ & $R$ &
$r_{\rm eff}^{\rm max}$ & $\dot{M}$   \\
   & &  & &  kHz & kHz & km & km & $\dot{M}_{\rm e}$  \\
\hline
(A)  &   0.2 & L  & 1.12[1.00] & 1.755[1.754] & 1.756[1.756] & 11.4[11.4] &
18.7[18.7] & 50.9[30.7]  \\
  &  &       U  & 1.55[1.92] & 1.755[1.755] & 1.756[1.771] & 11.4[11.4] &
18.7[18.7] & 86.4[103.9]  \\
\hline
(B)  &   0.2 & L  & 1.24[1.04] & 1.112[1.109] & 1.122[1.119] & 15.5[15.3] &
23.5[23.3] & 61.2[37.7] \\
  &  &       U  & 1.72[2.11] & 1.113[1.114] & 1.137[1.159] & 15.7[15.7] &
23.6[23.7] & 106.3[130.8]  \\
\hline
(C)  &   0.2 & L  & 1.27[1.08] & 0.968[0.966] & 0.971[0.970] & 17.2[17.0] &
25.5[25.3] & 67.1[40.4]  \\
  &  &       U  & 1.76[2.17] & 0.968[0.968] & 0.979[1.000] & 17.2[17.3] &
25.6[25.6] & 116.6[140.1]  \\
\hline
(D)  &   0.2 & L  & 1.35[1.14] & 0.740[0.738] & 0.742[0.741] & 20.6[20.4] &
30.0[29.7] & 78.8[47.5]  \\
  &  &       U  & 1.88[2.31] & 0.740[0.740] & 0.748[0.765] & 20.7[20.7] &
30.1[30.1] & 136.9[168.5]  \\
\hline
(A)  &   0.8 & L  & 1.55[1.34] & 1.748[1.678] & 1.760[1.756] & 11.0[9.7] &
18.4[18.1] & 10.6[4.5]  \\
  &  &       U  & 2.29[3.10] & 1.755[1.755] & 1.879[2.147] & 11.4[11.4] &
18.7[18.7] & 23.3[30.0]  \\
\hline
(B)  &   0.8 & L  & 1.71[1.50] & 1.096[0.955] & 1.148[1.129] & 14.6[12.7] &
22.6[20.7] & 13.1[5.2]  \\
  &  &        U  & 2.47[3.22] & 1.110[1.112] & 1.244[1.532] & 15.4[15.6] &
23.4[23.6] & 28.0[36.0]  \\
\hline
(C)  &   0.8 & L  & 1.76[1.53] & 0.953[0.798] & 0.984[0.974] & 16.0[13.8] &
24.2[21.9] & 14.0[5.6]  \\
  &  &        U  & 2.54[3.30] & 0.967[0.968] & 1.093[1.350] & 17.1[17.2] &
25.4[25.5] & 30.7[38.6]  \\
\hline
(D)  &   0.8 & L  & 1.87[1.61]  & 0.727[0.557] & 0.753[0.744] & 19.1[16.2] &
28.2[24.6] & 16.5[6.3]  \\
  &  &        U  & 2.69[3.40]  & 0.739[0.740] & 0.845[1.069] & 20.6[20.7] &
29.9[30.1] & 36.0[46.4]  \\
\hline
\end{tabular}
\caption{Observational constraints for various EOS models : (A),
(B), (C), (D) for the source GX 349+2.
Other specifications are same as in Table 4.2.}
\end{table*}

%% file: chap5.tex
\markright{Chapter 5}
\def\note #1]{{\bf #1]}}
\chapter{General Relativistic Spectra of Accretion Disks around 
Rotating Neutron Stars}

\section{Introduction}

Low Mass X-ray Binaries are believed to harbor black holes or weakly magnetized
neutron stars with an accretion disk. The X-ray emission arises from
the hot ($\approx 10^7$ K ) innermost region of the disk. In the
case of a neutron star there will be emission, in addition, from a boundary
layer between the accretion disk and neutron star surface. Since the
observed emission arises from regions close to a compact object, these sources are
possible candidates for studying strong field gravity.

In the standard theory (Shakura \& Sunyaev 1973), the accretion disk is
assumed to be an optically thick Newtonian one.
In this model, the local emergent flux (assumed to be a blackbody) is
equated to the energy dissipation at a particular radial
point in the disk.  The
observed spectrum is then a sum of black body components arising from
different radial positions in the disk. General relativistic effects modify
this Newtonian
spectrum in two separate ways. First, the local energy dissipation at
a radial point is different from the Newtonian disk, giving rise 
to a modified temperature profile. Second, the observed spectrum is 
no longer a sum of local spectra because of effects like Doppler 
Broadening, gravitational redshifts, and light-bending.
Modified spectra, incorporating these effects, but
with different approximations have been computed by several authors
(e.g. Novikov \& Thorne 1973; Asaoka 1989) for accretion disks around
rotating (Kerr) black holes. These computations confirm the expected
result, that the relativistic spectral shape differs from the Newtonian one
by around 10\%. Thus, for comparison with observed data with systematic and
statistical errors larger than 10\%, the Newtonian
approximation is adequate. Ebisawa, Mitsuda and Hanawa (1991) showed
that for typical data from the {\it Ginga} satellite, the relativistic spectrum cannot
be differentiated from the Newtonian disk spectrum. They also found
that the relativistic spectrum is similar in shape (at the sensitivity
level of {\it Ginga}) to the Comptonized model spectrum. Although, {\it Ginga} was
not sensitive enough to distinguish between the different spectra, better
estimates of fit parameters like accretion rate and mass of the compact object
were obtained when the data was compared to relativistic spectra rather
than the standard Newtonian one.

The present and next generation of satellites (e.g. {\it ASCA, RXTE,
Chandra, XMM, Constellation-X})
with their higher sensitivity and/or larger effective area than {\it Ginga} are
expected to differentiate between relativistic and Newtonian spectra from low
mass X-ray binaries (LMXB) and black hole systems. However, as pointed out
by Ebisawa, Mitsuda and Hanawa (1991), the presence of additional components
(e.g. boundary layer emission from the neutron star surface) and smearing
effects due to Comptonization may make the detection ambiguous.
Nevertheless, the detection of strong gravity effects on the spectra from
these sources will be limited by the accuracy of theoretical modeling of
accretion disk spectra rather than limitations on the quality of the
observed data. Thus, it is timely to develop accurate relativistically
corrected spectra for comparison with present and future observations.
Apart from
the importance of detecting strong gravity effects in the spectra of these
sources, such an analysis may also shed light on the geometry and dynamics
of innermost regions of accretion disks.

 Novikov \& Thorne (1973) and Page \& Thorne (1974)
computed the spectra of accretion disks around rotating (Kerr) black
holes. This formalism when directly applied to rotating neutron stars
provides only a first order estimate: the absence of an internal solution
in the case of Kerr geometry makes it difficult to obtain, in a
straightforward manner, the coupling between the mass and the angular
momentum of the central accretor.
On one hand, this coupling depends on the equation of state of neutron
star matter, and on the other hand, it depends on the proper 
treatment of rotation within general relativity.
Equilibrium configurations of rapidly rotating neutron stars for
realistic equations of state have been computed in Chapter 2.
One crucial feature in all these calculations is that the
space--time geometry is obtained by numerically and self--consistently
solving the Einstein equations and the equations for hydrostatic
equilibrium for a general axisymmetric metric.
With the aim of modeling spectra of LMXBs, here we attempt to
compute the spectrum of accretion disks around rotating neutron
stars within such a space--time geometry. This is particularly important
since LMXBs are old (population I) systems and the central
accretor in these systems are expected to have large rotation rates
(Bhattacharya \& van den Heuvel 1991 and references therein).

In section 5.2, we describe the spectrum calculation method,
without considering the light-bending effect. This effect is 
taken into account in section 5.3. We display the results in section 
5.4 and draw conclusions in section 5.5.

\section{Calculation of the Spectrum: Without Light-Bending Effect}

The disk spectrum is expressed as:
\begin{eqnarray}
F(E_{\rm ob}) & = & (1/E_{\rm ob})\int I_{\rm ob}(E_{\rm ob}) d\Pi_{\rm ob}
\end{eqnarray}
\noindent where the subscript `ob' denotes the quantity in observer's frame,
the flux $F$ is expressed in photons/sec/cm$^2$/keV, $E$ is photon energy
in keV, $I$
is specific intensity and $\Pi$ is the solid angle subtended by the source
at the observer.

As $I/E^3$ remains unchanged along the path of a photon (see for e.g.,
Misner et al. 1973),
one can calculate $I_{\rm ob}$, if $I_{\rm em}$ is known (hereafter, the
subscript `em' denotes the quantity in emitter's frame). We assume the
disk to emit like a diluted blackbody, so
$I_{\rm em}$ is given by
\begin{eqnarray}
I_{\rm em} & = & (1/f^4) B(E_{\rm em},T_{\rm c})
\end{eqnarray}
\noindent where $f$ is the color factor of the disk assumed to
be independent of radius (e.g. Shimura \& Takahara 1995). $B$ is the
Planck function and $T_{\rm c}$ (the temperature in the central plane of
the disk) is related to the effective temperature $T_{\rm eff}$ through
the relation $T_{\rm c}= f T_{\rm eff}$. 
The effective temperature,
$T_{\rm eff}$ is a function of the radial coordinate $r$
and for a rotating accretor is given by Eq. (3.1).

The quantities $E_{\rm ob}$ and $E_{\rm em}$ are related through the
expression $E_{\rm em}$ = $E_{\rm ob} (1 + z)$, where $(1 + z)$ contains the
effects of both gravitational redshift and Doppler shift. For a
general axisymmetric metric (representing the space--time geometry
around a rotating neutron star), the factor $(1 + z)$ is expressed as
(see for example, Luminet 1979)
\begin{eqnarray}
1 + z & = & (1 + \Omega_{\rm K} b \sin{\alpha} \sin{i}) {(-g_{tt} - 2
\Omega_{\rm K} g_{t\phi} - {\Omega_{\rm K}}^2 g_{\phi\phi})}^{-1/2}
\label{eq:1pz}
\end{eqnarray}
\noindent where the $g_{\mu\nu}$'s are the metric coefficients and $t$
and $\phi$ are the time and azimuthal coordinates. In the above
expression (which includes light--bending effects), $i$ is the
inclination angle of the source $(i = 0$ implies face--on), $b$ the impact
parameter of the photon relative to the line joining the source
and the observer and $\alpha$ the polar angle of the position of the
photon on the observer's detector plane.
For the sake of illustration and simplicity in calculations, here we
neglect light-bending.  We thus write
$b \sin{\alpha}$ = $r \sin{\phi}$ and
\begin{eqnarray}
d\Pi_{\rm ob} & = & r dr d\phi \cos{i} \over D^2
\end{eqnarray}
\noindent where $D$ is the distance of the source from the observer.

For our purpose here, we compute constant gravitational mass
sequences (as described in Chapter 3) whose rotation rates vary from zero to the centrifugal mass
shed limit (where gravitational forces balance centrifugal forces).
For realistic neutron stars, the inner radius $r_{\rm in}$ may be located
either at the marginally stable orbit or the surface of the neutron
star depending on its central density and rotation rate
(see Chapter 3), having important
implications for the gravitational energy release as well as the
temperature profiles of accretion disks.
The procedure of calculating $T_{\rm eff}$ for rapidly rotating neutron
stars considering the full effect of general relativity is given 
in Chapter 3. In Fig. 3.2, it is shown that the difference between Newtonian temperature profile and general
relativistic temperature profile is substantial at the inner portion of the
disk.  As will be shown herein, it turns out that this is the major reason
for the difference between Newtonian and general relativistic spectra at
high energies.

To summarize this section, we calculate the accretion disk spectrum
using Eq. (5.1), taking the radial integration limits as
$r_{\rm in}$
and $r_{\rm out}$ and the azimuthal integration limits as 0 and 2$\pi$.
We choose a very large value ($\approx 10^5$ Schwarzschild radius) for
$r_{\rm out}$.

\section{Calculation of the Spectrum: With Light-Bending Effect}

We describe the calculation of the spectrum, considering light 
bending effect, in a separate section, because it involves very 
time consuming numerical computations. Whether such computations are 
worthwhile to do, we will see that in the next section. Light-bending 
calculations use the Eqs. (5.1), (5.2) and (5.3). But the elementary 
solid angle will be given by 

\begin{eqnarray}
d\Pi_{\rm ob} & = & b\;db\;d\alpha \over D^2 \label{eq:Piob}
\end{eqnarray}

\noindent where $b$, $\alpha$ and $D$ are same parameters, as 
mentioned in section 2. As we consider the effect of gravity on 
photon's path, here we need to trace the photon's trajectory 
numerically. The procedure is described below.

For a configuration, described by $M$ and $\Omega_{\rm *}$ (and thus
 specified by a set of $g_{\mu \nu}$), we obtain $\Omega_{\rm K}$.
 To calculate the spectrum for a given value of $i$ with light
 bending effects, we backtrack the photon's path from the observer
 to the disk, using standard ray tracing techniques
 (e.g. Chandrasekhar 1983) and the relevant boundary conditions.
For convenience, we use $\mu$ ($= \cos\theta$) instead of $\theta$ and
$s$
(= $\bar r/(A + \bar r$)) instead of $r$ as the coordinates. Here 
$\bar r$ is the quasi--isotropic radial coordinate. 
Consequently, the metric (2.3) becomes 

\begin{eqnarray}
dS^2 & = & -e^{\rm {\gamma + \rho}} dt^2 + e^{\rm {2\alpha}} ((A^2/(1-s
)^4)
ds^2 \nonumber \\
 &   & + A^2 (s/(1-s))^2 (1/(1-\mu^2)) d\mu^2) \nonumber \\
 &   & + e^{\rm {\gamma - \rho}}
A^2 (s/(1-s))^2 (1-\mu^2) (d\phi - \omega dt)^2
\end{eqnarray}

\noindent Here, $A$ is a known constant of the dimension of distance.
Now it is quite straightforward to calculate the geodesic equations for
photons, which are given below.

\begin{eqnarray}
dt/d\lambda & = & e^{\rm {-(\gamma + \rho)}} (1 - \omega L)
\end{eqnarray}

\begin{eqnarray}
d\phi/d\lambda & = &  e^{\rm {-(\gamma + \rho)}} \omega (1 - \omega L)
\nonumber\\
 &   & + L/(e^{\rm {\gamma - \rho}} A^2 (s/(1-s))^2 (1-\mu^2))
\end{eqnarray}

\begin{eqnarray}
(ds/d\lambda)^2 & = &  e^{\rm {-2\alpha}} ((1-s)^4/A^2)
(e^{\rm {-(\gamma + \rho)}} (1 - \omega L)^2 \nonumber \\
 &   & - L^2/(e^{\rm {\gamma - \rho}}
A^2 (s/(1-s))^2 (1-\mu^2))) \nonumber \\
 &   & - s^2 (1-s)^2 (1/(1-\mu^2)) y^2
\end{eqnarray}

\begin{eqnarray}
d\mu/d\lambda & = &  y
\end{eqnarray}

\begin{eqnarray}
dy/d\lambda & = & -2 (\alpha_{\rm ,s} + (1/(s(1-s)))) y (ds/d\lambda) \
\nonumber \\
 &   & + \alpha_{\rm ,\mu} (1/(s(1-s)))^2 (1-\mu^2) (ds/d\lambda)^2 \nonumber \\
 &   & - (\alpha_{\rm ,\mu} + (\mu/(1-\mu^2))) y^2 \nonumber \\
 &   & + ((1/2) e^{\rm {\gamma - \rho - 2\alpha}} (\gamma_{\rm ,\mu} -
\rho_{\rm ,\mu}) (1-\mu^2)^2 \omega^2 \nonumber \\
 &   & - e^{\rm {\gamma - \rho - 2\alpha}} \mu (1-\mu^2) \omega^2 \nonumber \\
 &   & + e^{\rm {\gamma - \rho - 2\alpha}} (1-\mu^2)^2 \omega \omega_{\rm ,\mu}
 \nonumber \\
 &   & - (1/2) e^{\rm {\gamma + \rho - 2\alpha}} (\gamma_{\rm ,\mu} +
\rho_{\rm ,\mu}) \nonumber \\
 &   & ((1-s)/(A s))^2 (1-\mu^2)) (dt/d\lambda)^2 \nonumber \\
 &   & + e^{\rm {\gamma - \rho - 2\alpha}} (1-\mu^2)^2 (-\omega_{\rm ,\mu}
- \omega (\gamma_{\rm ,\mu} - \rho_{\rm ,\mu}) \nonumber \\
 &   & + 2 \omega (\mu/(1-\mu^2))) (d\phi/d\lambda) (dt/d\lambda) \nonumber \\
 &   & + e^{\rm {\gamma - \rho - 2\alpha}} (1-\mu^2)^2 ((1/2)
(\gamma_{\rm ,\mu} - \rho_{\rm ,\mu}) \nonumber \\
 &   & - \mu/(1-\mu^2)) (d\phi/d\lambda)^2
\end{eqnarray}

\noindent where, $\lambda$ is the affine parameter, $L$ is the negative of the ratio of
the $\phi$-component of the angular momentum and the $t$-component of the
angular momentum of photon and a comma followed by a variable as subscript to
a quantity, represents a derivative of the quantity with respect 
to the variable.

We cover the disk between radii $r_{\rm in}$ and
 $r_{\rm mid}=1000 r_{\rm g}$;  $r_{\rm in}$ being the radius
 of the inner edge of the disk and $r_{\rm g}$ the Schwarzschild
 radius (increasing $r_{\rm mid}$ has no significant effect
 on the spectrum).
 Beyond $r_{\rm mid}$, we ignore the
 effect of light-bending i.e., we take
 $b \sin\alpha = r \sin\phi$ ($\phi$ is the azimuthal
 angle on disk plane) and
 $d\Pi_{\rm ob}$ = $(r\;dr\;d\phi\;\cos i)\;/\;D^2$
 (see section 5.2).

We have performed several consistency checks on our results:
(1) by switching off the light-bending effect (i.e. by considering
    flat space-time while backtracking the photon's path), we see
    that the spectrum matches very well with that computed by ignoring
    light-bending effects (calculated by an independent code --
    section 5.2).   Also, in this case,
    the analytically calculated values of several quantities
    on the disk plane (e.g. $r$, $\phi$, $d\phi/dt$, $d\theta/dt$
    etc.) are reproduced satisfactorily by our numerical method,
(2) an increase in the number of grid points on the ($b$,$\alpha$)
    plane do not have any significant effect on the computed spectrum,
(3) the spectrum matches very well with the Newtonian spectrum
    (Mitsuda et al. 1984) at low energy limit.
This would imply that for higher frequencies, our spectrum is
correct to within 0.2\% to 0.3\%.

\section{The Results}

\nopagebreak
\begin{figure}[h]
\psfig{file=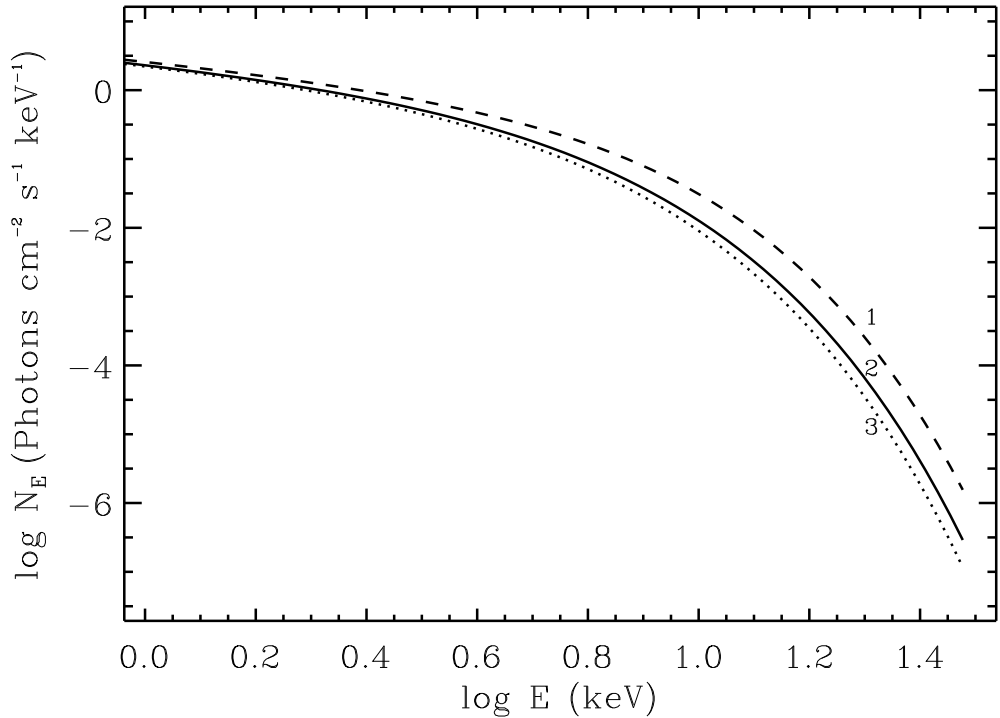,width=15 cm}
\caption{
Effect of general relativity: spectra from accretion
disk around a neutron star of mass $1.4 M_\odot$. All the
curves are for EOS model (B), $\Omega_{\rm *} = 0$, $D = 5$~kpc,
$i = 60^{\circ}$, $\dot M = 10^{18}$~g/sec and $f = 2$.
Curve (1) corresponds to the Newtonian case,
curve (2) to the general relativistic case including the
effect of light-bending and
curve (3) to the general relativistic case without considering
the effect of light-bending.
}
\end{figure}

\nopagebreak
\begin{figure}[h]
\psfig{file=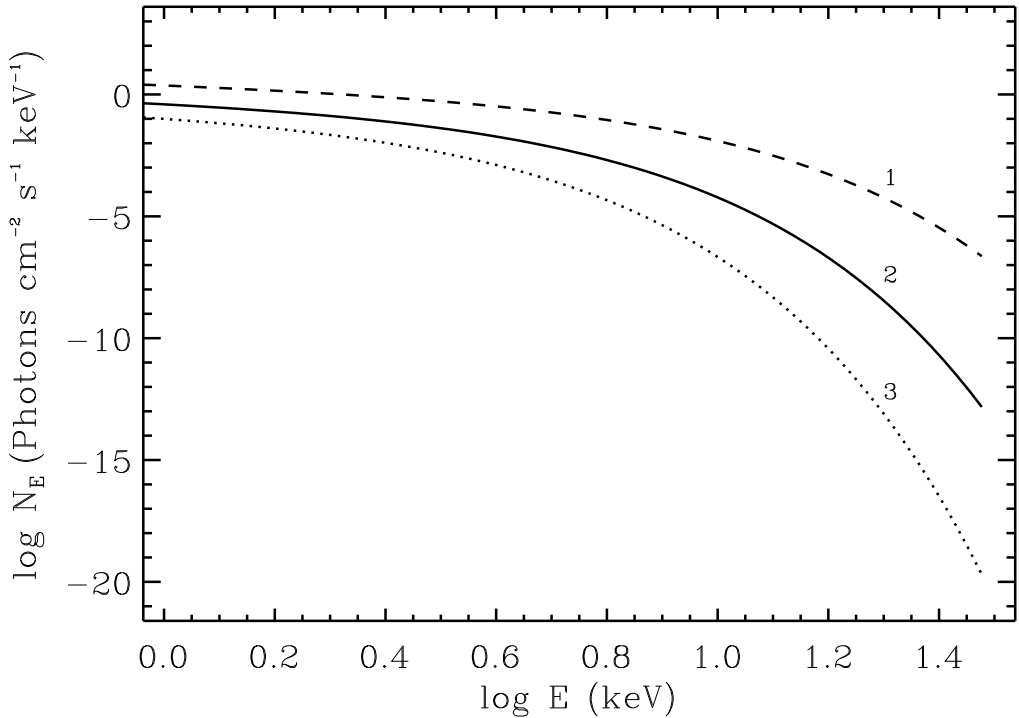,width=15 cm}
\caption{Accretion rate dependence: general relativistic
spectra including light-bending effects from accretion
disk around a neutron star of mass-shed limit configuration
($\Omega_{\rm *} = 7001$~rad/s).
Curve (1) corresponds to $\dot M = 10^{18}$~g/sec,
curve (2) to $\dot M = 10^{17}$~g/sec and
curve (3) to $\dot M = 2 \times 10^{16}$~g/sec.
The values of all the other parameters are as in Fig. 5.1.
}
\end{figure}

\nopagebreak
\begin{figure}[h]
\psfig{file=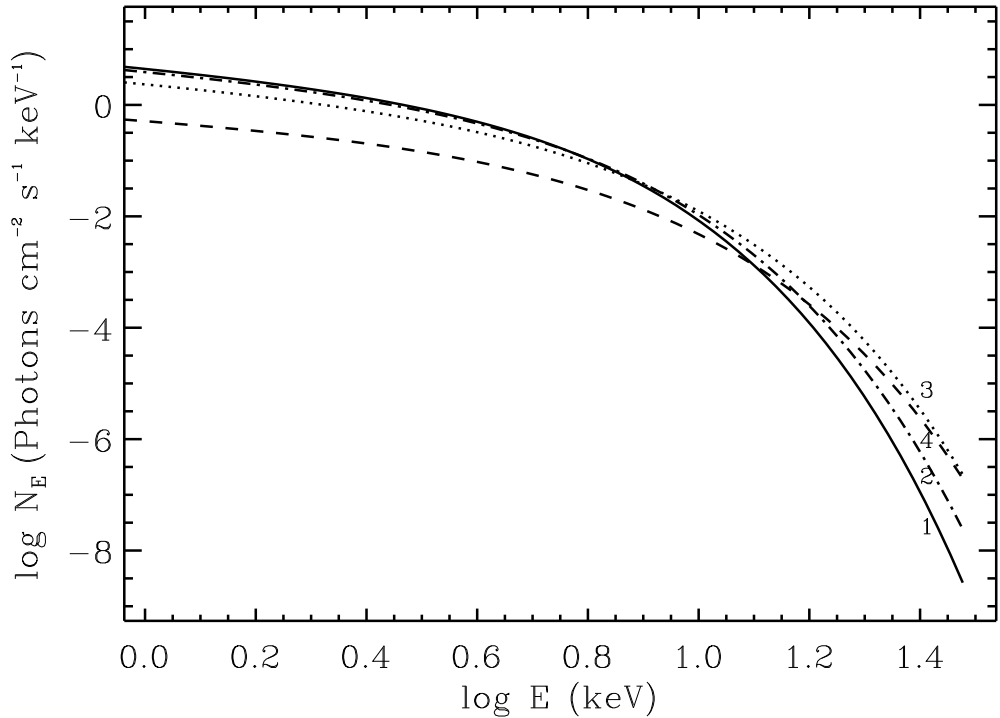,width=15 cm}
\caption{Inclination angle dependence: general relativistic
spectra including light-bending effects from accretion disk
around a neutron star of mass-shed limit configuration
($\Omega_{\rm *} = 7001$~rad/s).
Curve (1) corresponds to $i = 0^{\circ}$,
curve (2) to $i = 30^{\circ}$,
curve (3) to $i = 60^{\circ}$ and
curve (4) to $i = 85^{\circ}$.
The values of all the other parameters are as in Fig. 5.1.}
\end{figure}

\nopagebreak
\begin{figure}[h]
\psfig{file=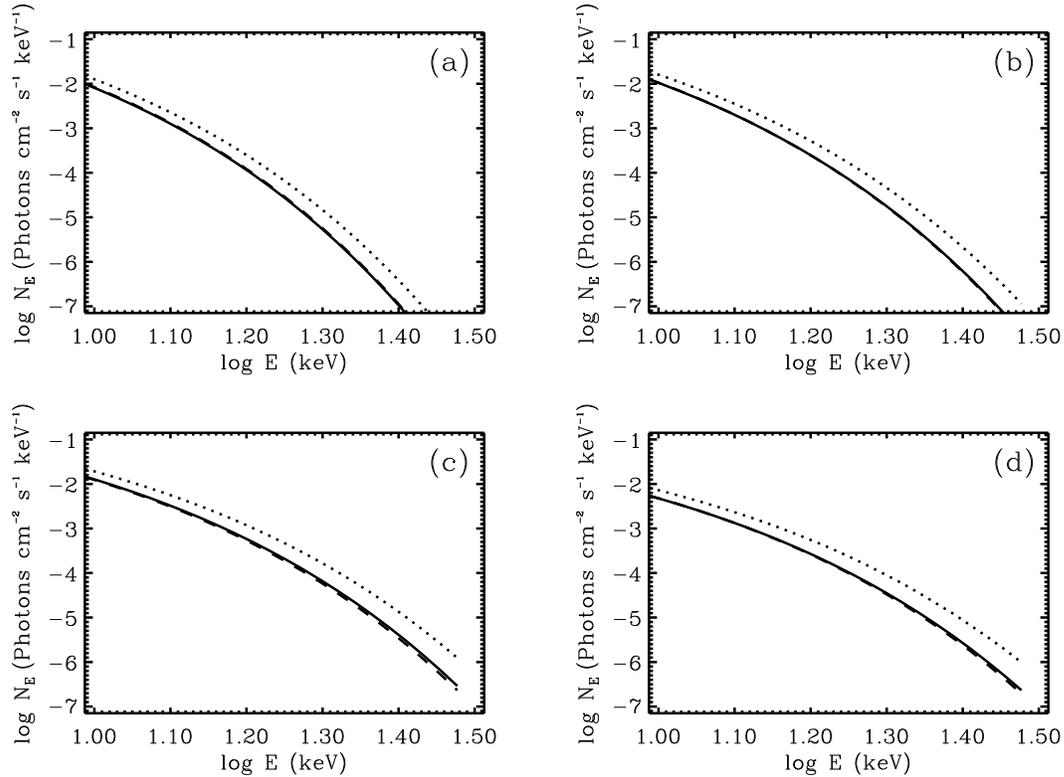,width=15 cm}
\caption{Rotation rate dependence: general relativistic spectra
including light-bending effects from accretion disk around a
neutron star.
Panel (a) corresponds to $i = 0^{\circ}$,
panel (b) to $i = 30^{\circ}$,
panel (c) to $i = 60^{\circ}$ and
panel (d) to $i = 85^{\circ}$.
In each panel, the solid curve corresponds to $\Omega_{\rm *} = 0$ rad/s,
the adjacent dashed curve corresponds to $\Omega_{\rm *} = 7001$ rad/s
(the mass-shed limit)
and the dotted curve corresponds to $\Omega_{\rm *} = 3647$ rad/s.
The values of all the other parameters are as
in Fig. 5.1.}
\end{figure}

\nopagebreak
\begin{figure}[h]
\psfig{file=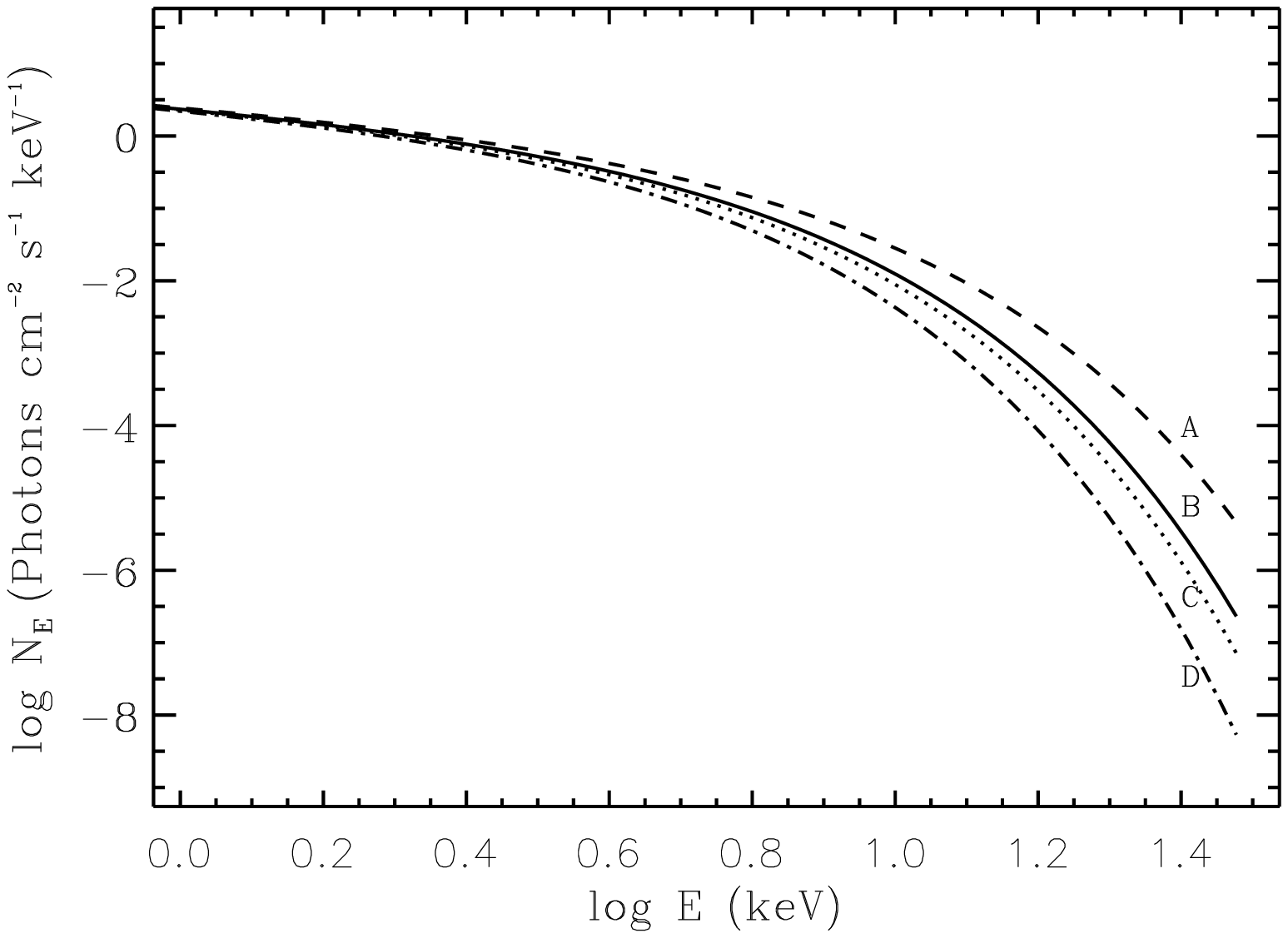,width=15 cm}
\caption{EOS model dependence: general relativistic spectra
including light-bending effects from accretion disk
around a neutron star of mass-shed configuration.
EOS model (A) is for $\Omega_{\rm *} = 11026$~rad/s,
model (B) is for $\Omega_{\rm *} = 7001$~rad/s,
model (C) is for $\Omega_{\rm *} = 6085$~rad/s and
model (D) is for $\Omega_{\rm *} = 4652$ rad/s. The values
of all the other parameters are as in Fig. 5.1.}
\end{figure}

We calculate the general relativistic spectrum from the accretion
disk around rapidly rotating neutron star, taking into account the
light-bending effect.  The spectrum is calculated as a function
of 6 parameters : $M$, $\Omega_{\rm *}$, distance of the source ($D$),
inclination angle ($i$) (for face-on, $i = 0^{\circ}$),
accretion rate ($\dot M$) and color factor $f$,
for each of the chosen EOS. Our results are displayed
in Figs. 5.1 to 5.5. In all the displayed spectra, we have
assumed $M = 1.4 M_{\odot}$ (canonical mass for neutron stars),
$D = 5$ kpc and $f = 2.0$.

In Fig. 5.1, we have plotted the Newtonian spectrum and
GR spectra with (LBGR) and without (NLBGR) light-bending
effect, keeping the values of all the parameters same. At
10 keV, the Newtonian flux is almost 2.5 times the LBGR flux.
This is quite expected, because in the inner parts of the disk,
Newtonian temperature is considerably higher than the GR temperature
(see Fig. 3.2). LBGR flux is about
50\% higher than NLBGR flux at 10 keV. This is because light-bending
causes the disk to subtend a larger solid angle at the observer than
otherwise. Thus the general effect of light-bending is to increase
the observed flux.

According to Shimura \& Takahara (1995), the thin blackbody
description of the accretion disk, as adopted in this paper,
is valid for $0.1 {\dot M}_{\rm e} < \dot M < {\dot M}_{\rm Edd}$,
where ${\dot M}_{\rm e} \equiv L_{\rm Edd}/c^2$. Here $L_{\rm Edd}$
is the Eddington luminosity and ${\dot M}_{\rm Edd}$ is the
Eddington accretion rate (see Chapter 4 for description). For the purpose of demonstration,
 we have taken three different values of $\dot M$ in this range
(for the mass-shed configuration) and plotted the corresponding
spectra in Fig. 5.2. As is expected, we see that the high energy
part of the spectrum is more sensitive to the value of $\dot M$.
It is seen that the spectra for different values of $\dot M$ are
easily distinguishable.

The inclination angle $i$ is a very important parameter in
determining the shape of the spectrum and its overall normalisation.
In Fig. 5.3, we have plotted the spectra for four inclination angles,
for the mass-shed configuration. We see that the observed flux at low
energies is higher for lower values of $i$.  This is simply due to
the projection effect (proportional to $\cos i$). But at higher
energies ($> 10$~keV) this trend is reversed mainly because Doppler
effect becomes important. The most energetic photons mainly come
from the inner portion of the disk, where the linear speed of
accreted matter is comparable to the speed of light. The net
effect of Doppler broadening is a net blue shift of the
spectrum, as a larger amount of flux comes from the
blue-shifted regions than from the red-shifted regions.
This is a monotonic trend, but it will be noticed from Fig. 5.3
that the curve for $i=85^{\circ}$ overcomes that for
$i=60^{\circ}$ only at the edge of the figure, i.e., at
energies $\geq 30$~keV.  This is due to the fact that
between these two inclinations the difference in the $\cos i$
factor is severe, and the blueshift overcomes this only at
high energies.

In Fig. 5.4, we have four panels for four inclination
angles. In each panel, we have shown spectra for 3 different
$\Omega_{\rm *}$ (corresponding to non-rotating, intermediate and the
mass-shed configurations).  With the increase of $\Omega_{\rm *}$, disk
temperature profile does not vary monotonically (see Fig. 3.3a).
Hence the behavior of the
spectrum is also non-monotonic with $\Omega_{\rm *}$. For non-rotating
and mass-shed configurations (for the assumed values of other
parameters) the temperature profiles are very similar.
As a result, the plotted spectra for these two cases lie
almost on top of each other.  However, for $i = 0^{\circ}$
the flux corresponding to the mass-shed configuration
is slightly higher than that for $\Omega_{\rm *}=0$, while
the case is opposite at higher inclinations.  This is a
result of the inclination dependence of the $(1+z)$ factor
given in Eq. 5.3.

In Fig. 5.5, we have compared the spectra for the four
EOS models adopted by us, for configurations at the
respective mass-shed limits (which correspond to different
values of $\Omega_{\rm *}$ because of the EOS dependence of the stellar
structure). The values of all other parameters have been kept
the same. We see that the total flux received varies
monotonically with the stiffness parameter, and is higher for
the softer EOS.  This effect has been mentioned in Chapter 3.
We see that at high energies
the fluxes for different EOS are considerably different.
Therefore, fitting the observed spectra of LMXBs with our
model spectra, particularly in hard X--rays, may provide a
way to constrain neutron star EOS. However one must
remember that these computations have been made assuming
that the magnetic field of the compact object does not
limit the inner boundary of the accretion disk.  In the
presence of a magnetic field strong enough to do so,
appropriate modifications must be taken into account
for the expected flux at high energies.

\section{Conclusion}

In this chapter we have computed the observed radiation
spectrum from accretion disks around rapidly rotating
neutron stars using fully general relativistic disk
models.  This is the first time such a calculation has
been made in an exact way, without making any approximation
in the treatment of either rotation or general relativity.
In computing the observed spectrum from the disk, we
explicitly include the effects of Doppler shift,
gravitational redshift and light-bending for an appropriate
metric describing space--time around rapidly rotating
neutron stars.  We find that the effect of
light-bending is most important in the high-energy ($> 3$~keV)
part of the observed spectrum.  Photons at these high energies
originate close to the central star, and hence their trajectories
are most affected by the light-bending effect.  Depending on the
viewing angle, this can enhance the observed flux at $\sim 10$~keV
by as much as $250$\% compared to that expected if light-bending
effects are neglected.

It is to be noted that in this work we have neglected the effect of
irradiation of the disk. Miller \& Lamb (1996) have discussed such
effects on a test particle moving towards a slowly rotating neutron
star. A strongly irradiated disk may not remain thin and the
radiation force may relocate the position of inner edge of the disk.
In addition to that, fractions of angular momentum and energy of the
accreted matter may be transferred to the irradiating photons, resulting
in a redistribution of emitted flux in the disk. These effects will change
our calculated spectrum to some extent. 
Therefore, we aim to modify our calculation in the direction of the
work of Miller \& Lamb (1996).
However, it is to be noted that for rapidly rotating neutron stars,
boundary layer emission is small and hence the effect of irradiation may
not be important. 

The calculations presented here deal only with the
multicolor blackbody disk.  In reality, there will be additional
contributions to the observed spectrum from the boundary layer
as well as a possible accretion disk corona, both of which are likely
to add a power-law component at high energies (Popham \& Sunyaev
2000, Dove et al. 1997).  On the other hand, the spectra
presented in Figs 5.2, 5.3 and 5.5 should remain essentially
unaffected by boundary layer contribution, as these are for
neutron stars rotating near the mass-shed limit for which the
boundary layer luminosity will be negligible.  For slowly rotating
neutron stars, the disk component of the spectrum can be obtained
by fitting and removing the contribution of the boundary layer,
provided a good model for the boundary layer spectrum is available.
Popham \& Sunyaev (2000) have made an attempt to compute the boundary
layer spectrum in the Newtonian approximation.  General Relativistic
modifications need to be included in these calculations to get
a realistic estimate of the spectrum of the boundary layer.  We
plan to address this issue in a future work.
In the slow rotation case, the spectrum of
the disk itself may be somewhat modified by the presence of a
boundary layer if it extends beyond the disk inner radius assumed in
our computations here, thus curtailing the inner edge of the disk.

In addition to the contribution of the boundary layer, the
possible contribution of an accretion disk corona to the emergent
spectrum could also be significant. To be able to costrain
the EOS models of Neutron Stars using the observed spectrum,
this contribution must also be accurately estimated.  We have
not attempted to estimate this in the present work, where
we restrict ourselves to thin
blackbody and non--magnetic accretion disks in order to
understand the effect of the EOS models describing neutron stars
on the spectrum of the accretion disk alone. We view this as
the first step in accurately modeling of the spectra of accreting
neutron stars including the effects of general relativity and
rotation.  We may mention that the radiation originating in
the accretion disk corona would also be modified by the
gravitational redshift and light-bending effects, and the
technique presented by us here will be useful also in that context.

The comparison of the non-rotating limit of our results with
those of the fitting routine {\it GRAD} in the X--ray spectral
reduction package {\it XSPEC} (Ebisawa et al. 1991), shows
that the latter model overpredicts the high-energy
component of the flux by a large factor.
With the help of K. Ebisawa \& T. Hanawa we have been able
to trace this disagreement to certain simplifying approximations
made in the {\it GRAD} code, as well as a couple of incorrect
expressions being used there.
Conclusions based on the use of the {\it GRAD} routine may
therefore need to be revised in the light of the new calculations
presented here.

The computation of the complete spectrum in the manner
presented here is rather time-consuming and therefore not
quite suited to routine use.  Therefore, in order to make
our results available for routine spectral
fitting work, we need to present a series of approximate
parametric fits to our computed spectra. We do it to some extent
in the next chapter.

The spectra presented here will find use in
constraining the combined parameter set of the mass, the
rotation speed and, possibly, the EOS, particularly of
weakly magnetised, rapidly rotating neutron stars.  The
relevant signatures are most prominent in hard X--rays,
above $\sim 10$~keV.  Sensitive observations of hard X--ray
spectra of LMXBs, therefore, are needed to fully utilise the
potential of these results.

%% file: chap6.tex
\markright{Chapter 6}
\def\note #1]{{\bf #1]}}
\chapter{Functional Approximation of Disk Spectra}

\section{Introduction}

The observed spectrum for luminous LMXBs can be well-fitted by the 
sum of a multi-color blackbody spectrum (presumably from the accretion
disk) and a single temperature blackbody spectrum (presumably from the
boundary layer) (see Mitsuda et al. 1984). The multi-color spectrum 
can be calculated if the temperature profile of the accretion disk is
known. Such a calculation should include the general relativistic 
effect, as near the surface of the neutron star, accretion flow is 
governed by the strong gravity. As is argued in Chapter 2, 
the effect of rapid rotation should also be taken into account.
We have calculated such a spectrum for thin accretion disk in 
Chapter 5. The computation has been done both ignoring and considering
the light-bending effect. 

Our model spectra, when fitted to the observational data, can in 
principle constrain EOS and the values of the source parameters.
However, computation of the spectra is numerically time 
consuming and hence
direct fitting to the observational data is impractical. For the sake of
ease in modeling, we present in this chapter, a simple empirical
analytical expression that describes
the numerically computed spectra. As shown later, the same expression
(which has three parameters including normalization) can also
describe the Newtonian spectra. In particular, the value of one
of the parameters (called $\beta$-parameter here) 
determines whether the spectrum is relativistically corrected or not. 
This will facilitate comparison
with observational data since only this $\beta$-parameter has to be
constrained to indicate the effect of strong gravity in the 
observed spectrum.

Here, for fitting, we consider spectra, without light-bending effect,
as the light--bending--calculation takes a huge amount of time.
However, we fit the analytical function to a few 
light--bending--spectra, and show that the general conclusion remains
the same, if the inclination angle $(i)$ is not too large.

In section 6.2, we describe the method of functional approximation of
the computed spectra. We display the results in section 6.3 and give 
a conclusive discussion in section 6.4.

\section{Functional Approximation Method}

In order to facilitate comparison with observations, we introduce a simple
analytical expression which empirically describes the computed relativistic
(and Newtonian) spectra.
\begin{eqnarray}
S_{\rm f} ( E ) & = & S_o E_a^{-2/3} \mbox{\Huge (}{E\over E_a}\mbox{\Huge )}^{\gamma} 
\exp\mbox{\Huge (}-{E\over E_a}\mbox{\Huge )} 
\end{eqnarray}
where, $\gamma = -(2/3)(1+E\beta/E_a)$, $E_a$, $\beta$ and $S_o$ are
parameters and $E$ is the energy of the photons in keV. $S_{\rm f}
(E) $ is in units of photons/sec/cm$^2$/keV. To compare this empirical
function with the computed spectra, we use a reduced $\chi^2$ technique.
In particular, we define a function
\begin{eqnarray}
\chi^2 & = & {1\over N}\sum_{i = 1}^{N} \mbox{\Huge [}{S_{\rm c}(E_i)-S_{\rm f}(E_i)
\over 0.1  S_{\rm c}(E_i)}\mbox{\Huge ]}^2
\end{eqnarray}
where $S_{\rm c}(E)$ is the computed spectra. The spectra are divided
into $N$ logarithmic energy bins. We have chosen the range of energy
used in calculating $\chi^2$ to be dependent on the location of the
maximum of the energy spectrum $(E S_{\rm c} (E))$ which is typically
at 2 keV. The minimum energy is set to be  one hundredth of this value
(typically 0.02 keV) while the maximum is set at ten times (typically
20 keV). $\chi^2$ is fairly insensitive to the number of energy bins;
we take $N = 200$. For each $S_{\rm c}(E)$ the best-fit parameters
($E_a$, $\beta$ and $S_o$) are obtained by minimizing $\chi^2$.

The $S_o$ parameter in Eq. (6.1), is the normalization
factor and is independent of the relativistic effects. It depends only
on the mass of the star ($M$), accretion rate ($\dot M$), distance to
the source ($D$), color factor (f) and inclination angle (i) i.e
$S_o \propto \dot M^{2/3} f^{-4/3} M^{1/2} D^{-2} \cos{i}$.  The $E_a$
parameter (which is in units of keV) describes the high energy cutoff
of the spectrum. Its dependence on the space-time metric 
and inclination
angle is complicated but it scales as $E_a \propto \dot M^{1/4} f$. The
$\beta$-parameter depends only on the space-time metric (and the inclination angle), 
but not on accretion rate, distance to the source or color factor. This makes
the $\beta$-parameter useful as a probe into the underlying space-time
metric.

\section{The Results}

\nopagebreak
\begin{figure}[h]
\psfig{file=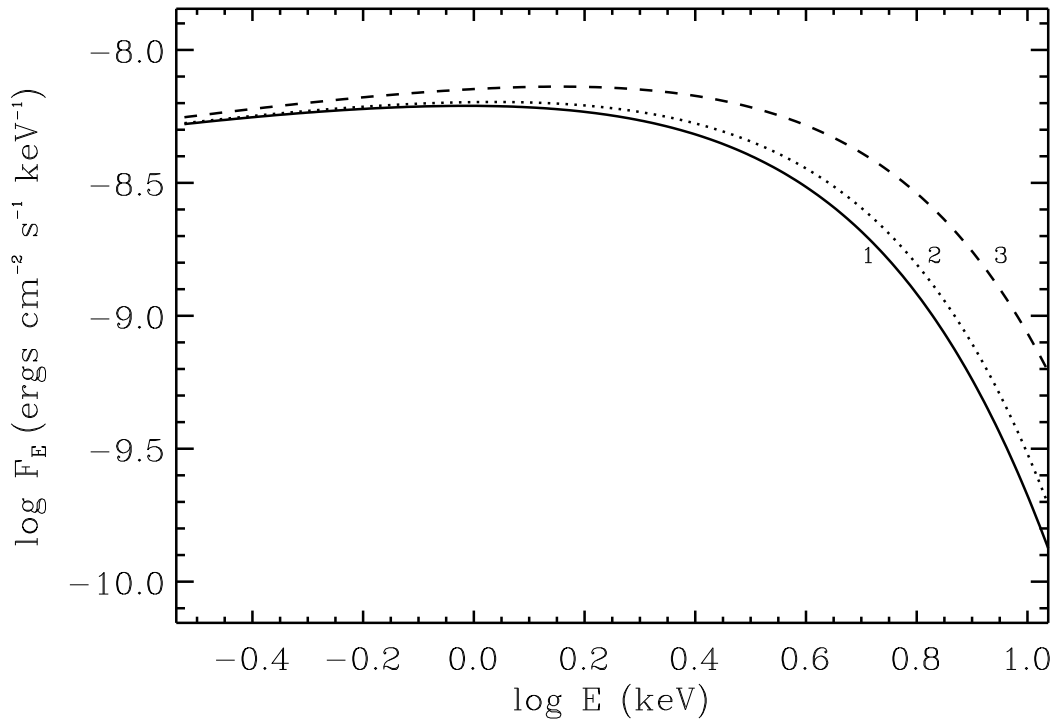,width=15 cm}
\caption{General relativistic spectrum (solid line) for a neutron star
configuration with mass $M = 1.4 M_\odot$, spin rate $\Omega_* = 0$,
distance to the source $D = 5$ kpc, inclination angle $i = 30^{\circ}$,
accretion rate $\dot M = 10^{18}$ g s$^{-1}$ and color factor $f = 2$. Dashed
line: the spectrum expected from a source with the same disk parameters
but without the relativistic effects (Newtonian spectrum). Dotted
line: The spectrum for the same disk parameters but without the effect
of Doppler and gravitational red-shifts (i.e. $z$ is set to zero).
The EOS model (B) is used here.}
\end{figure}

\nopagebreak
\begin{figure}[h]
\psfig{file=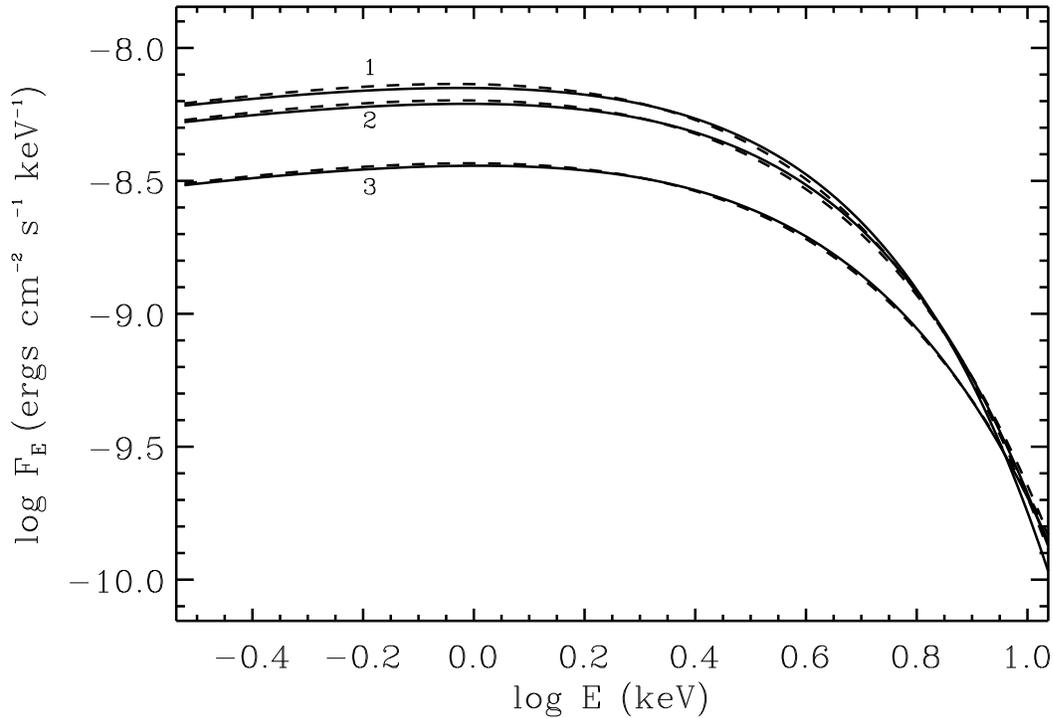,width=15 cm}
\caption{Relativistic spectra for three different
inclination angles ($i = 0^{\circ},30^{\circ},60^{\circ}$) with rest of the parameters same
as in Fig. 6.1 (solid lines). Dashed lines: empirical fit to
the relativistic spectra using Eq. (6.1). The minimum
$\chi^2 = 0.073, 0.049$ and $0.026$ for $i = 0^{\circ}, 30^{\circ} and 60^{\circ}$ respectively.}
\end{figure}

\nopagebreak
\begin{figure}[h]
\psfig{file=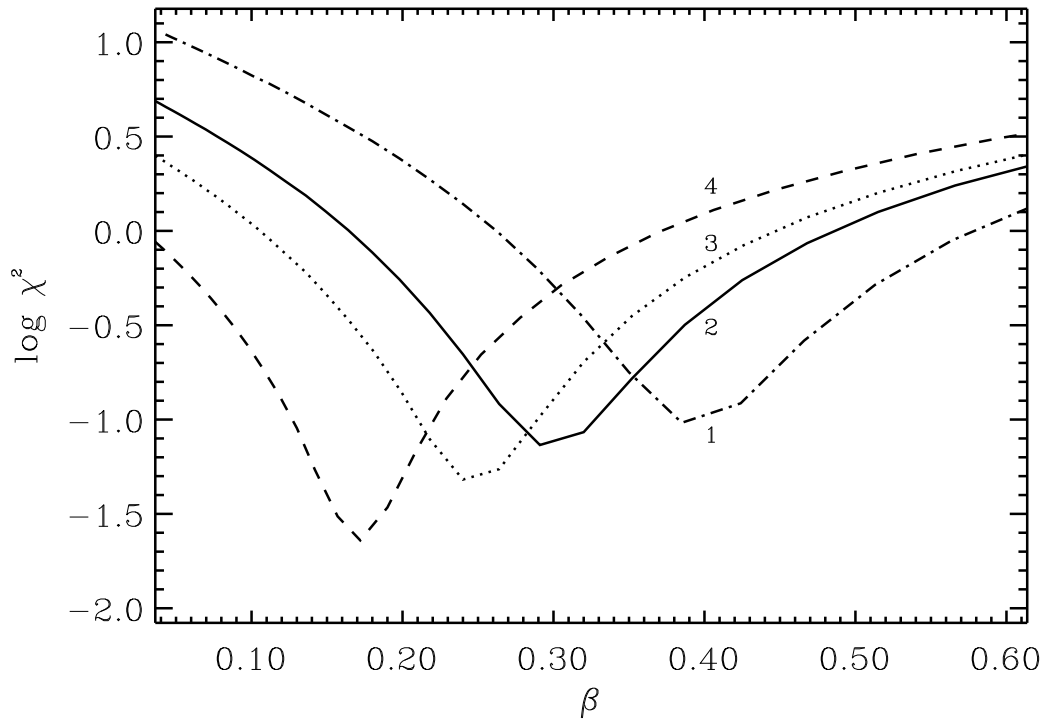,width=15 cm}
\caption{Variation of minimum $\chi^2$ (i.e. minimized with respect to
parameters $E_a$ and $S_o$) with parameter $\beta$. Curves marked 2, 3
and 4 correspond to the spectra shown in Fig. 3
for $i = 0^{\circ}, 30^{\circ}$ and $60^{\circ}$ respectively. Curve marked 1 is for the
Newtonian spectra shown in Fig. 6.1.}
\end{figure}

\nopagebreak
\begin{figure}[h]
\psfig{file=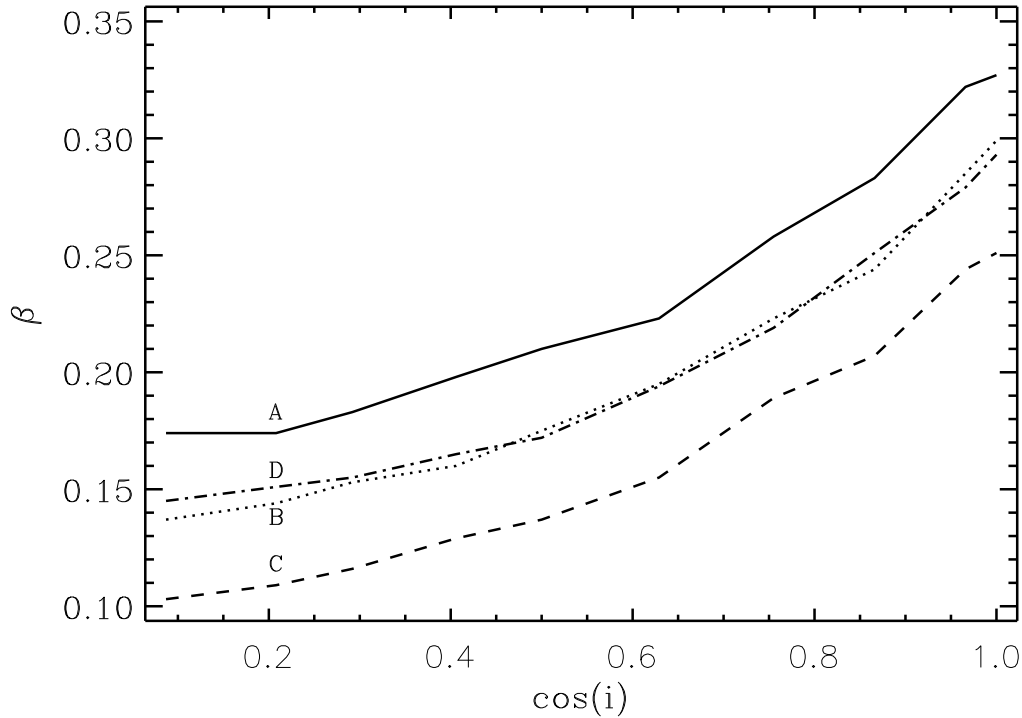,width=15 cm}
\caption{Variation of the best-fit $\beta$-parameter with inclination
angle for different equations of states (each curve is marked by the
corresponding EOS model). 
The values of the other parameters are as in Fig. 6.1.}
\end{figure}

\nopagebreak
\begin{figure}[h]
\psfig{file=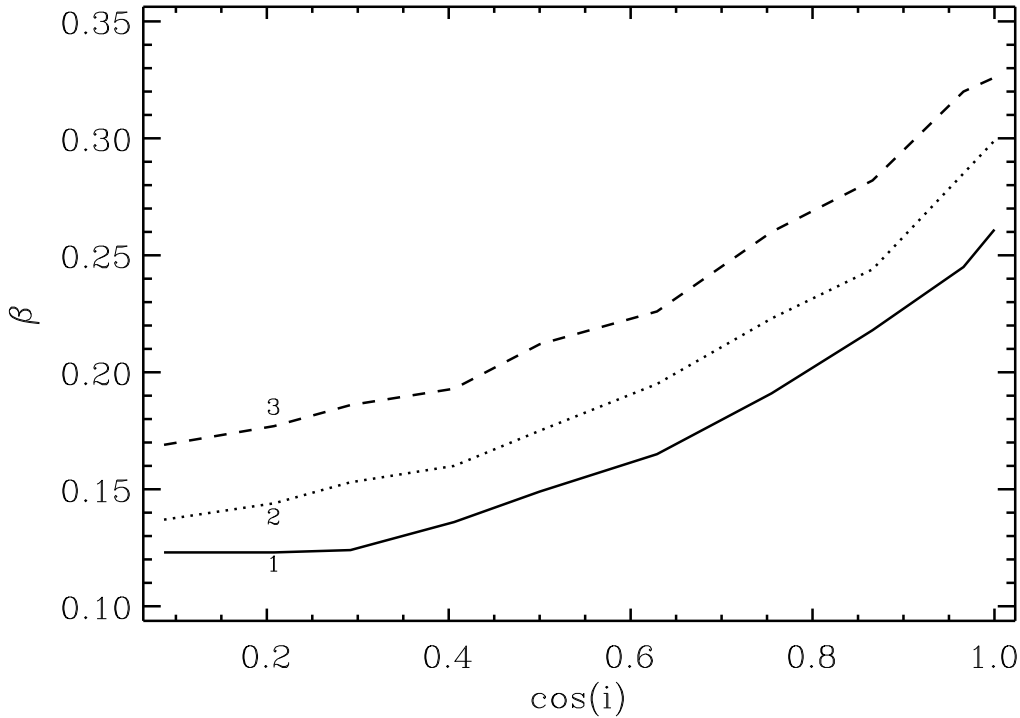,width=15 cm}
\caption{Variation of the best-fit $\beta$-parameter with inclination angle
for different neutron star masses. Curve 1: $M = 1.0 M_\odot$,
curve 2: $M = 1.4 M_\odot$,
curve 3: $M = 1.788 M_\odot$.
The values of the other parameters are as in Fig. 6.1.}
\end{figure}

\nopagebreak
\begin{figure}[h]
\psfig{file=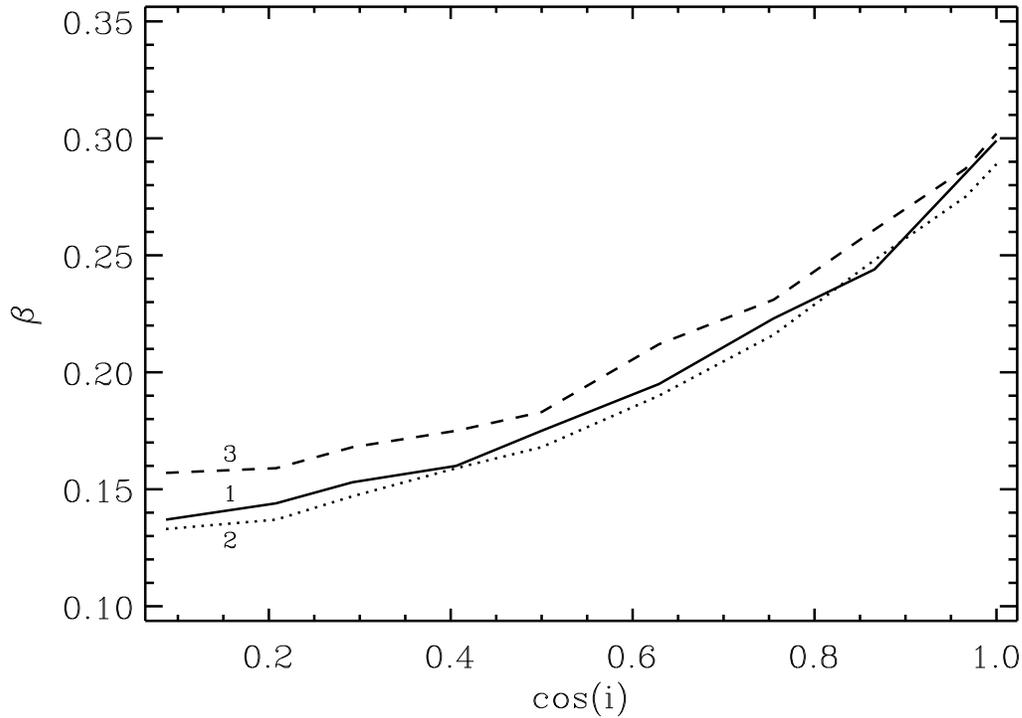,width=15 cm}
\caption{Variation of the best-fit $\beta$-parameter with inclination
angle for different neutron star spin rates. Curve 1:
$ \Omega_* = 0$ radians s$^{-1}$, curve 2: $ \Omega_* = 2044$ radians s$^{-1}$, 
curve 3: $ \Omega_* = 7001$ radians s$^{-1}$ (mass-shed limit).
The values of the other parameters are as in Fig. 6.1.}
\end{figure}

\nopagebreak
\begin{figure}[h]
\psfig{file=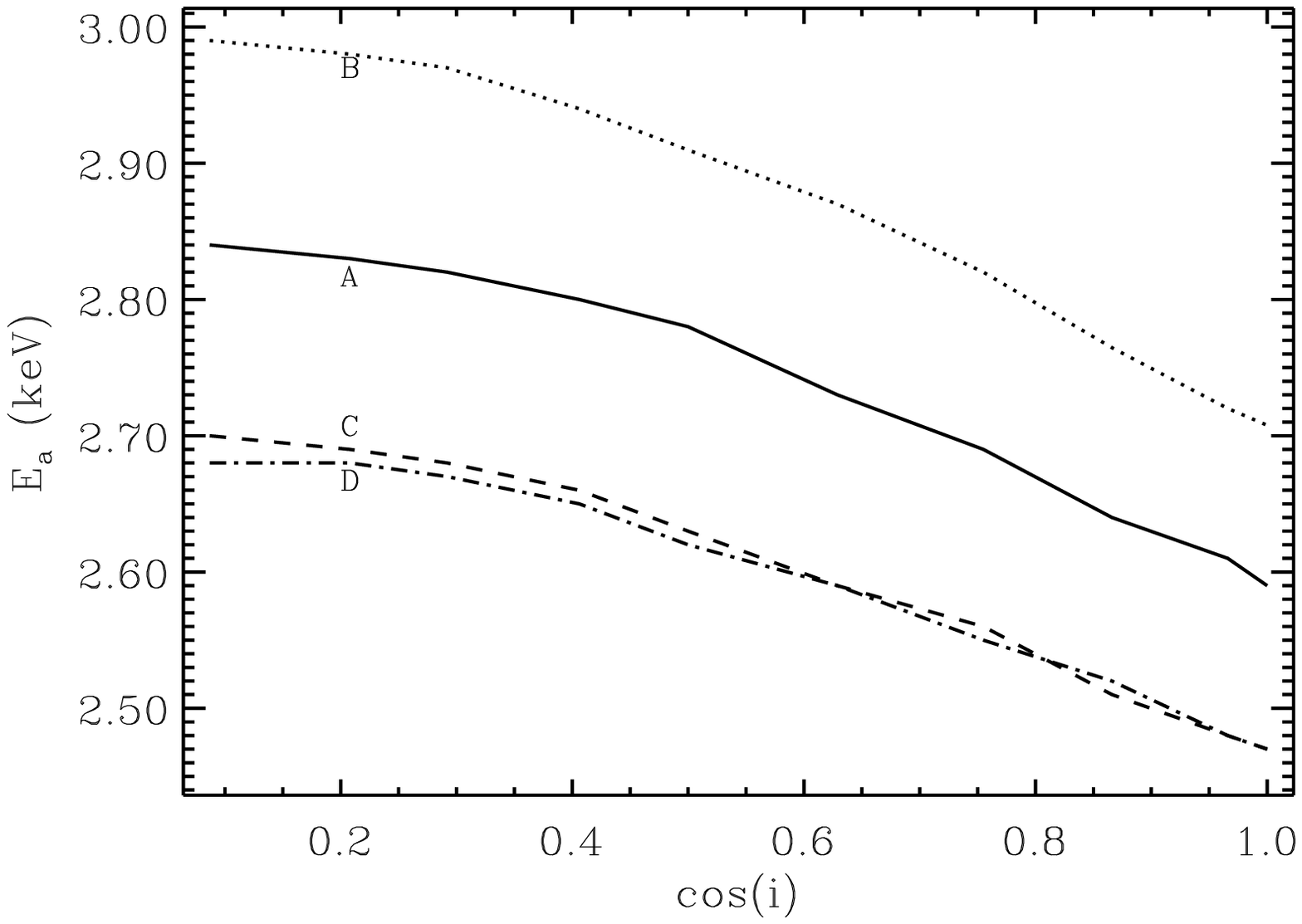,width=15 cm}
\caption{Variation of the best-fit $E_a$-parameter with inclination
angle for different equations of state. The curves correspond
to the same equation of states as listed in Fig. 6.4.
The values of the other parameters are as in Fig. 6.1.}
\end{figure}
 
To illustrate the differences between the relativistic and
Newtonian spectra, we show in Fig. 6.1, the computed relativistic
spectrum (solid line) and the Newtonian spectrum (dashed line) for
the same parameters. 
It is to be noted that, here we plot energy flux $(E S_{\rm c} [E])$, 
instead of photon flux $(S_{\rm c} [E])$ (as is the case in Fig. 5.1).
This is, because, $E S_{\rm c} [E]$ is used to choose the energy range
for fitting. However, as mentioned in the previous section, we fit 
$S_{\rm c} [E]$ by the analytical function. 
The Newtonian spectrum is the spectrum
expected from a standard non-relativistic disk (Shakura
\& Sunyaev 1973) but with the specific intensity and the effective temperature
modified by the color factor (Eqs. 3.1 and 5.2).
In order to isolate the different contributions,
we have also plotted in Fig. 6.1, the theoretical spectrum arising
from relativistic temperature profile but without the effect of
Doppler/gravitational red-shift (dotted line).  The relativistic
spectrum is under-luminous compared to the Newtonian one at high
energies -- this is primarily because of the difference in the
radial temperature profile (see Chapter 3). The difference
between the two spectra is nearly 50\% at 2 keV.  We emphasize here,
that such high difference is only true when both the spectra are
calculated for the same disk parameters. If the Newtonian spectra
is calculated for slightly different values of disk parameters (e.g.
accretion rate, inclination angle, distance to the source) the
average discrepancy between the two spectra will be less (Ebisawa,
Mitsuda and Hanawa 1991).

Fig. 6.2 shows the relativistic spectra for three different inclination
angles (solid lines) and the corresponding empirical fits using
Eq. 6.1 (dashed lines). The minimum $\chi^2$ obtained
while fitting these spectra was $ < 0.1$, which means that the average
discrepancy is less than 3\%. This is also true for other disk parameters
and EOS models considered in this work. Thus the empirical function
(Eq. 6.1) is a reasonable approximation to the computed
relativistic spectra. It also describes the Newtonian spectra to a
similar degree of accuracy.

We show in Fig. 6.3, the variation of minimum $\chi^2$
(i.e. minimized w.r.t. to parameters $E_a$ and $S_o$ only) as a
function of the $\beta$-parameter for the three spectra shown in
Fig. 6.2 and for the Newtonian one. For the Newtonian case the minimum
$\chi^2$ occurs for $\beta \approx 0.4$ while it is lower
for the relativistic cases. For example, consider the relativistic
spectrum for parameters listed in Fig. 6.1 and for $i = 30^{\circ}$ (line
marked as 3 in Fig. 6.3). If this spectrum is fitted with the empirical
function the minimum $\chi^2 = 0.05$ (corresponding to an average
discrepancy of 2\%) and the best-fit $\beta$-parameter is
$\beta \approx 0.25$. For a Newtonian $\beta$-parameter value of
$\approx 0.4$, the minimum $\chi^2$ increases to 0.1, corresponding
to an average discrepancy of more than 3\%. Thus the empirical function
can resolve the difference between the Newtonian and the relativistic
one at the 10\% level. For an observed spectrum fitted using the
empirical function, if the best-fit range of $\beta$-parameter
excludes the Newtonian value of 0.4, that would strongly indicate
that the spectrum has been modified by strong gravitational effects.
To show the robustness of this result we display in Figs. 6.4, 6.5 and 
6.6, the
variation of the best-fit $\beta$-parameter with $i$, for different EOS
models, masses and spin rates of the central object respectively. For
all these cases the best-fit $\beta$-parameter is less than 0.4.
However, for very high value of $i$ $(i ~\gsim ~85^{\circ})$, fitting to the
light--bending--spectra gives $\beta > 0.4$. Therefore, for 
$i ~\gsim ~85^{\circ}$, Newtonian and general relativistic spectra can not
be distinguished by this method. But for upto moderately high 
values of $i$ (for which light--bending--spectra still gives 
$\beta < 0.4$), this method is very effective.
Parameter $E_a$ is useful to determine the accretion rate. However,
it also depends on the metric and inclination angle. We show this
dependence in Fig. 6.7.

\section{Summary and Discussion}

In this chapter, 
a simple empirical function has been presented which describes
the numerically computed relativistic spectra well. This will
facilitate direct comparison with observations. The empirical
function (Eq. 6.1) has three parameters including normalization.
Another important advantage of this function is that it also
describes the Newtonian spectrum adequately, and the value of one
of the parameters ($\beta$-parameter) distinguishes between the two.
In particular, the best-fit $\beta$-parameter $\approx 0.4$ for the
Newtonian case, while it ranges from $0.1$ to $0.35$ for relativistic
case depending upon the inclination angle, EOS, spin rate and mass
of the neutron stars. However, as mentioned in section 6.3, this 
method is effective for upto moderately high values of $i$.

In principle, for sufficiently high quality data, the effects of
strong gravity on the disk spectrum can be detected using 
this empirical
function as a fitting routine and constraining the $\beta$-parameter.
However, it must be emphasized that there are several reasons
why this may not be possible. There could be systems which have
additional components in the X--ray spectra; for example boundary
layer emission from the neutron star surface. Uncertainties
in modeling these additional components may lead to a wider range
in the best-fit $\beta$-parameter. Thus accurate spectra of the
boundary layer (with relativistic corrections) is also needed for
modeling these systems. Moreover, X--rays could be emitted from hotter
regions (e.g an innermost hot disk or a corona) giving rise to a
Comptonized spectra instead of the sum of local emission assumed
here. In this case, the empirical fit will probably not describe
the observational data well. It has been assumed here that
the color factor is independent of radius.
Shimura and Takahara (1995) have shown from numerical computation that this
could be the case for an accretion disk in a
Schwarzschild metric. Apart from the fact that
this was done for Schwarzschild metric, their numerical
computation depends on the vertical structure of the
disk which in turn depends on the unknown viscosity
mechanism in the disk. If the color
factor has a radial dependence,
the spectral shape might change, which may be confused to be a
relativistic effect.

Despite these caveats the method described in this chapter will be
a step forward in the detection of strong gravity effects in the
spectra of X-ray binaries. Future comparison with high quality
observational data, will highlight the theoretical requirements
that have to be met, before concrete evidence for strong gravity
are detected in these systems and the enigmatic region around
compact objects is probed.

%% file: chap7.tex
\markright{Chapter 7}
\def\note #1]{{\bf #1]}}
\chapter{Disk Temperature Profiles for Strange Stars: A Comparison 
With Neutron Stars}

\section{Introduction}

We have mentioned in the earlier chapters that low mass 
X--ray binaries (LMXBs) are believed to contain
either neutron stars (NSs) or black holes accreting from
an evolved or main sequence dwarf companion that fills its
Roche--lobe.  The proximity of the companion in these systems
cause matter to spiral in, forming an accretion disk around the
central accretor.  Observations of LMXBs can provide vital
clues of the structure parameters of the accretors and, in
particular for NSs, this can lead to constraining the property
of the high density matter composing their interiors.
Therefore, the estimation of the radius of the central accretor in
SAX J1808.4-3658 and 4U 1728-34 (Li et al. 1999a; Li et al. 1999b;
Burderi \& King 1998; Psaltis \& Chakrabarty 1999) indicating
the object to be more compact than stars composed of high
density nuclear matter, acquires significance.  These results
moot alternate suggestion about the nature of the central accretors
in at least some of the LMXBs.

In this regard, the {\it strange matter hypothesis}, formulated
by Bodmer (1971) and Witten (1984) (see also Itoh 1970;
Terazawa 1979), has received much attention recently.
The hypothesis suggests strange quark matter (SQM, made up of u, d and
s quarks), in equilibrium with weak interactions, to be the actual
ground state of strongly interacting matter rather than $^{56}$Fe.
If this were true, under appropriate conditions, a phase transition
within a neutron star (e.g. Olinto 1987; Cheng \& Dai 1996;
Bombaci \& Datta 2000) could convert the entire
system instantaneously into a conglomeration of strange matter
or, as is commonly referred to in literature, strange stars (SSs).
Here we consider only {\it bare} strange stars, i.e., we neglect the 
possible presence of a crust of normal (confined) matter above the 
deconfined quark matter core (see e.g. Alcock, Farhi \& Olinto 1986). 

It is of fundamental interest - both for particle physics and
astrophysics - to know whether strange quark matter exists.  Answering
this question requires the ability to distinguish between SSs
and NSs, both observationally as well as theoretically and this
has been the motivation of several recent calculations
(Xu et al. 2001; Gondek-Rosinska et al. 2000;
Bombaci et al 2000; Zdunik 2000; Zdunik et al 2000a;
Zdunik et al 2000b; Datta et al 2000; Stergioulas et al 1999;
Gourgoulhon et al 1999; Xu et al 1999; Gondek \& Zdunik 1999;
Bulik et al. 1999; Lu 1998; Madsen 1998).  One of the most basic
difference between SSs
and NSs is the mass--radius relationship
(Alcock, Farhi \& Olinto 1986): while for NSs, this is
an inverse relationship (radius decreasing for increasing mass),
for SSs there exist a positive relationship (radius increases with
increasing mass).
In addition to this difference, due to SSs being
self--bound objects, there also exists the possibility of having
configurations with arbitrarily small masses; NSs on the other
hand, have a minimum allowed mass (e.g. Shapiro \& Teukolsky 1983;
Glendenning 1997; and more recently, Gondek et al. 1997; Gondek et
al. 1998; Goussard et al. 1998; Strobel et al 1999;
Strobel \& Weigel 2001).
Nevertheless, it must be remarked that for a value of gravitational
mass equal to $1.4 ~\msun$ (the canonical mass for compact
star candidates), the difference between the  predicted
radii of nonrotating configurations of SS and NS amounts,
   at most,
only to about 5 km; a value
that cannot be directly observed.  There arises, therefore, a
necessity to heavily rely on models of astrophysical phenomena
associated with systems containing a compact star to
estimate the radius: for isolated pulsars, models of glitches
(e.g. Datta \& Alpar 1993; Link et al. 1992) have been used in
the past for making
estimates of the structure parameters and for compact stars in
binaries, such estimates have been made by appropriately modeling
photospheric expansion in X--ray bursts (van Paradijs 1979;
Goldman 1979) and more recently by constraining the inner--edge
of accretion disks and demanding that the radius of the compact
star be located inside this inner--edge (Li et al. 1999a; Li et al. 1999b;
Burderi \& King 1998; Psaltis \& Chakrabarty 1999).  In particular,
the work by Li et al. (1999a; 1999b) suggest strange stars as possible
accretors.  However, these calculations did not include the full
effect of general relativity. Even on inclusion of these effects
(Bombaci et al 2000), the results for at least one source: 4U 1728-34,
remain unchanged.
There have also been contradictory reports on the existence of
strange stars: for example, calculations of magnetic field evolution
of SSs over dynamical timescales, make it difficult to explain
the observed magnetic field strengths of isolated pulsars
(Konar 2000).
On the other hand, Xu \& Busse (2001) show that SSs
may possess magnetic fields, having the observed strengths.
These magnetic fields, these authors argue, originate due to dynamo
effects.
In our analysis here, we ignore the effects of magnetic field.

In this chapter, we calculate constant gravitational mass
equilibrium sequences of rotating SSs, considering the full effect
of general relativity. We solve Einstein field equations
and the equation for hydrostatic equilibrium simultaneously
for different SS equations of state (EOS) models, using the same 
procedure as described in Chapter 2. We compare our
theoretical results with those obtained for NSs (Chapter 3).
In addition, we calculate the radial profiles of effective
temperature in accretion disks around SSs (same procedure as 
described in Chapter 3). These profiles are
important inputs in accretion disk spectrum calculations,
crucially depending on the radius of the
inner edge of the accretion disk. This radius is determined by the
location of $r_{\rm orb}$ with respect to that of the surface (R)
of the star,  both of which are sensitive to the EOS, through the
rotation of the central object.  In particular, we notice that
$r_{\rm orb}$ increases with stellar angular momentum ($J$) beyond
a certain critical value (a property not seen
in either rotating black holes or neutron stars). We trace this
behavior to the dependence of $dr_{\rm orb}/dJ$ on the rate of
change of the radial gradient of the Keplerian angular velocity at
$r_{\rm orb}$ with respect to $J$. The prospect of using
the temperature profiles for calculation of accretion disk
spectrum and subsequent comparison with observational data,
therefore, gives rise to the possibility of constraining SS
EOS, and eventually to distinguish between SSs and NSs.

In section 7.2, we discuss the equations of state used in this 
chapter. We display the results in section 7.3 and give a summary
of the chapter in section 7.4.

\section{Equation of State}

For strange quark matter we use two phenomenological models for the EOS.
The first one is a simple EOS (Farhi \& Jaffe 1984)
based on the MIT bag model for hadrons.
We begin with the case of massless, non-interacting ({\it i.e.} QCD
structure constant $\alpha_{\rm c} = 0$) quarks and with a bag constant
$B = 60$ MeV/fm$^3$ (hereafter EOS H).
Next, we consider a finite value for the mass of the strange quark within
the same MIT bag model EOS. We take $m_{\rm s} = 200$~MeV and
$m_{\rm u} = m_{\rm d} = 0$,
$B = 60$ MeV/fm$^3$, and $\alpha_{\rm c} = 0$ (EOS G).
To investigate the effect of the bag constant, we take (almost) the largest
possible value of $B$ for which SQM is still the ground state of strongly
interacting matter, according to the strange matter hypothesis.
For massless non-interacting quarks this gives $B = 90$ MeV/fm$^3$ (EOS F).
The second model for SQM is the EOS given by Dey et al. (1998),
which is based on a different quark model than the MIT bag model.
This EOS has asymptotic freedom built in, shows confinement at zero
baryon density, deconfinement at high density, and, for an appropriate
choice of the EOS parameters entering the model, gives absolutely
stable SQM according to the strange matter hypothesis.
In the model by Dey et al. (1998), the quark interaction is
described by a screened inter--quark vector potential originating from
gluon exchange, and by a density-dependent scalar potential which restores
chiral symmetry at high density (in the limit of massless quarks).
The density-dependent scalar potential arises from the density dependence
of the in-medium effective quark masses $M_{\rm q}$, which are taken to depend
upon the baryon number density $n_{\rm B}$ according to
$M_{\rm q} = m_{\rm q} + 310 {\rm MeV} \times \mbox{sech}
\big(\nu \displaystyle{{n_{\rm B}}\over{n_0}}\big)$,
where $n_0 $ is the normal nuclear matter density,
${\rm q} (= {\rm u},{\rm d}, {\rm s})$ is the flavor index, and
$\nu$ is a parameter.   The effective quark mass $M_{\rm q}(n_{\rm B})$
goes from its constituent masses at zero density, to its current mass
$m_{\rm q}$,  as $n_{\rm B}$ goes to infinity.
Here we consider a parameterization of the EOS by
Dey et al. (1998), which corresponds to the choice $\nu = 0.333$
for the parameter entering in the effective quark mass,
and we denote this model as EOS E.

For NSs, we use three representative equations of state which span
a wide range of {\it stiffness}. These are EOS models A, B and D, as
mentioned in section 2.4.

A list of the designation along with the salient features of the EOS
models used here is provided in Table 7.1.

\begin{table*}[ht]
\begin{center}

\begin{tabular}{ccc} \hline \hline
  EOS label     & compact star &   EOS model   \\
\hline

  E  &  SS    &  Dey et al. (1998), ~their model SS1  \\
  F  &  SS    &  Farhi and Jaffe (1984),
                 $B = 90~{\rm MeV/fm}^3$, $m_s = 0$   \\
  G  &  SS    &  Farhi and Jaffe (1984),
                 $B = 60~{\rm MeV/fm}^3$, $m_s = 200~{\rm MeV}$ \\
  H  &  SS    &  Farhi and Jaffe (1984),
                 $B = 60~{\rm MeV/fm}^3$, $m_s = 0$ \\
\hline
  A  &  NS    &  Pandharipande (1971), ~hyperonic matter  \\
  B  &  NS    &  Baldo et al.  (1997), ~nuclear  matter  \\
  D  &  NS    &  Sahu et al.   (1993), ~nuclear  matter  \\
 \hline \hline
\end{tabular}
\end{center}
\caption{The list of EOS models used in this chapter.}
\end{table*}

\nopagebreak
\begin{figure}[h]
\psfig{file=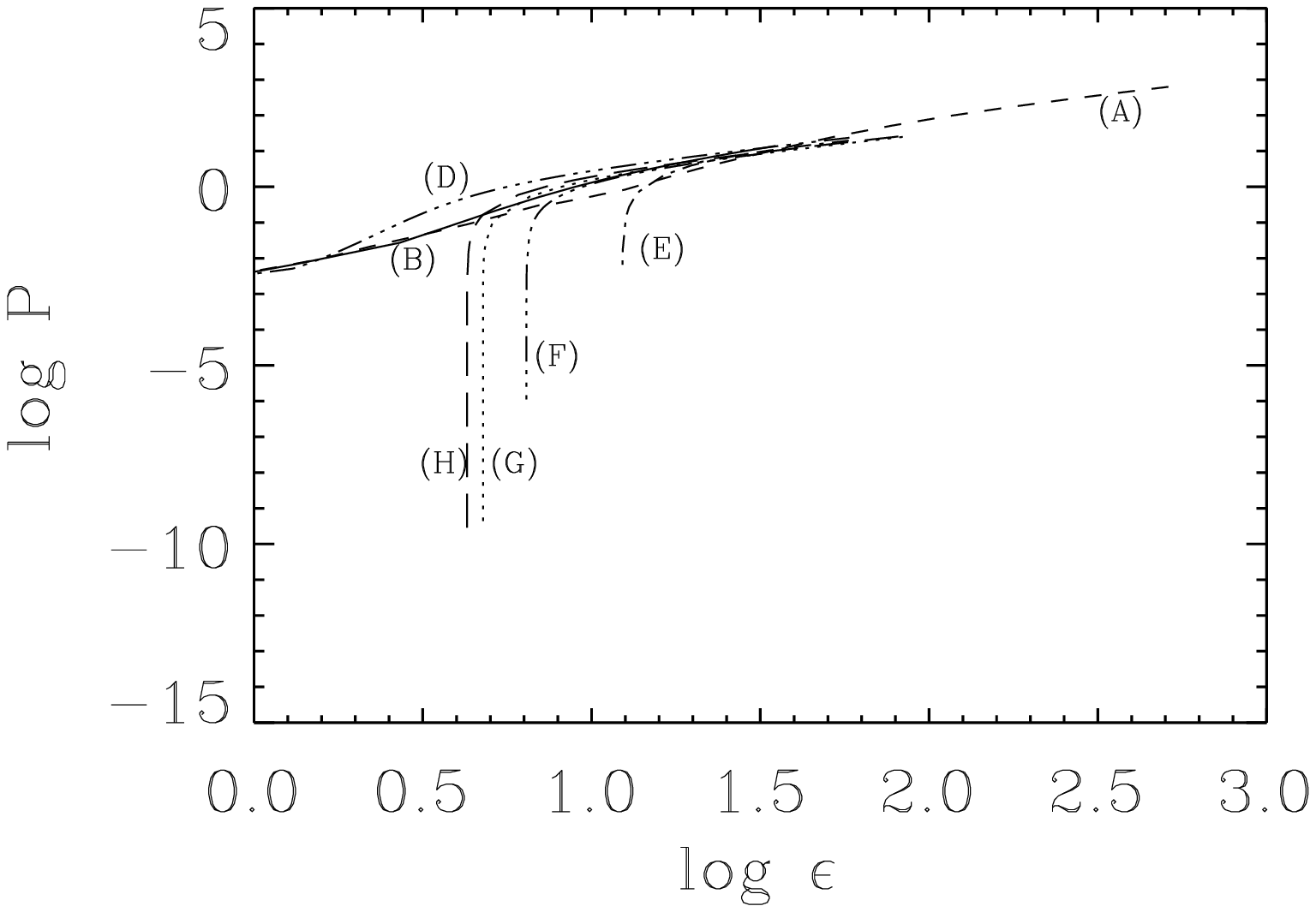,width=15 cm}
\caption{Logarithmic plot of pressure vs. matter density for the
EOS models used here.  The density and pressure are in units of
$1.0\times10^{14}$~g cm$^{-3}$ and $(1.0\times10^{14})$~c$^2$~cgs respectively.}
\end{figure}

\nopagebreak
\begin{figure}[h]
\psfig{file=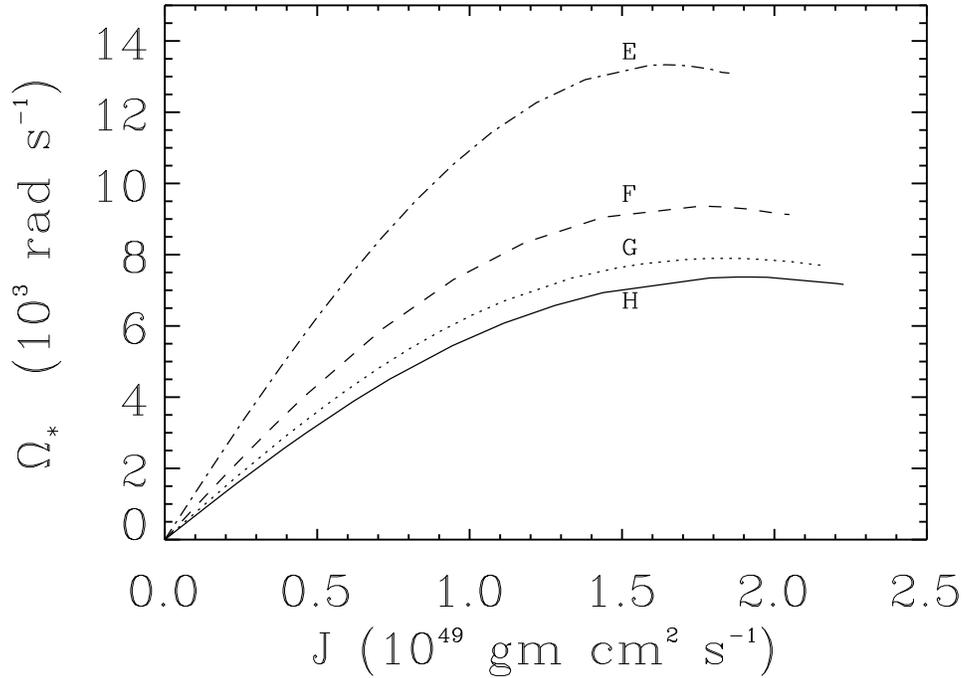,width=15 cm}
\caption{Angular speed ($\Omega_{\rm *}$) as a function of total angular
momentum ($J$) for strange star. The curves are labelled by the
nomenclature of Table 7.1 and are for a fixed gravitational mass
($M=1.4~\msun$) of the strange star.}
\end{figure}

\nopagebreak
\begin{figure}[h]
\psfig{file=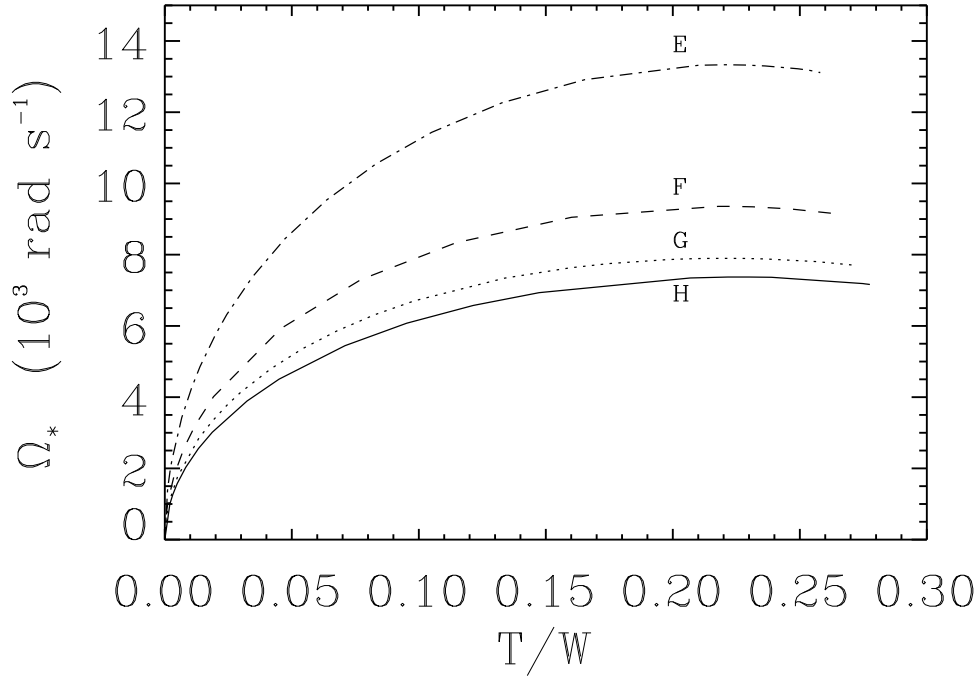,width=15 cm}
\caption{Angular speed ($\Omega_{\rm *}$) as a function of the ratio of
rotational kinetic energy and gravitational binding energy ($T/W$) for strange
star. Curve labels have the same meaning as in Fig. 7.2.}
\end{figure}

\nopagebreak
\begin{figure}[h]
\psfig{file=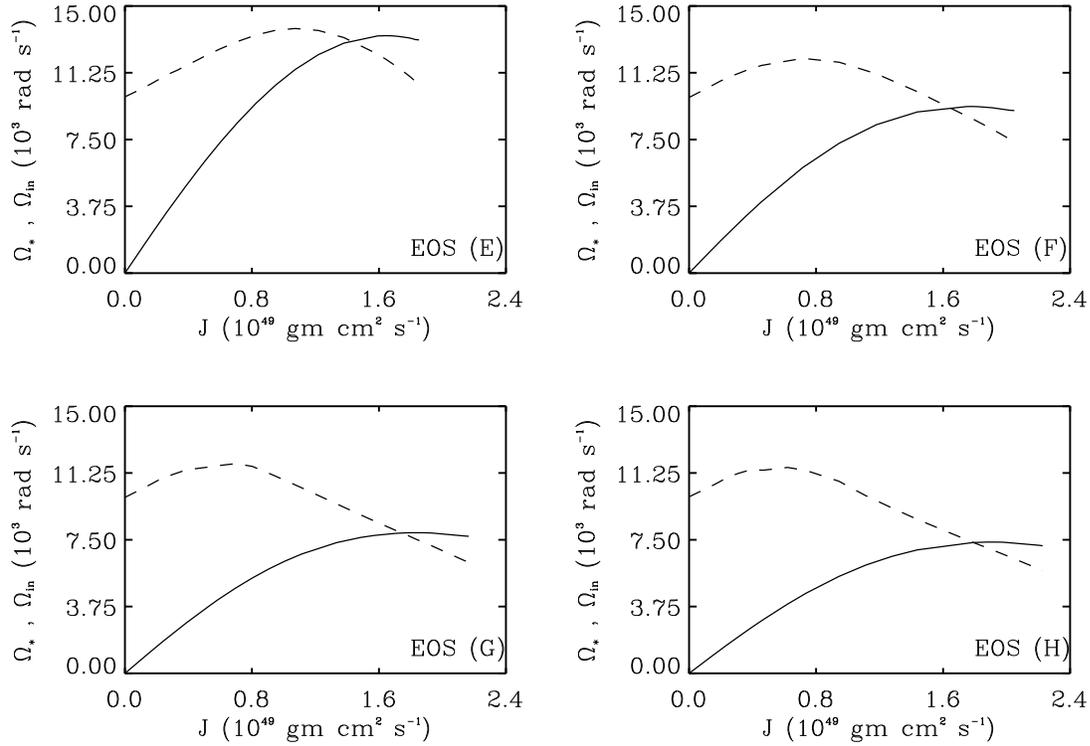,width=15 cm}
\caption{Angular speed ($\Omega_{\rm *}$) of the strange star (solid curve)
and the Keplerian angular speed ($\Omega_{\rm in}$) of a test particle at
the inner edge of the disk (dashed curve) as functions of total angular
momentum ($J$) of strange star. The curves are for a fixed gravitational
mass ($M=1.4~\msun$) of the strange star. Different panels are for different
SS EOS models.}
\end{figure}

\nopagebreak
\begin{figure}[h]
\psfig{file=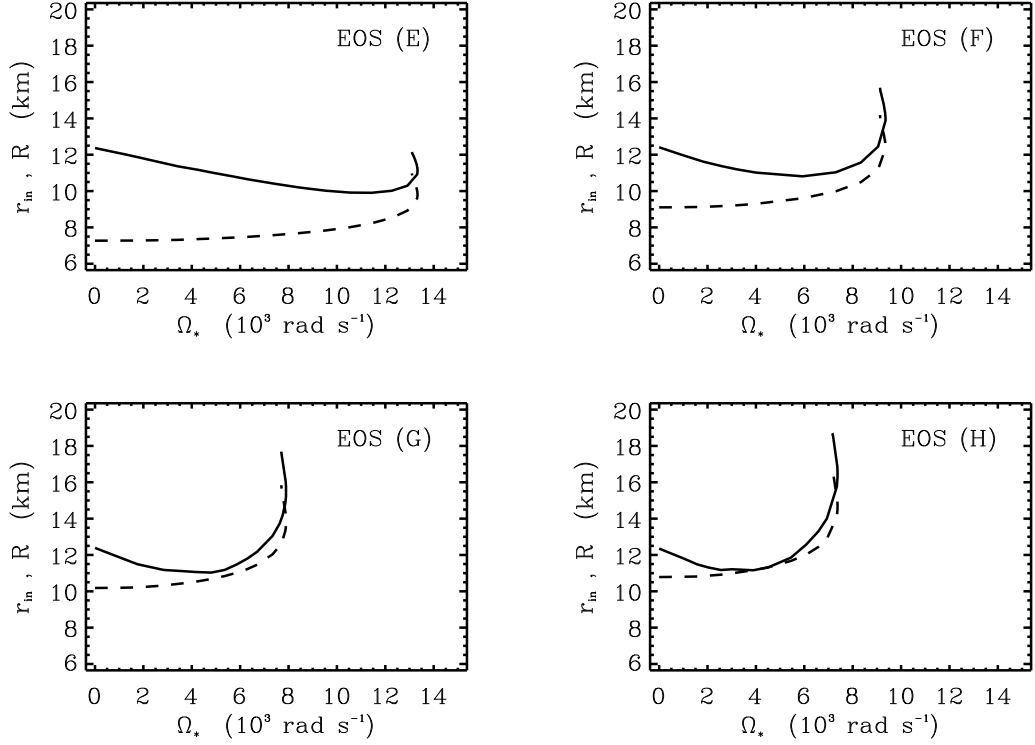,width=15 cm}
\caption{Disk inner edge radius ($r_{\rm in}$, solid curve) and strange star
radius ($R$, dashed curve), as functions of angular speed
$\Omega_{\rm *}$ for various EOS models. The curves are for a fixed
gravitational mass ($M=1.4~\msun$) of the strange star.}
\end{figure}

\nopagebreak
\begin{figure}[h]
\psfig{file=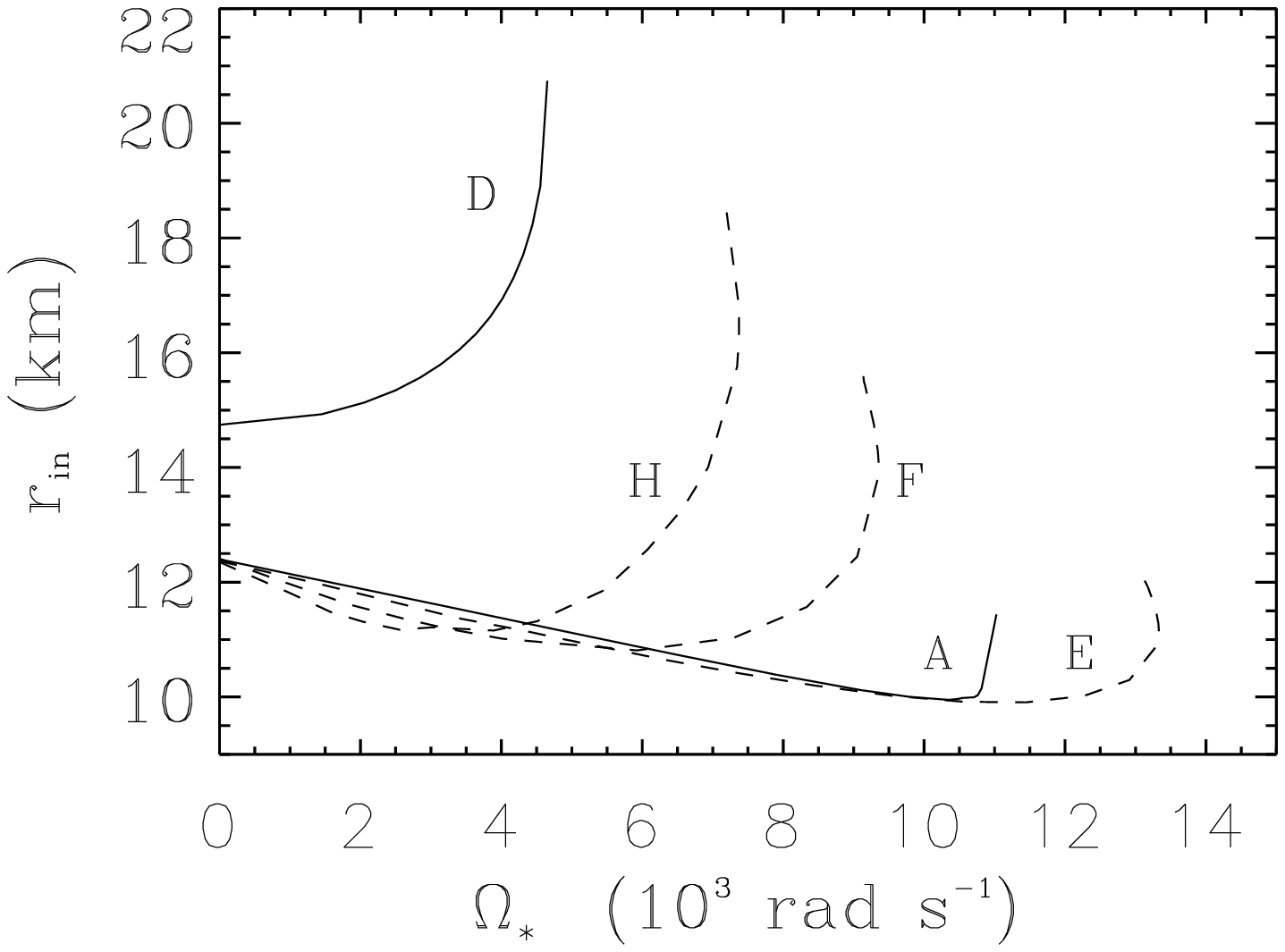,width=15 cm}
\caption{Disk inner edge radius ($r_{\rm in}$) as a function of
angular speed $\Omega_{\rm *}$ of the compact star. The curves
have their usual meaning.}
\end{figure}

We also display the qualitative variations in these EOS models in
a log--log plot of Fig. 7.1.  The differences between SS and NS EOS
are plainly evident, especially at lower pressures.

\section{The Results}

We have calculated the structure parameters and the disk temperature
profiles for rapidly rotating, constant gravitational mass sequences
of SSs in general relativity.  The results for SS are compared with
those for NS.  For illustrative purposes here, we have chosen the value
of gravitational mass to be $1.4~\msun$.

Fig. 7.2 depicts the variation of $\Omega_{\rm *}$ with the total
angular momentum ($J$) for constant gravitational mass and for the four SS
EOS. The curves extend from the static limit to the mass-shed limit.
The striking feature here is that, although $J$ increases monotonically
from slow rotation to mass-shed limit, $\Omega_{\rm *}$ shows a
non-monotonic behavior: maximum value of $\Omega_{\rm *}$ (i.e.
$\Omega_{\rm *}^{\rm max}$) occurs at a value of $J$ lower than that
for mass--shed limit. Although this seems to be a generic feature for SS EOS,
$\Omega_{\rm *}$ is always a monotonic function of $J$ for constant
gravitational mass NS sequences and hence constitutes an essential
difference between SS and NS (see section 7.4 for discussions).  Our
calculations show that at maximum $\Omega_{\rm *}$, the ratio of
rotational kinetic energy to total gravitational energy: $T/W$ approaches
the value of $0.2$ (see next paragraph).  It has been pointed out by
Gourgoulhon et al., 1999 that such high values of
$T/W$ make the configurations unstable to triaxial instability.
It can also be noticed that for stiffer EOS, the star possesses
a higher value of $J$ at mass-shed limit ($\Omega_{\rm *}^{\rm max}$
also occurs proportionately at higher $J$).

In order to expand upon the results of Fig. 7.2, we plot
$\Omega_{\rm *}$ vs. $T/W$ for various SS EOS in Fig. 7.3.
It is seen that for all SS EOSs, $T/W$ becomes greater than 0.25 at
mass-shed limit, while for NS EOSs it is usually between 0.1 and 0.14
(Cook et al. 1994). Interestingly, for all SS EOSs,
$\Omega_{\rm *}^{\rm max}$ occurs at about the same value of $T/W$
($\approx 0.2$).

Fig. 7.4 displays $\Omega_{\rm *}$ and $\Omega_{\rm in}$ (i.e. the
Keplerian angular speed of a test particle at $r_{\rm in}$) 
against $J$.
The four panels are for the four SS EOS we use. We notice the
interesting behavior that $\Omega_{\rm *}$ and $\Omega_{\rm in}$
curves cross each other at a point near $\Omega_{\rm *}^{\rm max}$.
For rotating NS configurations, since the equality $r_{\rm in} = R$
is almost always (except for very soft EOS models: Fig. 3.1)
achieved for rotation rates well below that at
mass-shed limit (for $M = 1.4~\msun$), always
$\Omega_{\rm *} \leq \Omega_{\rm in}$ (the equality is achieved only
at mass-shed limit).  On the other hand, for SSs, $r_{\rm orb}$
is almost always greater than $R$ (as explained in the next paragraph)
and when the star approaches Keplerian angular speed at the equator,
$\Omega_{\rm *}$ becomes greater than $\Omega_{\rm in}$.

Fig. 7.5 is a plot of the variation of $r_{\rm in}$ and $R$ with
$\Omega_{\rm *}$ for four SS EOSs. We see that the behavior of $R$
is monotonic from slow rotation to the mass-shed limit, even though
that of $\Omega_{\rm *}$ is not.  As mentioned earlier for all
$\Omega_{\rm *}$ from static limit upto mass-shed limit,
$r_{\rm in} > R$ for 3 SS EOSs. Only for the stiffest SS EOS,
that we have chosen, does the disk touch the star (for an
intermediate value of $\Omega_{\rm *}$).  This is distinct
from the case of NS (see Fig. 3.1).
The reason for such a behavior is the non--monotonic
variation of $r_{\rm orb}$ with $J$ for SSs (contrary to the case
of NSs and black holes); this is discussed further in the next section.

In Fig. 7.6, we plot the variation of $r_{\rm in}$ with $\Omega_{\rm *}$
for three SS EOSs and two NS EOSs: for each case, our softest EOS and our
stiffest EOS have been chosen. In addition, we display the corresponding
results for EOS model F too. It is clear that in the
$r_{\rm in}$--$\Omega_{\rm *}$ space, there exists a region that is
spanned by both NS and SS configurations. Interestingly,
however, there also exists certain regions occupied exclusively by
either SS or NS configurations. The possible observational
consequences of this result is discussed in the next section.

Fig. 7.7 displays the radial profiles of temperature: (i) assuming a
purely Newtonian accretion disk and (ii) considering general
relativistic accretion disks for (a) SS (EOS H) and (b) NS (EOS B),
each represented by two configurations: the non--rotating and mass
shed for $M=1.4~\msun$. We also display the temperature profile
(curve 5) for a SS configuration of $M=1~\msun$, described by EOS (A)
(the constraints obtained by Li et al. 1999a; 1999b) and having
a period $P=2.75$~ms (the mass and period corresponding to that
inferred for the source 4U 1728-34: M\'{e}ndez \& van der Klis 1999).
It must be
remembered that in this figure (and the next), curve 5 represents
the temperature profile for
a different $M$ value than the rest of the curves and is displayed
in the same figure, only for illustrative purposes.  From this
figure we see that for $M=1.4~\msun$,
the Newtonian value of temperature is about 25\% higher than the
general relativistic value near the inner edge of the disk. This
shows the importance of general relativity and rotation near the
surface of the star.
The difference between the effects of SS EOS and NS EOS on temperature
profiles (at the inner portion of the disk) is also prominent at
mass-shed limit (due to the difference in rotation rates for these
two configurations).  Such differences in temperature profiles are also
expected to show up in the calculations of spectra at higher energies.

In the panel (a) of Fig. 7.8, we display the temperature profiles
for configurations (as in Fig. 7.7) composed of SS EOS (H) 
(curves 1--4),
represented by  different $\Omega_{\rm *}$ (corresponding to
$\Omega_{\rm *} = 0$, for minimum $r_{\rm in}$,
$\Omega_{\rm *} = \Omega_{\rm *}^{\rm max}$ and mass-shed limit);
curve (5) is the same as in Fig. 7.7.  The
behavior of temperature profiles is non-monotonic with $\Omega_{\rm *}$.
The panel (b) shows the temperature profiles at mass-shed for various
SS EOS along with curve (5).  Here the temperature profiles show monotonic
behavior with the stiffness of EOS.  The behavior of the temperature
profiles in
both the panels are similar to those calculated for NSs (Chapter 3).
Notice the substantial difference in the maximum
temperature; sufficiently sensitive observations are, therefore,
expected to complement the findings of Li et al. (1999a; 1999b).

The variations of $E_{\rm D}$, $E_{\rm BL}$, the ratio
$E_{\rm BL}/E_{\rm D}$ and $T_{\rm eff}^{\rm max}$ with $\Omega_{\rm *}$
are displayed in Fig. 7.9.  Each plot contains curves corresponding to
all the SS EOS models considered here. The behavior of
all the curves are similar to those for any NS EOS (see Fig. 3.5).
The only difference being that due to the
non-monotonic behavior of $\Omega_{\rm *}$ from slow rotation to
mass-shed limit for SS EOSs, making the curves turn inward at the
terminal (mass-shed) rotation rate.

In Fig. 7.10, we make a comparison between SSs and NSs for the same
quantities displayed in Fig. 7.9. 
We have used three SS EOSs and two NS EOSs models (the softest 
and the stiffest for each case and the EOS model F).
In all the panels, SS and NS both are seen to have
their own exclusive regions in the high and low $\Omega_{\rm *}$
parameter space respectively.  This is especially prominent for
$E_{\rm BL}$ and $E_{\rm BL}/E_{\rm D}$.
We also notice that for SS, at $\Omega_{\rm *}=\Omega_{\rm *}^{\rm max}$,
the values of $E_{\rm BL} \approx 0.05$ and
$E_{\rm BL}/E_{\rm D} \approx 1.0$ for all EOS.
On the contrary, for neutron stars,
both $E_{\rm BL}$ and $E_{\rm BL}/E_{\rm D}$ become $\approx 0$ at
$\Omega_{\rm *}^{\rm max}$ ($= \Omega_{\rm ms}$).

\section{Summary and Discussion}

\nopagebreak
\begin{figure}[h]
\psfig{file=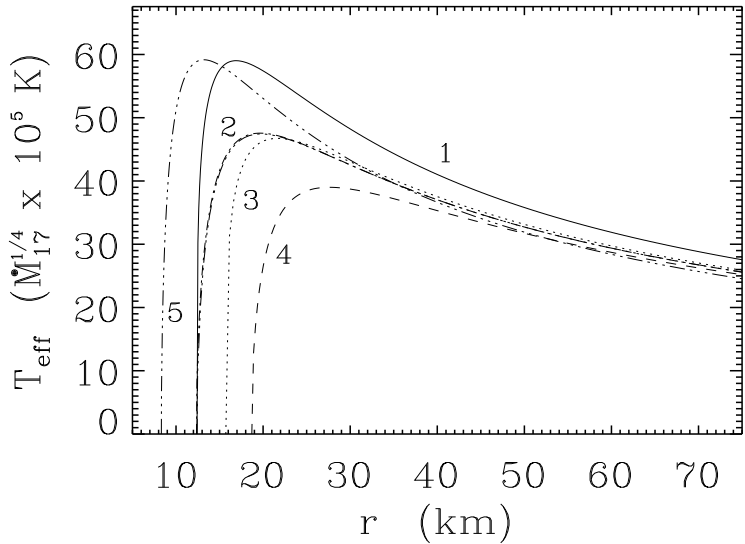,width=12 cm}
\caption{Accretion disk temperature profiles: Curve (1)
corresponds to the Newtonian case, curve (2) to the Schwarzschild case
(coincident curves for NS EOS model B and SS EOS model H), curve (3) to a
neutron star (EOS model B) rotating at the centrifugal mass-shed limit and
curve (4) to a strange star (EOS model H) rotating at the centrifugal
mass-shed limit. For curve (1) it is assumed that, $r_{\rm in}=6 G M/c^2$.
The curves (1--4) are for a fixed gravitational mass ($M=1.4~\msun$) of
the compact star. Curve (5) corresponds to a configuration that has
$M=1~\msun$ and $\Omega_{\rm *}$ corresponding to a period $P=2.75$~ms
(inferred for 4U 1728-34; see text) and described by EOS model E.
In this and all subsequent figures, the temperature is expressed in units
of $\dot{M}_{17}^{1/4} \times 10^5$~K, where $\dot{M}_{17}$ is the steady state mass
accretion rate in units of $10^{17}$~g~s$^{-1}$.}
\end{figure}

\nopagebreak
\begin{figure}[h]
\psfig{file=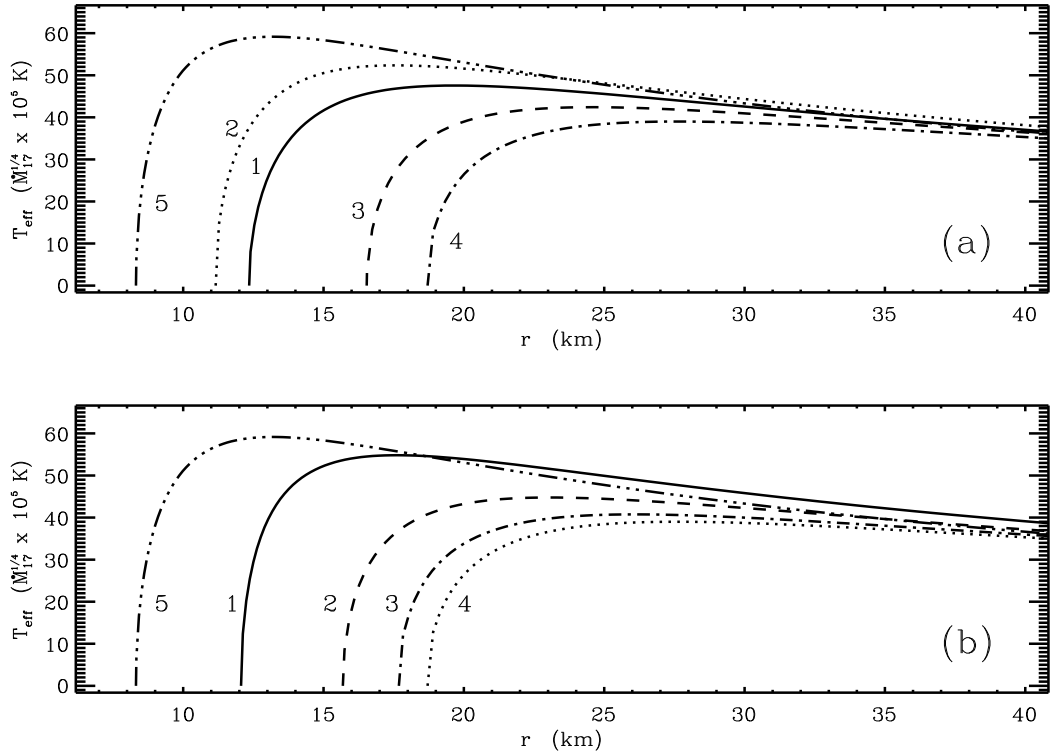,width=15 cm}
\caption{Temperature profiles incorporating the effects of rotation of the
strange star. The plots correspond to (a) EOS model H and an assumed strange
star mass of $M=1.4~\msun$ (curves 1--4) for rotation rates:
$\Omega_{\rm *}=0$ (curve 1),
$\Omega_{\rm *}=3.891\times10^3$~rad~s$^{-1}$ (curve 2),
$\Omega_{\rm *}=7.373\times10^3$~rad~s$^{-1}$ (curve 3),
$\Omega_{\rm *}=7.163\times10^3$~rad~s$^{-1}= \Omega_{\rm ms}$ (curve 4),
(b) the same assumed mass and $\Omega_{\rm *}=\Omega_{\rm ms}$ for
the four EOS models (E):curve 1, (F):curve 2, (G):curve 3 and (H):curve 4.
In both panels, curve (5) is the same as that in Fig. 7.7.}
\end{figure}

\nopagebreak
\begin{figure}[h]
\psfig{file=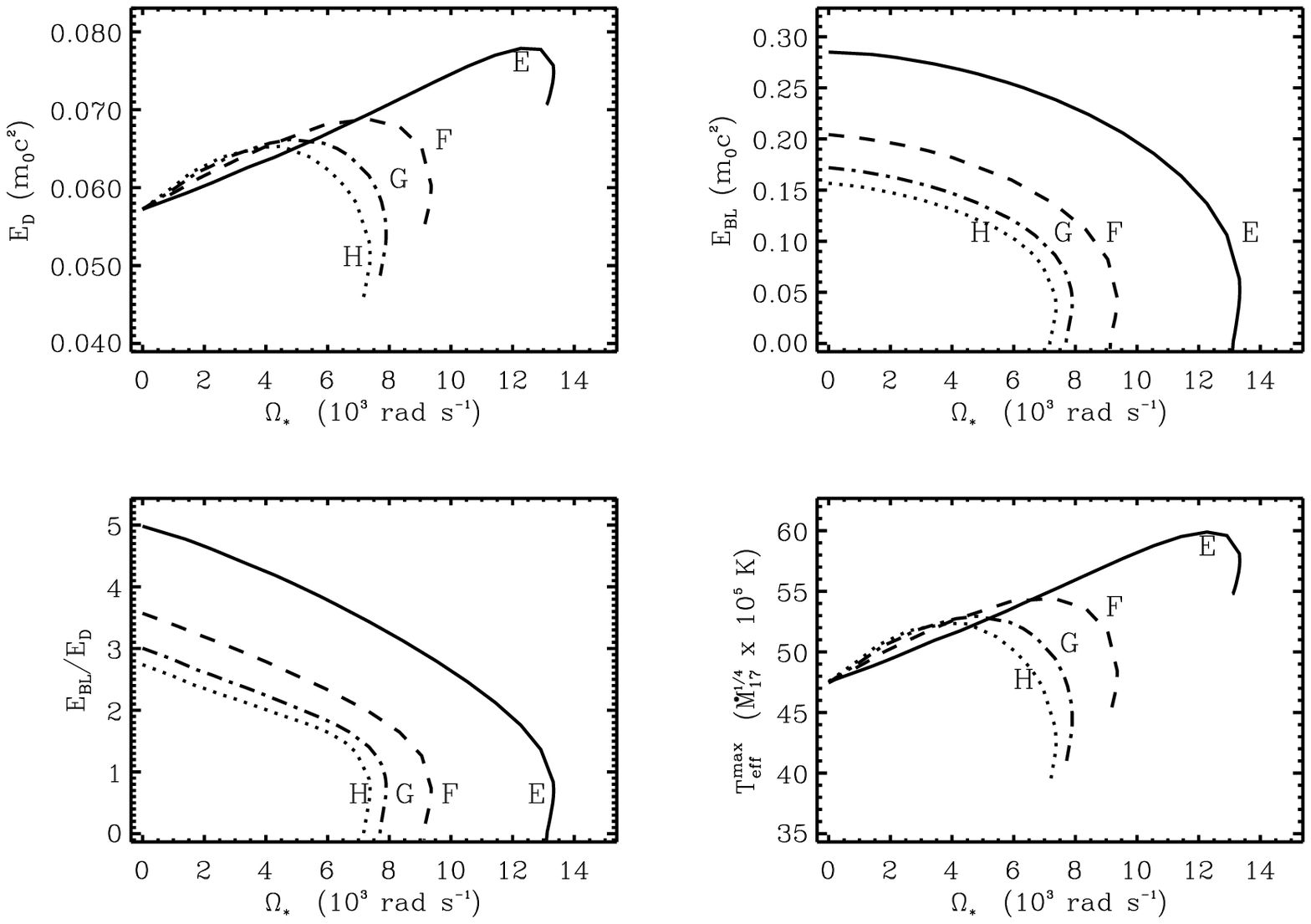,width=15 cm}
\caption{The variations of the $E_{\rm D}$, $E_{\rm BL}$,
$E_{\rm BL}/E_{\rm D}$ and $T_{\rm eff}^{\rm max}$ with
$\Omega_{\rm *}$ for a chosen strange star mass value of $1.4~\msun$ for
the four SS EOS models. The curves have the same significance as Fig. 7.3.}
\end{figure}

\nopagebreak
\begin{figure}[h]
\psfig{file=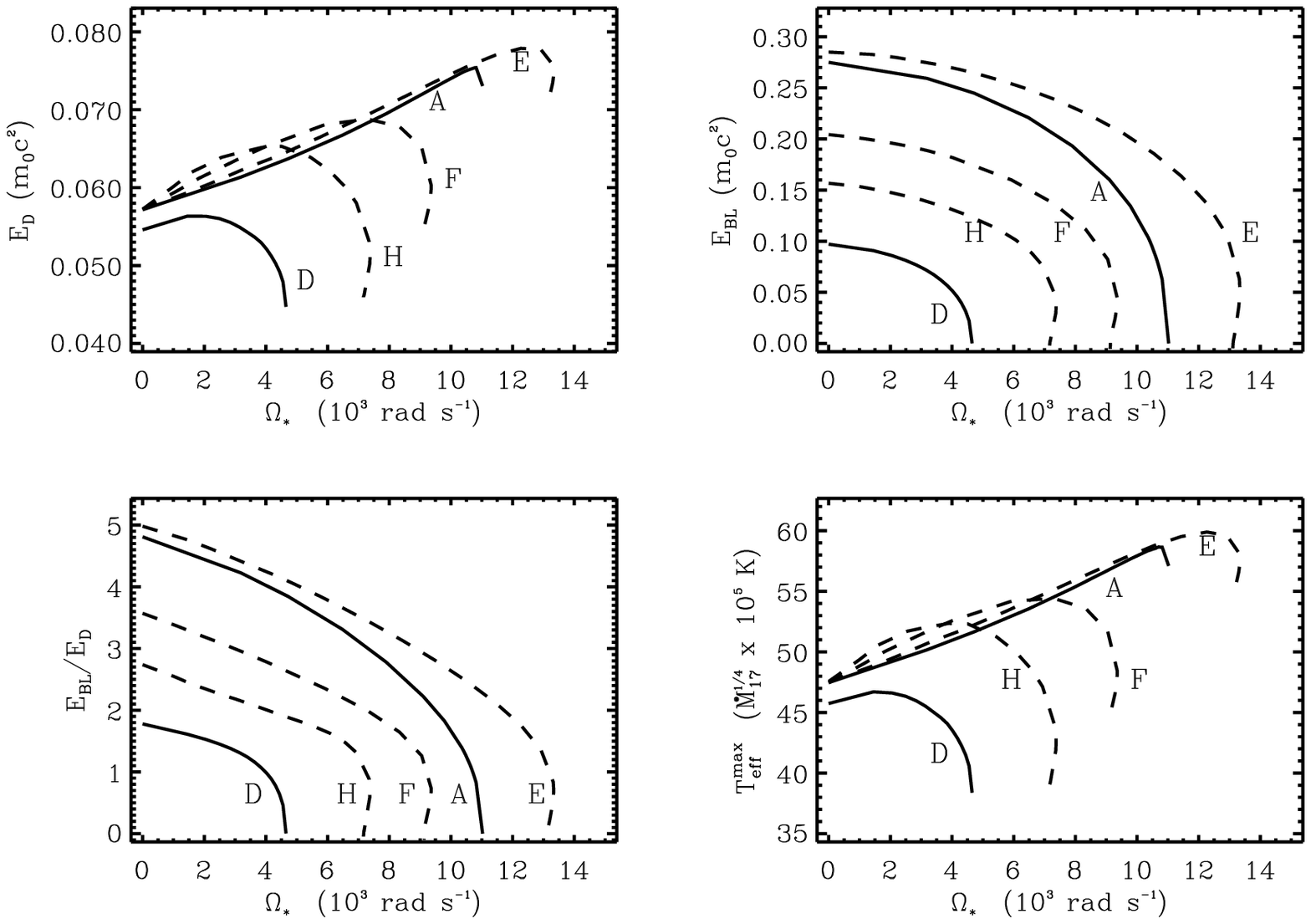,width=15 cm}
\caption{Same as Fig. 7.9, except the fact that here two NS EOS models and three
SS EOS models are used. The curves have the same significance as Fig. 7.6.}
\end{figure}

In this chapter we have calculated the structure parameters and the disk
temperature profiles for rapidly rotating SSs (for constant gravitational
mass sequence with $M = 1.4~\msun$) and compared them with those for
NSs with the aim of finding possible ways to distinguish between
the two.
For the sake of completeness, we have compared the properties of these two
types of stars all the way from slow rotation to mass-shed limit.

The striking feature of SSs is the non--monotonic behavior of
$\Omega_{\rm *}$ with $J$ such that $\Omega_{\rm *}^{\rm max}$
occurs at lower value of $J$ than that of the mass--shed limit.
Hence the other SS structure parameters become non-monotonic
functions of $\Omega_{\rm *}$.  This behavior is observed even
for the constant rest mass sequences of SSs (e.g. Gourgoulhon et al
1999; Bombaci et al. 2000).  In contrast, for NSs, the structure
parameters are all monotonic functions of $\Omega_{\rm *}$.
An implication of the non--monotonic behavior of $\Omega_{\rm *}$ with
$J$ is that if an isolated sub--millisecond pulsar
is observed to be spinning up, it is likely to be an SS rather
than an NS.

Because of higher values of $T/W$ (\greq 0.2), SSs are more
prone to secular instabilities compared to NSs at rapid rotation
(Gourgoulhon et al. 1999).  Our calculations show that at
$\Omega_{\rm *}^{\rm max}$, $T/W > 0.2$.

Another important feature of SS gravitational mass sequence
(in contrast to the corresponding NS sequences) is the crossing
point in $\Omega_{\rm *}$ and $\Omega_{\rm in}$. This feature
has important implication in models of kHz QPOs: for example
if $\Omega_{\rm *}$ is greater than $\Omega_{\rm in}$, the
beat--frequency models ascribing higher frequency to
Keplerian frequencies will not be viable.

It can be noted from Fig. 7.4, that with the increase in stiffness
of the EOS models, $J_{\rm cross}$ increases and
$\Omega_{\rm *,cross}$ (the subscript ``cross'' corresponds to
the point $\Omega_{\rm in} = \Omega_{\rm *}$) decreases monotonically.
It is also seen that in general all the quantities vary monotonically
with the stiffness for both SS and NS EOSs (see Chapter 3).

For SSs, the inner--edge of the accretion disk rarely touches the
surface of the star (even for maximum rotation rates), while for
rapidly rotating NSs, the accretion disk extends upto the stellar
surface for almost all rotation rates. Since the inner accretion
disk boundary condition is different for both these cases, we expect
important observable differences (both temporal and spectral) in
X--ray emission (from the boundary layer and the inner accretion
disk) from SSs and NSs.

A brief note on the variation of $r_{\rm orb}$ with specific angular
momentum is in order here. As mentioned earlier, beyond a certain
value of the angular momentum, the radius of the ISCO increases with
increasing angular momentum -- a property not seen either in the case
of NSs or black holes.
The reason for this can be traced to the radial gradient of the
angular velocity of the particles at the marginally stable orbit
and the analysis is described as follows:

For the metric described by Eq. (2.3), the second derivative of
the effective potential is given by Eq. (2.47).
Simplification of this, using the other equations of motion 
(Bardeen 1970), yield
\begin{eqnarray}
r^2 (1-\tilde{v}^2) \tilde{V}_{, r r} & = & - X \left[\frac{r \Omega_{,r}}{\Omega-\omega} + \right.
\left. \frac{1-\tilde{v}^2}{2 \tilde{v}^2} X\right]
\end{eqnarray}

\noindent where $\Omega$ is the angular speed of the particle and 
$X=\tilde{v}^2 (2 + r \gamma_{,r}  - r \rho_{,r}) + r(\gamma_{,r} + \rho_{,r})$.
The marginal stability criterion, therefore, yields the rate of change of
the marginal stable orbit, with respect to 
$j$ $(= J/M^2)$ as:
\begin{eqnarray}
r_{{\rm orb}, j} & = & r_{\rm orb}
\left\{\frac{(\Omega-\omega)_{,j}}{\Omega-\omega} \right.
\left. -\frac{\Omega_{,rj}}{\Omega_{,r}}  \right.
\left.+ 2 \frac{\tilde{v}_{,j}}{\tilde{v}(\tilde{v}^2-1)} \right.
\left.+\frac{X_{,j}}{X}\right\} \label{eq:rmsj}
\end{eqnarray}
where the terms in the bracket are to be evaluated
at $r_{\rm orb}$.  We calculate the four terms in the
bracket in Eq. (7.2) and find that
the second term dominates the net rate of change of
$r_{\rm orb}$ with $j$. This implies that at the value of
$j$ where $r_{{\rm orb},j}$ changes sign, although
the first three terms are observed to change sign, the net sign
is only dependent on that of $\Omega_{,rj}$ at ISCO.

From Fig. 7.6 we see that for $\Omega_{\rm *}$ in the range
(0, 4028) rad s$^{-1}$ (the second quantity in the range
is the rotation rate of PSR 1937+21: Backer et al. 1982,
the fastest rotating pulsar observed so far), a major
portion of the $r_{\rm in}$-$\Omega_{\rm *}$ space is
occupied exclusively by NS. So if $r_{\rm in}$ can be
determined independently from observations (for example,
by fitting the soft component of the observed spectrum
by the XSPEC model ``diskbb'' available in XANADU: see
for example Kubota et al 1998, or, from the
observed kHz QPO frequencies), there is a fair chance of
inferring the central accretor to be an NS rather than an SS
(provided the mass of the central accretor is known by other
means). This is also applicable to $E_{\rm BL}$ and
$E_{\rm BL}/E_{\rm D}$ (Fig. 7.10).  It is also to be noted that
Li et al. (1999 a; 1999b) did a similar search in the $M-R$
parameter space and concluded the millisecond X--ray
pulsar SAX J1808.4-3658 and the central accretor in
4U 1728-34 to be likely SSs.
If, indeed this is true, then it is possible to constrain the
{\it stiffness} of the equation of state of SQM (Bombaci 2000),
and to exclude EOS models (like EOS G and EOS H) stiffer 
than our EOS model F.

Calculation of the accretion disk spectrum involves
the temperature profiles as inputs. The spectra of
accretion disks, incorporating the full effects of
general relativity for NSs (Chapters 5, 6)
show sensitive dependence on the EOS of high density matter.
However, the similarity in the values of the maximum disk temperature
implies an indistinguishability between the spectra of SSs
and those of NSs in general. Nevertheless, just as $E_{\rm BL}$ and
other quantities show that NS exclusively occupy certain
regions in the relevant parameter space, we expect that
it will be possible to make a differentiation between these
two compact objects by modeling the boundary layer emission.
If as mentioned in previous paragraph, we exclude EOS models
stiffer than F, then from Fig. 7.10, we see that a fairly
accurate measurement of $E_{\rm BL}$ ($L_{\rm BL}/\dot{M}c^2$)
and $\Omega_{\ast}$
can indicate whether the central accretor is an NS or an SS if the
corresponding point falls outside the strip defined by curves
F and A.

The current uncertainties in theoretical models of boundary
layer emission and the variety of cases presented by models
of rotating compact objects, calls into order a detailed
investigation into these aspects of LMXBs -- especially with
the launch of new generation X-ray satellites (having better
sensitivities and larger collecting areas) on the anvil.

%% file: chap8.tex
\markright{Chapter 8}
\def\note #1]{{\bf #1]}}
\chapter{Summary and Conclusion}

\section{Introduction}

The X--ray binary systems consist of two stars, rotating around each other.
One of them (primary) 
is a compact star (neutron star, strange star or black hole) and
the other one (secondary companion) is a main-sequence star or an evolved
star (red sub-giant, blue super-giant or white dwarf). When the companion star
fills its Roche-lobe, matter from its surface flows towards the compact star.
Due to initial angular momentum, this matter can not fall radially; rather it
follows a spiral path and forms a disk. Such a disk is called an accretion 
disk. Due to viscous dissipation, energy is radiated from the disk. As the 
temperature of the inner portion of the disk is very high $(\sim 10^6$ K), 
X--rays are generated in this region. If the compact star has a hard surface 
(i.e., if it is not a black hole), the inflowing matter hits this surface and 
another component of X--ray is produced in a thin layer, called the boundary 
layer. However, it is to be remembered that the disk accretion is not the only
mechanism for accretion process. Such a process can also happen from the 
wind of the companion star.
 
There are two classes 
of X--ray binary systems: HMXBs and LMXBs. The secondary companion in 
an HMXB is a high mass star (generally, O or B type). As mentioned in Chapter 1, the age
of such a system is rather low $(\sim 10^7$ yrs). Most of the energy coming 
out of such systems is in the visible range. On the other hand, an LMXB consists 
of a low mass $(\lsim 1 \msun)$ companion star, with the age typically 
$\sim 10^9$ yrs and most of its radiated energy is in X--rays.

For our work, we have chosen LMXB systems with neutron stars or strange stars 
as the central accretors. These systems offer several advantages over the HMXBs 
in understanding the properties of the compact stars. For example, most of the 
energy (in X--rays) from such systems come from the inner regions of 
accretion disks. The motion of matter in these regions is expected to be 
influenced by the mass--radius relation and the total angular momentum of the 
compact star. Therefore, the analysis of X--ray spectra from these systems
may shed light on the properties of the compact stars. Moreover, the accretion disks
of such systems may extend very close to the stellar surface, as the 
magnetic fields of the primary stars in LMXBs are expected to be decayed to 
lower values $(\sim 10^8$ G; see Bhattacharya \& Datta 1996 and 
Bhattacharya \& van den Heuvel 1991). This ensures that the observed spectra 
can actually reflect some properties of the compact stars. Besides, accretion 
via wind is negligible in LMXBs, which may make the spectral calculation
for such systems simpler than that for HMXBs.

LMXBs exhibit many complex spectral and temporal behavior. It is a challenge
to explain these phenomena using theoretical models. However, most of the 
existing models for spectral fitting are Newtonian. 
But near the surface of a compact star,
the accretion flow is expected to be governed by the laws of
general relativity due to the presence of strong gravity. Therefore
general relativistic models should be used for the
purpose of fitting to get the correct best-fit values of the parameters.
Besides, the principal motivation behind the study of the spectral and
temporal behaviors of compact star LMXBs is to understand the properties
of very high $(\sim 10^{15}$ g cm$^{-3})$ density matter at the compact star
core (van der Klis 2000).
This is a fundamental problem of physics, which can not be addressed
by any kind of laboratory experiment. The only way to answer this question
is to assume an equation of state (EOS) model for the compact star core, to
calculate the structure parameters of the compact star and then to
calculate an appropriate spectral model. By fitting such models (for different
chosen EOSs) to the observational data, one can hope to constrain the existing
EOS models and hence to understand the properties of high density matter.
However, general relativistic calculation is essential to calculate the
structure parameters of a compact star and therefore to constrain the EOS
models.

It is expected that the compact stars in LMXBs are rapidly rotating due to
accretion-induced angular momentum transfer.
LMXBs are thought to be the progenitors of milli-second (ms) radio pulsars
(Bhattacharya \& van den Heuvel 1991) like PSR 1937+21 with $P \sim
1.56$~ms (Backer et al. 1982). The recent discovery of ms $(P \sim 2.49$~ms)
X--ray pulsations in XTE J1808-369 (Wijnands \& van der Klis 1998) has
strengthened this hypothesis. Therefore it is necessary to
calculate the structure of a rotating compact star considering the full effect
of general relativity. This was done by Cook et al.
(1994) and the same
procedure was used by Thampan \& Datta (1998)
to calculate the luminosities of the disk and the boundary layer.

In our work, we have calculated the structure parameters of a rapidly rotating 
neutron star and the metric coefficients in and around it. Then we have 
computed disk temperature profiles and disk spectra for various EOS models 
and many $(M, \Omega_{\rm *})$ combinations. We have considered the accretion 
disk to be geometrically thin and radiating locally like 
a blackbody. This may be the true case for 
luminous LMXBs, as shown by Mitsuda et al. (1984). These authors showed 
that the observed spectra of Sco X-1, 1608-52, GX 349+2 and GX 5-1, 
obtained with the {\it Tenma} satellite, can be well-fitted with the sum of a
multicolor blackbody spectrum (possibly from the accretion disk) and a single temperature
blackbody spectrum (believed to come from the boundary layer). 
Apart from the temperature profile, we have also 
calculated the disk luminosity and the boundary layer luminosity. Comparing the 
theoretical values of the luminosities and the 
disk temperature with the fitted values (fitted to 
{\it EXOSAT} data), we have constrained several properties of five LMXB sources: 
Cygnus X-2, XB 1820-30, GX 17+2, GX 9+1 and GX 349+2. We have also fitted the calculated
spectra with an analytical function and tried to distinguish between Newtonian 
and general relativistic spectra.

It has been known for many years that the neutron star may in fact be a `hybrid star' 
consisting of ordinary nuclear matter in the outer parts and quark matter in the 
central regions. This will be the case if {\it strange quark matter} (SQM; see Chapter 
7 for discussion) is metastable at zero pressure, being stabilized relative to 
hadronic matter by the high pressure within a neutron star (Baym \& Chin 1976; Chapline
\& Nauenberg 1976; Freedman \& McLerran 1978). If SQM is absolutely stable at the 
zero pressure, an even more intriguing possibility opens up, namely the existence of 
strange stars consisting completely of SQM (Witten 1984; Haensel, Zdunik \& Schaeffer 
1986; Alcock, Farhi \& Olinto 1986). 

The identification of a strange star will prove that the so called {\it strange matter 
hypothesis} is true. According to this hypothesis, strange quark matter, in 
equilibrium with respect to the weak interactions, could be the actual ground state of
strongly interacting matter rather than $^{56}$Fe. This is a fundamental problem 
of physics, which may be solved only if it is possible to distinguish between a 
neutron star and a strange star.

Strange stars are expected to behave quite differently from neutron stars due to 
an unusual equation of state. But, even then, it is very difficult to distinguish between 
them. For example, more massive neutron stars have the lower values of radii, while 
this relationship is opposite for strange stars. Nevertheless, for a value of 
gravitational mass equal to $1.4 \msun$ (the canonical mass for compact star 
candidates), the difference between the predicted radii of nonrotating configurations 
of a strange star and a neutron star comes out to be, 
at most, only about 5 km. It is very difficult to observe such a small 
value directly. Another distinction between strange stars and neutron stars was for a 
long time believed to be a much more rapid cooling of SQM due to neutrino emitting 
weak interactions involving the quarks (Alcock et al. 1986). Thus a strange star 
was presumed to be much colder than a neutron star of similar age, a signature 
potentially observable from X--ray satellites. But recently the story has been 
complicated considerably by the finding that ordinary neutron $\beta$-decay may be 
energetically allowed in nuclear matter (Lattimer et al. 1991), so that the cooling 
rate may be comparable to that of SQM. There are other possible ways (for example, 
study of pulsar glitches, oscillation and maximum rotation rate of the stars) 
to distinguish between these two kinds of stars (see Madsen 1998 for discussions).

In Chapter 7, we have computed the structure parameters of a strange star and 
calculated the corresponding disk temperature profiles and luminosities.
We have then compared these values with those for a neutron star and tried to 
distinguish between them.

In sections 8.2 and 8.3, we discuss the conclusions from the calculations of 
disk temperature profile and spectrum respectively. We give the summary of the work
with strange stars in section 8.4. In section 8.5, 
we discuss the future prospects and in 
section 8.6, we give the final conclusion.

\section{Disk Temperature Profile}

We have calculated (in Chapter 3) 
the temperature profiles of (thin) accretion disks around
rapidly rotating neutron stars (with low surface magnetic fields), taking
into account the full effects of general relativity. 
We have also computed the corresponding disk luminosities and boundary layer 
luminosities. All of these have been calculated as functions of $M$ and 
$\Omega_{\rm *}$ for various EOS models. 

It is important to notice (as shown in Fig. 3.3a) that the disk
temperature profiles do not have a monotonic behavior with respect to $\Omega_{\rm *}$.
This is a result of two mutually opposing effects: (1) the energy flux emitted from 
the disk increases with $\Omega_{\rm *}$ and (2) the nonmonotonic 
nature of the dependence of $r_{\rm in}$ on $\Omega_{\rm *}$ (see Fig. 3.1). Such a 
behavior of temperature profiles is reflected on disk spectra and one has to be careful
when trying to constrain $\Omega_{\rm *}$ by spectral fitting.

We have then considered a model (in Chapter 4)
for the spectrum of the X--ray emission from the disk, parameterized by
the mass accretion rate, the color temperature and the rotation rate
of the neutron star. We derive constraints on these parameters for the
LMXB sources: Cygnus X-2, XB 1820-30, GX 17+2, GX 9+1 and GX 349+2, 
using the estimates of the maximum temperature in
the disk along with the disk and boundary layer luminosities, taking the
spectrum inferred from the {\it EXOSAT} data. 

Our calculations suggest that the
neutron stars in Cygnus X-2 and GX 9+1 rotate close to the centrifugal
mass--shed limit. The LMXB source GX 349+2 also contains a rapidly rotating 
neutron star. This is in accord with the belief that LMXBs are the progenitors
of milli-second radio pulsars. Such a result also shows that the inclusion of 
rapid rotation in the neutron star structure-parameter-calculation is very important.
However at this point it should be mentioned that using a scalar theory of gravitation,
Papaloizou \& Pringle (1978) have concluded that 
a neutron star rotating with a frequency close to $1$~kHz may be unstable
to the radiation of gravitational waves by non-radial stellar modes. 

We have also discussed the possible constraints on the neutron star equation of state.
We could not actually rule out any EOS model, but looking at the QPO frequencies, 
gravitational masses $(M)$ and color factors $(f)$ for the sources, we have tried to 
conclude what kind EOS models could best represent the true EOS of a neutron star (see 
Chapter 4). 
According to our results, soft and stiff EOS models are unfavored, i.e., the EOS 
models with intermediate stiffnesses are supported.

In our work, we have not tried to model the observational temporal behaviors of the sources, 
and in particular, the QPO observations. Our results do not tally with the simple 
beat-frequency model. However, a pure beat-frequency model has been called into 
question because of several observations (see section 4.4 for a brief discussion).

In our analysis, we have assumed that the boundary layer does not affect the inner
region of the disk. This approximation will be valid when the boundary layer 
luminosity is smaller than the disk luminosity and the boundary layer extent is small 
compared to the radius of the star. This has been shown to be true for the chosen 
LMXB sources (see section 4.4).

It is to be remembered that in our calculations, we have neglected the effect of 
neutron star's magnetic field on the accretion flows. Therefore, our results are
valid if the Alf\'{v}en radius is less than the radius of the inner edge of the 
accretion disk. As we have seen in Chapter 4, this requires the upper limit of the 
surface magnetic field of the neutron star to be ~$\sim 10^8$~G, which is a reasonable 
value for LMXBs.

\section{Disk Spectrum}

We have computed X--ray spectra (in Chapter 5), as seen by a distant observer,
from the accretion disk around a rapidly rotating neutron star.
Our calculations have been carried out in a fully general relativistic
framework, with exact treatment of rotation. We have taken into
account the Doppler shift, gravitational redshift and light
bending effects in order to compute the observational spectrum. For this purpose,
we have computed the differential 
equations of motion for photons (Eqs. 5.7 - 5.11) 
in a space-time specified by the metric 
given by Eq. (2.3). Then the paths of the photons have been backtracked for the
calculation of the disk spectrum (see section 5.3 for the description of the 
procedure). We have calculated the spectrum as a function of $M$ and 
$\Omega_{\rm *}$ for various EOS models.

We have found that light-bending significantly modifies the high-energy
part of the spectrum. It can be seen (see Fig. 5.1) 
that the inclusion of light-bending effect enhances the predicted flux from the disk.
This is because, due to light-bending, the disk subtends 
a larger solid angle to the observer than it otherwise would.
We also see (from Fig. 5.2) that the spectrum, specially the high energy part, is very
sensitive to the accretion rate. This may be useful for constraining the accretion rate
by fitting the observational data with our model.

The inclination angle $i$ is a very important parameter in determining the shape of 
the spectrum. For lower energies, the observed flux is higher for lower values
of $i$, while this effect is opposite at higher energies due to Doppler blue shift.
We can also notice (from Fig. 5.4) that the behavior of the disk spectrum is not 
monotonic with $\Omega_{\rm *}$. This is expected from the non-monotonic behavior of 
the disk temperature profile. We can also expect to constrain EOS models by spectral
fitting, as the disk spectrum is fairly sensitive to the chosen EOSs. However, as the 
spectrum is a function of a large number of free parameters, it is very difficult 
to constrain the equations of state in a decisive way. But, it may actually be possible
with the data of new generation X--ray satellites with very good spectral resolution. 

The calculations presented here deal only with the thin Keplerian 
blackbody disk. In reality,
there may be other X--ray emitting components (boundary layer, accretion disk corona 
etc.) present in the LMXB source. 
In addition to that the disk may not be thin, Keplerian or a blackbody. Our results
will change for such cases. For example, our temperature profile and hence the 
spectrum will not be valid for a non-Keplerian disk. The effect of such uncertainty
of the nature of the source may be more important than the effect of general relativity
and rapid rotation. However, there is no competition between these two kinds
of effects. General relativistic modifications should be considered to calculate the 
spectra from all the X--ray emitting components to have the full general relativistic 
spectrum of a source. 
However, as this is a first step for this kind of work, we choose the simplest system,
i.e., a thin blackbody disk.

As mentioned in section 5.5, our results for non-rotating neutron stars did not 
match with those of the spectral fitting routine {\it GRAD}. With the help of 
Ebisawa \& Hanawa ({\it private communication}) we traced this mismatch to certain
simplifying approximations, as well as a couple of errors made in the {\it GRAD} code.

The computation of our model spectrum is rather time-consuming and therefore not 
quite suited to routine use.
To facilitate direct comparison with observations, we have presented
a simple empirical function which describes the numerically
computed relativistic spectra well. This empirical function (which
has three parameters including normalization) also describes the
Newtonian spectrum adequately. Thus the function can in principle be
used to distinguish between the two. In particular, the best-fit
value of one of the parameters ($\beta$-parameter) is $\approx 0.4$
for the Newtonian case, while it ranges from $0.1$ to $0.35$ for
relativistic case depending upon the inclination angle (if $i$ is not too high), 
EOS, spin rate and mass of the neutron star. Constraining this parameter by
fits to future observational data of X-ray binaries may, therefore, indicate
the effect of strong gravity in the observational spectrum.

\section{Strange Star}

We have computed the temperature profiles of accretion disks
around rapidly rotating strange stars, using constant gravitational
mass equilibrium sequences of these objects, considering the full
effect of general relativity. We have also calculated the corresponding disk 
luminosity, the boundary layer luminosity and the bulk structure 
parameters for the strange stars. These results have been compared with those for
neutron stars.

The striking feature of strange stars is the non-monotonic behavior of $\Omega_{\rm *}$
with $J$ such that $\Omega_{\rm *}^{\rm max}$ occurs at a lower value of $J$ than 
that of the mass-shed limit. For neutron stars, such a thing never occurs. Therefore, 
if an isolated sub-millisecond pulsar is observed to be spinning up, it is likely to 
be a strange star rather than a neutron star.

Because of higher values of $T/W$ $(\gsim 0.2)$, the tendancy of strange stars 
to be unstable to triaxial instability is larger than neutron stars in rapid 
rotation. Another interesting aspect of strange stars is that their gravitational
mass sequences (in contrast to the corresponding neutron star sequences) for 
$\Omega_{\rm *}$ and $\Omega_{\rm in}$ (see Chapter 7 for the meanings of the symbols used) 
cross at some point. This may have 
importance in kHz QPO modeling, as mentioned in Chapter 7. It can also be noticed that 
for strange stars, the accretion disk seldom touches the surface of the star (even 
for very high rotation rates), while for rapidly rotating neutron stars, 
the disk almost always extends upto the stellar surface. This may have important
observational effects as mentioned in Chapter 7.

Beyond a certain critical value
of stellar angular momentum ($J$), we observe the radius ($r_{\rm orb}$)
of the innermost stable circular orbit (ISCO) to increase with
$J$ (a property seen neither in rotating black holes nor in rotating neutron
stars). The reason for this is traced to the crucial dependence of
$dr_{\rm orb}/dJ$ on the rate of change of the radial gradient of the
Keplerian angular velocity at $r_{\rm orb}$ with respect to $J$.

The temperature profiles obtained are
compared with those of neutron stars, as an attempt to provide
signatures for distinguishing between the two. We show that when
the full gamut of strange star equation of state models, with
varying degrees of stiffness are considered, there exists a substantial
overlap in properties of both neutron stars and strange stars. However, we also notice that
neutron stars and strange stars exclusively occupy certain parameter
spaces. This result implies the possibility of distinguishing these
objects from each other by sensitive observations through future
X--ray detectors.

\section{Future Prospects}

Our work may be considered as a first step in understanding the effect of both 
general relativity and rapid rotation on the properties of accretion disk. For the sake
of simplicity, we have assumed the accretion disk to be thin and a blackbody. Now, 
as a future project, this work may be extended to the other kinds of disks (optically 
thin, geometrically thick, non-Keplerian etc.). The effect of rapid rotation can 
also be estimated for the other possible X--ray emitting regions, such as, boundary 
layer, disk corona etc., following the same procedure mentioned here.

We observe broadened iron lines from LMXBs such as Cygnus X--2, GX 349+2 etc.
An intrinsically narrow iron line emitted by an accretion
disk around a compact star is believed to be broadened and
skewed by Doppler effect and gravitational redshift. As a
result, the fitting of the line components of the observed
spectra by proper theoretical model should reveal the nature
of the flow of matter near the compact star's surface and help
us to constrain the equation of state of the compact star. It is, therefore, very important 
to calculate the structure of the broadened line as a function of the
compact star's angular speed and other parameters for various EOS models, using the 
metric given by Eq. (2.3).  

So far we have not considered the effect of the magnetic
field (of the compact star) on the accretion disk, assuming that
it has been decayed to a very small value $(\approx
10^8$ G). But the magnetic field may play some role in determining
the flow of the accreted matter near the surface of the compact
star. It is, therefore, instructive to look into this problem and include 
the effect of magnetic field in calculating the accretion disk spectrum.

In order to calculate the full spectrum of an LMXB source, it is necessary to compute 
the spectrum of the boundary layer. One should model this spectrum
(specially during the Type I Burst) using the metric coefficients
suitable for a rapidly rotating compact star. Addition of this spectrum
to the accretion disk spectrum (we have already calculated) may give
the full spectrum from LMXBs. 

For a compact star which is not rotating close to the mass-shed limit, the boundary
layer luminosity is fairly high. The inner region of the disk around such a compact
star may, therefore, be radiation pressure dominated. It is necessary to calculate
the spectrum from such a disk for the space-time geometry appropriate for a rapidly 
rotating compact star.

Once the spectrum of an LMXB is calculated considering all the major effects
described above, one can, in principle, fit the observational data by this model spectrum
and constrain the equation of state and the parameters of the source effectively.
However, computation of such a general relativistic spectrum is time-consuming
and rather unsuitable for the fitting procedure. Therefore, it is important
do a series of parametric fits to this spectrum for making it available for 
routine spectral fitting work.

\section{Final Conclusion}

The main purpose of the study of the properties of an LMXB with a neutron star or a 
strange star as the central accretor is to understand the properties of very high 
density matter and to address the question of the existence of strange quark matter.
These can be achieved only by fitting the observed spectral and temporal behavior
of such sources with appropriate theoretical models. It is not possible to 
incorporate all the important factors in such a model in a single work and one should
proceed step by step. In our work, we have included two major factors, namely, 
the effects of general relativity (essential for constraining EOS) 
and rapid rotation, which we think is a step forward towards our aim.

%% file: biblio.tex
\markright{}
\def\note #1]{{\bf #1]}}


%% file: pub.tex
\begin{center}
{\Huge List of Publications}
\end{center}
\thispagestyle{empty}

\vspace{-0.4cm}
\begin{center}
\noindent {\bf Appeared or In Press (In Refereed Journals)}
\end{center}
\vspace{-0.4cm}

\noindent
[1] Study of accretion discs around rapidly rotating neutron
stars in general relativity and the implications for four
Low Mass X--ray Binaries, Sudip Bhattacharyya, 2001, Astronomy 
and Astrophysics, in press (astro-ph/0112178).

\noindent
[2] Limits to the mass and the radius of the compact star in 
SAX J1808.4--3658 and their implications, Sudip Bhattacharyya, 2001, 
Astrophysical Journal Letters, {\bf 554}, 185.

\noindent
[3] Temperature profiles of accretion discs around rapidly rotating 
strange stars in general relativity: a comparison with neutron stars, 
Sudip Bhattacharyya, Arun V. Thampan \& Ignazio Bombaci, 2001, 
Astronomy and Astrophysics, {\bf 372}, 925.

\noindent
[4] General relativistic spectra of accretion discs around rapidly 
rotating neutron stars: Effect of light bending, 
Sudip Bhattacharyya, Dipankar Bhattacharya \& Arun V. Thampan, 
2001, Monthly Notices of the Royal Astronomical Society, {\bf 325}, 989.

\noindent
[5] General relativistic spectra of accretion disks around rotating 
neutron stars, Sudip Bhattacharyya, Ranjeev Misra \& Arun V. Thampan, 
2001, Astrophysical Journal, {\bf 550}, 841.

\noindent
[6] Temperature profiles of accretion disks around rapidly rotating 
neutron stars in general relativity and implications for Cygnus X-2, 
Sudip Bhattacharyya, Arun V. Thampan, Ranjeev Misra \& Bhaskar Datta, 
2000, Astrophysical Journal, {\bf 542}, 473.

\noindent
[7] Neutron stars and accretion disks in low mass X-ray binaries, 
Sudip Bhattacharyya, 1999, Indian Journal of Physics, {\bf 73B}, 889.

\vspace{-0.4cm}
\begin{center}
\noindent {\bf Submitted}
\end{center}
\vspace{-0.4cm}

\noindent
[1] Correlated X--ray timing and spectral behavior of GX 349+2: RXTE PCA data, 
Vivek K. Agrawal \& Sudip Bhattacharyya, 2001, Astronomy and Astrophysics, 
submitted (astro-ph/0112545).

\vspace{-0.4cm}
\begin{center}
\noindent {\bf Appeared or In Press (In Conference Proceedings)}
\end{center}
\vspace{-0.4cm}

\noindent
[1] Millisecond X--Ray Pulsar SAX J1808.4--3658: Limits to the Mass and the
Radius of the Compact Star, Sudip Bhattacharyya, 2001, in the proceedings 
of `5th Hellenic Astronomical Conference', Crete, Greece, September 20-22, 
2001 (astro-ph/0112175).

\noindent
[2] General Relativistic Spectra from Accretion Disks around Rapidly
Rotating Neutron Stars, Sudip Bhattacharyya, Dipankar Bhattacharya, 
Ranjeev Misra \& Arun V. Thampan, 2001, in the proceedings of `The 
Physics of Cataclysmic Variables and Related Objects', Goettingen, 
August 5-10, 2001 (astro-ph/0112136).